\documentclass{amsart}
\usepackage[margin=1.25in]{geometry}
\usepackage{amsmath}
\usepackage{amssymb}
\usepackage{mathrsfs}
\usepackage{graphicx}
\usepackage{booktabs}
\usepackage{url}

\usepackage{bbm}
\usepackage{mathtools}
\usepackage{stmaryrd}
\usepackage{enumitem}
\usepackage{tikz}
\usepackage{tikz-cd}
\usetikzlibrary{positioning}
\usepackage{proof}
\usepackage{tabularx}
\usepackage{thm-restate}
 \usetikzlibrary{arrows,arrows.meta}
  \usepackage{xcolor}
  \usepackage{wrapfig}

\usepackage{pifont}
\newcommand{\proofpen}{\raisebox{.2ex}{{\scriptsize\ding{46}}}}
\newcommand{\prooficon}[1]{\hyperref[proof:#1]{\proofpen}\;}
\newcommand{\proofappendix}[2][]{%
  \phantomsection\label{proof:#2}%
  \begingroup
    \renewcommand{\prooficon}[1]{%
      \hyperref[#1]{\raisebox{.2ex}{{\scriptsize$\hookleftarrow$}}}}%
    \csname #2\endcsname*%
  \endgroup}

\newcommand{\prooflink}[1]{%
  \begin{proof}\hyperref[proof:#1]{\textnormal{See the appendix.}}\end{proof}%
}


 \theoremstyle{plain}
 \newtheorem{theorem}{Theorem}[section]

 \theoremstyle{definition}
 \newtheorem{definition}[theorem]{Definition}
 \newtheorem{example}[theorem]{Example}
  \theoremstyle{remark}

\usepackage[colorlinks=true,allcolors=blue]{hyperref}
\usepackage{cleveref}

\makeatletter
\newsavebox{\@brx}
\newcommand{\llangle}[1][]{\savebox{\@brx}{\(\m@th{#1\langle}\)}%
  \mathopen{\copy\@brx\mkern2mu\kern-0.9\wd\@brx\usebox{\@brx}}}
\newcommand{\rrangle}[1][]{\savebox{\@brx}{\(\m@th{#1\rangle}\)}%
  \mathclose{\copy\@brx\mkern2mu\kern-0.9\wd\@brx\usebox{\@brx}}}
\makeatother

\newcommand{\define}{:=}

\newcommand{\imp}{\mathop{\supset}}

\newcommand{\N}{\mathbb{N}}

\newcommand{\pHA}{\text{\bf pHA}}
\newcommand{\ef}{\mathcal{E\!F}}

\newcommand{\pole}{\ensuremath{{\bot\!\!\!\bot}}}

\newcommand{\xle}[1]{\xrightarrow{\!#1\!}}
\newcommand{\eid}{e_{\mathtt{id}}}

\newcommand{\ecomp}[2]{#1 \mathop{;} #2}

\newcommand{\etrue}{e_{\scriptscriptstyle\top}}

\newcommand{\epair}[2]{\llangle #1, #2 \rrangle}

\newcommand{\efst}{e_{\mathtt{fst}}}

\newcommand{\esnd}{e_{\mathtt{snd}}}

\newcommand{\elambda}[1]{\lambda #1}

\newcommand{\eeval}{e_{\mathtt{eval}}}

\newcommand{\pair}[2]{\left\langle #1,#2\right\rangle}
\newcommand{\Cont}{\mathrm{Cont}}
\newcommand{\cc}{\mathrm{cc}}
\newcommand{\mcacc}{\mca[\cc]}
\newcommand{\sepcc}{\Separator[\cc]}
\newcommand{\MCId}{\effshell[\mathrm{Id}]}
\newcommand{\MCCont}{\effshell[\Cont]}
\newcommand{\MCContcc}{\effshell[\Cont]^\cc}
\newcommand{\Appcc}{\App[\cc]}
\newcommand{\cps}[1]{\llbracket #1 \rrbracket}
\newcommand{\callcc}{\mathsf{call/cc}}
\newcommand{\throw}[1]{\mathsf{throw}_{#1}}

\newcommand{\A}{\mathcal{A}}
\newcommand{\D}{\mathcal{D}}

\newcommand{\T}{\mathcal{T}}

\newcommand{\Set}{\text{\bf Set}}
\newcommand{\op}{^{op}}

\newcommand{\C}{\mathscr{C}}

\newcommand{\automath}[1]{\relax\ifmmode{#1}\else{$#1$}\fi}
\newcommand{\autorule}[1]{\relax\ifmmode{\scriptstyle(#1)}\else$(#1)$\fi}

\newcommand{\holds}{\mathsf{holds}}

\newcommand{\UFam}{\ensuremath{\mathsf{UFam}}}

\newcommand{\tripcart}{\ensuremath{\mathbf{Trip}_{\mathsf{cart}}}}
\newcommand{\tripreg}{\ensuremath{\mathbf{Trip}_{\mathsf{reg}}}}
\newcommand{\tripgeom}{\ensuremath{\mathbf{Trip}_{\mathsf{geom}}}}

\newcommand{\efreg}{\ensuremath{\mathbf{EF}_{\mathsf{reg}}}}

\DeclareMathOperator{\Sub}{Sub}
\newcommand{\mono}{\hookrightarrow}

\newcommand{\ev}{\mathrm{ev}}

\newcommand{\E}{\mathcal{E}}

\newcommand{\Boxmod}{\mathsf{\Box}}

\newcommand{\MCcat}{\mathbf{MC}}

\newcommand{\efD}{\ef_{\mathbf{C},S}}
\newcommand{\efboxm}{\ef^{\Boxmod}_{\mathbf{C},S}}
\newcommand{\efboxmj}{\ef^{j\Boxmod}_{\mathbf{C},S}}
\DeclareMathOperator{\Psh}{Psh}

\newcommand{\Sep}[1]{\mathcal{S}_{#1}}

\newcommand\evCD{\mathsf{b}_{\epsilon}}
\newcommand\EFC{\mathcal{EF}_{P}}

\newcommand{\xleC}[1]{\xrightarrow{\!#1\!}_P}
\newcommand\topC{\top_P}
\newcommand\andC{\wedge_P}
\newcommand\arrowC{\supset_P}
\newcommand\negP{\neg_P}
\newcommand\coprodP{\coprod_P}
\newcommand\Hom{\mathrm{Hom}}

\newcommand\vdashE[1]{\vdash_{#1}}
\newcommand\curry[1]{\lambda F}

\newcommand\ComplR[1]{#1^c}
\newcommand\IntR[1]{\mathrm{Int}(#1)}

\newcommand\arrowE{\supset}

\newcommand\eBind{e_{\mathtt{b}}}
\newcommand\eR{e_{\mathtt{r}}}

\newcommand{\effshell}[1][]{\mathcal{MC}_{#1}}
\newcommand{\mca}[1][]{\mathcal{A}_{#1}}

\newcommand{\modality}[1][]{\lozenge_{#1}}
\newcommand{\App}[1][]{\circledast_{#1}}

\newcommand{\after}[3]{\left\{ #1\,\leftarrow\,#2\,\mid\,#3\right\} }

\newcommand{\Sim}[1][]{\gamma_{#1}}
\newcommand{\SRealizer}{\mathsf{r}}

\newcommand{\predB}{\psi}

\newcommand{\xlej}[1]{\xrightarrow{\!#1\!}_j}

\newcommand\efj{\mathcal{EF}_{j}}

\newcommand\topj{\top_j}
\newcommand\andj{\wedge_j}
\newcommand\arrowj{\supset_j}
\newcommand{\eflax}{\ensuremath{\mathbf{EF}_{\mathsf{lax}}}}
\newcommand{\efcd}{\ensuremath{\mathbf{EF}_{\mathsf{geom}}}}
\newcommand{\eevalj}{e_{\mathtt{eval,j}}}

\newcommand{\interp}[1]{#1}

\newcommand{\TotCorrect}[3]{\exists #1 . #2 \downarrow #1 \sqcap #3}
\newcommand{\TotCorrectIgnore}[3]{\exists #1 . #2 \downarrow #1}

\newcommand{\Separator}[1][]{\mathcal{S}_{#1}}

\newcommand{\codef}[1]{\mathsf{#1}}

\newcommand{\Entail}[4][]{\forall #2 . #3 \Longrightarrow^{#1} #4}

\newcommand{\EletRib}[2]{#1\Leftarrow#2}
\newcommand{\Elets}[2]{\mathbf{let}\; #1 \;\mathbf{in}\;#2}%
\newcommand{\Elet}[3]{ \Elets{\EletRib{#1}{#2}}{#3} }%
\newcommand{\Eret}[2][]{\mathbf{return}_{#1}\,#2}%
\newcommand{\eLam}[3][]{\Lambda_{#1} #2 . #3}%
\newcommand{\eApp}[3][]{#2 \cdot^{#1} #3}%

\newcommand{\eAfter}[4][]{\left\{ #2\,\leftarrow\,#3\,\mid\,#4\right\}^{#1} }

\newcommand{\cS}[0]{\codef{S}}
\newcommand{\cK}[0]{\codef{K}}

\newcommand{\code}[1][]{\codef{c}_{#1}}
\newcommand{\codeI}[1][]{\codef{c}_{#1}}
\newcommand{\codeII}[1][]{\codef{d}_{#1}}

\newcommand{\paren}[1]{\left( #1 \right)}

\newcommand{\ncatfont}[1]{\mathbb{\mathbf{{#1}}}}%

\renewcommand{\N}{\mathbb{N}}%
\newcommand{\R}{\mathbb{R}}%

\renewcommand{\mca}[1][]{\mathcal{A}_{#1}}%

\newcommand{\strength}[1][]{\mathsf{t}_{#1}}%
\newcommand{\id}[1][]{\mathsf{id}_{#1}}

\newcommand{\PreOrd}{\ncatfont{Preord}}%

\newcommand{\cpair}[2]{ \triangleleft #1 , #2 \triangleright }

\newcommand{\exqed}{\hfill\ensuremath{\triangleleft}}

\newcommand{\term}[1][]{E_{#1}}

\newcommand{\dimg}[1]{#1^{*}}

\setlist[description]{font=\normalfont\itshape\bfseries,
                      labelindent=1.5em,
                      leftmargin=1.5em,
                      labelsep=0.4em,
                      itemsep=3pt}

\title{Evidence-Carrying Transport Across Effects}

\author{F\'{e}lix Castro, Liron Cohen, Ariel Grunfeld, \'{E}tienne Miquey}

\begin{document}


\begin{abstract}

A program logic proved sound against one effectful semantics need not remain sound when the effect is re-implemented, yet the semantic relationship between reasoning before and after a change of effect is usually left implicit and re-established by hand.
This paper provides a uniform, proof-relevant account of it.
Working in realizability, where a notion of computation determines a model whose entailments carry explicit program evidence, we introduce simulations between effectful realizability cores, generalizing Longley's applicative morphisms to computation over an arbitrary strong monad.
A single simulation transports the source logic
along a ladder: it induces a morphism of the associated evidenced frames
and, functorially, of the induced triposes. These passages are $2$-fully-faithful, so the three levels agree and a change of effect
carries the entire higher-order program logic together with the evidence
witnessing each derivation.
A computationally dense simulation moreover computes a modality on the source logic, a nucleus
through which the transport factors and which marks exactly the reasoning stable under the change,  read off
from the translation rather than postulated.
This recovers known examples such as the double-negation modality of classical realizability, and examples such as sheafification arise
from a modal refinement of the  construction.
Overall, the framework
makes a change of effect a first-class object, one that can be composed,
factored through its modality, and shown in cases not to exist, opening a
path to transporting verified reasoning across  effects.
\end{abstract}

\maketitle

\section{Introduction}
\label{sec:intro}

There are two natural ways to model a nondeterministic program. One
returns the set of all possible outcomes, the other reads a
scheduler that resolves each choice and returns a single outcome per
schedule. These are the same computation observed differently, and
on the specifications one usually cares about they validate exactly
the same reasoning. Yet no monad morphism relates them: naturality
already forbids a map from the powerset monad to the reader monad,
since a permutation of outcomes, which such a map would have to
respect, has no canonical scheduler to be sent to.
This simple example is an instance of a pervasive situation. Effectful
programs rarely stay put: state is threaded as an argument,
nondeterminism is compiled to lists and backtracking, control is
passed to continuations~\cite{moggi1991notions}. Such changes of
effect are the daily business of compilation, program
transformation, and verification, and each comes with a standard
obligation: behaviour must be preserved.
This paper is about the
other obligation: does the
\emph{reasoning} survive?
In current practice the answer is
re-established by hand, one change of effect at a time, typically by a relational argument engineered for the specific pair of effects, e.g.~\cite{Reynolds83,BentonHur2009}, and when it
is positive, no structural map of the underlying monads explains
why. The change of effect itself is not a mathematical object: there
is no notion of morphism between effect disciplines along which
logic could be transported.
This paper provides one, and shows that
it uniformly transports everything: propositions, entailments, and the
programs witnessing them.

Answering such a question uniformly requires a setting in which
computation and the reasoning it supports are two readings of a
single object. That setting is realizability: a model of computation
determines a model of logic in which each proposition is a
specification on programs and each entailment is witnessed by a
program transforming realizers of the premise into realizers of the
conclusion. For \emph{pure} computation this construction is a
well-understood ladder. A partial combinatory algebra generates a
realizability tripos and hence a topos~\cite{Vanoosten_08}, with
evidenced frames recently isolated as the logical level through
which the passage factors~\cite{EF2021}. The ladder is moreover
functorial: Longley's applicative morphisms relate algebras at the
bottom and induce morphisms of the triposes above, computationally
dense ones induce geometric
morphisms~\cite{Longley95,Hofstra06relative,FabervanOosten2014}, and
nuclei on the algebra correspond to
subtriposes~\cite{Frey_Streicher_2021,tarantino2026}.

Computation, however, is rarely pure. The base of the ladder was
recently extended to effectful computation: monadic combinatory
algebras are combinatory algebras over an arbitrary strong monad
$M$, equipped with a modality specifying how effectful outcomes are
observed~\cite{MCA25}, and the logic such a core generates is again
an evidenced frame. Yet the effectful ladder this provides is
incomplete in two respects. First, it is anchored to $\Set$, whereas the
pure theory lives happily in any topos. More fundamentally, it has
objects but no morphisms: the changes of effect of the opening
paragraph are situations the theory cannot even state.

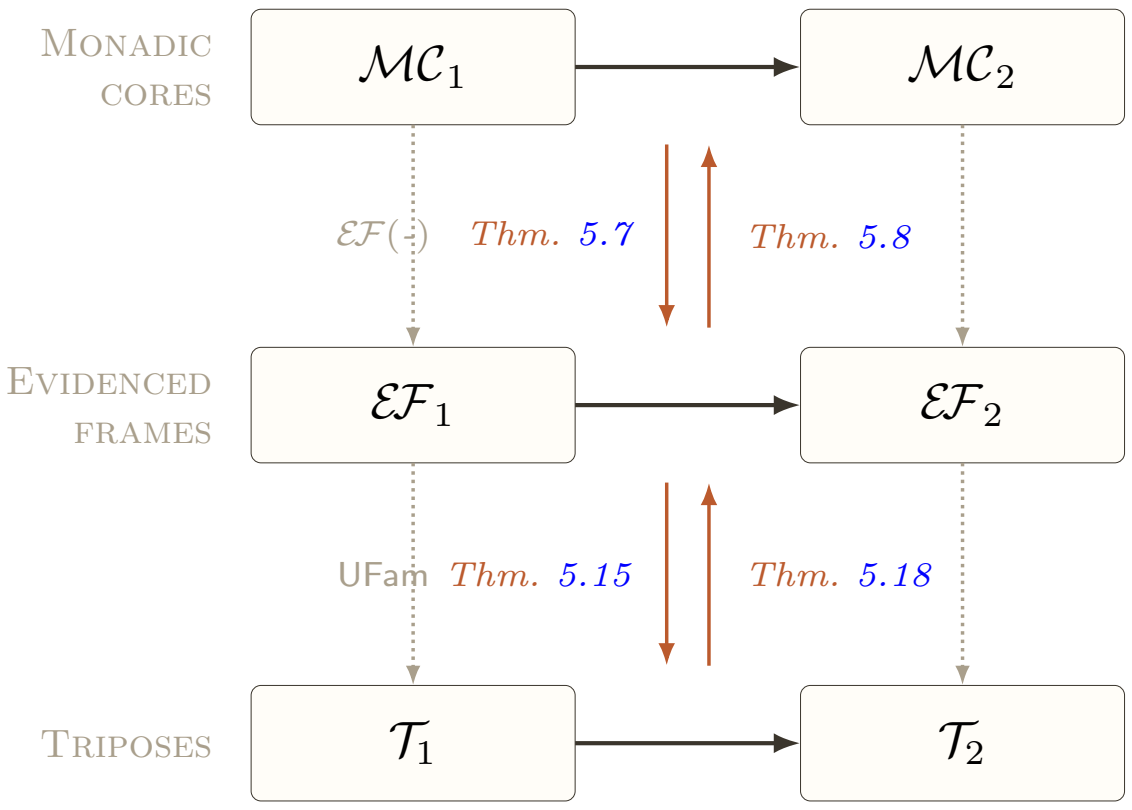
\definecolor{tlladderink}{HTML}{3A352C}
\definecolor{tlladderacc}{HTML}{BB5A2D}
\definecolor{tlladdergry}{HTML}{A99F8C}
\definecolor{tlladderfil}{HTML}{FFFDF8}

\begin{wrapfigure}{r}{0.50\linewidth}
  \centering
  \resizebox{\linewidth}{!}{%
  \begin{tikzpicture}[
    x=1cm, y=1cm,
    box/.style   ={draw=tlladderink, fill=tlladderfil, line width=.pt,
                   rounded corners=2pt, minimum width=2.3cm, minimum height=.82cm,
                   inner sep=2pt},
    rlab/.style  ={text=tlladdergry, font=\footnotesize\scshape, anchor=east},
    sim/.style   ={-{Latex[length=1.9mm]}, tlladderink, line width=.85pt},
    slab/.style  ={font=\itshape, midway, above=1pt, text=tlladderink},
    con/.style   ={-{Latex[length=1.4mm]}, tlladdergry, line width=.7pt,
                   densely dotted},
    clab/.style  ={font=\scriptsize\itshape, text=tlladdergry},
    rlab/.style = {text=tlladdergry, font=\footnotesize\scshape,
               anchor=east, align=right},
    thm/.style   ={-{Latex[length=1.6mm]}, tlladderacc, line width=.7pt},
    tlab/.style  ={font=\scriptsize\itshape, text=tlladderacc}
  ]
    \node[box] (mc1) at (0,0)      {$\effshell[1]$};
    \node[box] (mc2) at (3.9,0)    {$\effshell[2]$};
    \node[box] (ef1) at (0,-2.4)   {$\ef_1$};
    \node[box] (ef2) at (3.9,-2.4) {$\ef_2$};
    \node[box] (t1)  at (0,-4.8)   {$\mathcal{T}_1$};
    \node[box] (t2)  at (3.9,-4.8) {$\mathcal{T}_2$};

\node[rlab] at (-1.3,0)    {Monadic\\cores};
\node[rlab] at (-1.3,-2.4) {Evidenced\\frames};
\node[rlab] at (-1.3,-4.8) {Triposes};

    \draw[sim] (mc1) -- (mc2) node[slab]{};
    \draw[sim] (ef1) -- (ef2) node[slab]{};
    \draw[sim] (t1)  -- (t2)  node[slab]{};

    \draw[con] (mc1) -- (ef1) node[clab, pos=.5, xshift=-6pt]{$\ef(\text{-})$};
    \draw[con] (ef1) -- (t1)  node[clab, pos=.5, xshift=-6pt]{$\UFam$};
    \draw[con] (mc2) -- (ef2);
    \draw[con] (ef2) -- (t2);

    \draw[thm] (1.80,-0.55) -- (1.80,-1.85);
    \node[tlab, anchor=east] at (1.67,-1.2)
      {Thm.~\ref{thm:mca-simulation-induces-ef-simulation}};
    \draw[thm] (2.10,-1.85) -- (2.10,-0.55);
    \node[tlab, anchor=west] at (2.23,-1.2)
      {Thm.~\ref{thm:ef-simulation-induces-mca-simulation}};

    \draw[thm] (1.80,-2.95) -- (1.80,-4.25);
    \node[tlab, anchor=east] at (1.67,-3.6)
      {Thm.~\ref{thm:ef-simulation-induces-model-morphism}};
    \draw[thm] (2.10,-4.25) -- (2.10,-2.95);
    \node[tlab, anchor=west] at (2.23,-3.6)
      {Thm.~\ref{prop:UFamisom}};
  \end{tikzpicture}
}
  \label{fig:transport-ladder}
\end{wrapfigure}

This paper completes the ladder in both respects.
First, we rebuild
it over an arbitrary base topos $\E$ at every level: monadic cores,
their evidenced frames, and the induced triposes, so that presheaf
and sheaf models of computation become instances of the theory
rather than extensions of it.
Second, and centrally, we make the
effectful ladder functorial by equipping its bottom level with the
right notion of morphism.
A \emph{simulation} between monadic cores
is a relation on codes, compatible with effectful application in a
way that preserves every postcondition the source logic can express, generalizing  Longley's applicative morphism~\cite{Longley95}.
Our central result
is that this
bottom-level notion determines the morphisms at every level above:
the assignments from monadic cores to evidenced frames to triposes
are $2$-fully-faithful, so a simulation of cores, a regular morphism
of the induced evidenced frames, and a morphism of the generated
triposes each determine the others.
A change of effect thereby
becomes a first-class morphism, and it transports the entire
higher-order program logic a core generates. Because entailments in
an evidenced frame are witnessed by programs, this transport is
effective: the translator carried by a simulation turns source
evidence into target evidence, so a proof crosses a change of effect
by running a program on it, the way a compiler carries a program
across one.

The notion of morphism is the first place where the effectful
setting departs from the pure one. The obvious candidate, a morphism
of the underlying monads, is the wrong shape: as the opening example shows, such morphisms often do not exist between the monads that arise in practice, even when the two models are logically indistinguishable.
Simulations are accordingly relational and lax.
When a monad morphism does exist and agrees with codes and
observation, we obtain the \emph{algebraic} fragment of simulations, in which the global transport obligation reduces to local equations on the generating data.
A rich toolbox  relates effectful programs \textit{within} a fixed semantics: relators generalize applicative
bisimilarity to  algebraic effects~\cite{DalLago2017},
$\top\top$-lifting constructs logical relations for a given
monad~\cite{Katsumata2005}, modalities indexing admissible
observations yield logics for algebraic
effects~\cite{MatacheStaton2019}, and step-indexed logics such as Iris reason about programs of one fixed effectful
language~\cite{Jung2018}.
Simulations share the relational, lax
character of these tools but relate two \textit{distinct} semantics, carrying the entailment relation together with its evidence from one to the other.

The correspondence has consequences beyond transport. Among
simulations we single out the \emph{computationally dense} ones,
those whose transport $F$ admits a right adjoint $H$.
In the pure
case density is exactly what makes a map induce a geometric
morphism~\cite{Longley95,FabervanOosten2014}, and we prove the
effectful analogue.
Density also pays a dividend: the round trip
$HF$ is a \emph{nucleus} on the source logic, the realizability form
of a Lawvere-Tierney modality, and the transport factors through the
modal evidenced frame it cuts out.
The modality is thus not an axiom
we add but something the change of effect computes: it marks exactly
the fragment of the source logic that survives the change. In
a sense, the nucleus is the observation policy the target
enforces on the source, the abstraction barrier a re-implementation
imposes. Instantiating the construction recovers the double-negation
modality of classical realizability~\cite{OlivaStreicher08,Krivine09} and
representation independence as cases of one mechanism. Over $\Set$
we prove the converse, identifying the computable modalities with
the nuclei of the evidenced frame, and sheafification
arises from a modal refinement of the same construction that reaches
stage-dependent observation policies. Finally, making simulation
formal makes impossibility provable: a no-go theorem shows that no
arithmetic-preserving simulation relates angelic to demonic
observation, so the boundary of transport is itself a theorem.

\paragraph{Contributions}
Concretely, this paper:
\begin{itemize}
  \item lifts the effectful realizability ladder from $\Set$ to an
    arbitrary base topos $\E$
  \item introduces \emph{simulations between monadic cores},
    generalizing Longley's applicative morphisms to an arbitrary
    strong monad
  \item identifies the \emph{algebraic fragment}, the sub-case
    factoring through a strong monad morphism, which reduces the
    global transport obligation to local equations per effect
    operation
  \item establishes the \emph{transport correspondence}: the
    assignments from monadic cores to evidenced frames to triposes
    are $2$-fully-faithful functors, so a change of effect transports
    the higher-order program logic together with its evidence
  \item proves in $\Set$ that computationally dense simulations are exactly
    those admitting a right adjoint, that their round trip $HF$ is a
    nucleus through which the transport factors
\end{itemize}

\paragraph{Outline}
\Cref{sec:overview} previews the entire development on an illustrative
change of effect: continuation-passing.
\Cref{sec:premin}  fixes the internal-topos setting, and
\Cref{sec:mca-simulations} defines monadic
cores and their simulations, with the algebraic fragment
(\Cref{sec:algsim}).
\Cref{sec:ef-sims-main} establishes the transport correspondence across
evidenced frames and triposes.
\Cref{sec:transport-modalities} develops
the modality a change of effect computes.
\Cref{sec:related} discusses related work and \Cref{sec:conc} concludes.
Due to space constraints, most proofs are deferred to the appendix. The marker \proofpen{} next to
a statement links to its proof.

\section{A Worked Change of Effect: Continuation Passing}
\label{sec:overview}
We preview the whole development on a single, familiar change of effect:
implementing a language in a continuation-passing style~\cite{FelleisenFriedman87,SabryFelleisen92,Danvy_Filinski_1992}.
CPS translations are a textbook compilation step~\cite{Appel_1991}, and while they can be shown to preserve operational behavior and typing, this still does not provide methods to study the transport of specifications along CPS translation.

\paragraph{Running example: continuations}
Consider the usual $\lambda$-calculus which, through Curry-Howard, corresponds to intuitionistic logic.
A CPS translation $\cps{\cdot} : \Lambda \to \Lambda$ makes control flow explicit: every term expects a continuation reifying the future of the computation, so a value returns itself to the continuation,
$\cps{v} = \lambda k.\,k\,v$, which is exactly the unit $\eta(v)$ of
the \emph{continuation monad}.
So far, the source calculus $\Lambda$ and its image $\cps{\Lambda}$ are just two sides of the same coin, since they share the exact same computational expressivity.
Nonetheless, in the target, the continuation-passing discipline allows for more computational features in interaction with translated terms, such as backtracking.
Namely, it is possible to define a term in $\Lambda$ which, in interaction with terms CPS translated in $\cps{\Lambda}$, behaves as if it were the translation of $\callcc$. In other words, the source $\lambda$-calculus can safely be extended with control operators (such as $\callcc$~\cite{Krivine09}, $\mathsf{catch}$ and $\throw{}$~\cite{Crolard99}, etc.), yielding a calculus $\Lambda_C\supset \Lambda$ and such that the translation can be extended to $\cps{\cdot}:\Lambda_C \to \Lambda$.
In terms of the continuation monad, this means that the set of terms on which the continuation monad is defined might as well be extended to account for $\callcc$ and $\throw k$. Through Curry-Howard interpretations, such a calculus now corresponds to classical logic.
Summarizing, we mentioned three different calculi: (1) the usual $\lambda$-calculus $\Lambda$, which can be observed through (2) its CPS translation $\cps{\Lambda}$, and which can itself be extended to (3) the image of a calculus with control $\cps{\Lambda_C}$.

\paragraph{Monadic cores}
In our framework each calculus becomes a monadic core, that is,  a monadic combinatory algebra over a strong monad with a modality specifying how effectful
outcomes are observed~\cite{MCA25}.
In the realm of MCA, the continuation monad allows us to capture Krivine classical realizability~\cite{MCA25,Krivine09,OlivaStreicher08} by means of a modality expressing the orthogonality relation induced by a pole in Krivine's work (see \Cref{ex:continuation}).
The three calculi evoked earlier along the CPS translation correspond now to three different monadic cores:
\begin{itemize}
\item[(1)] a monadic core $\MCId$ defined by the identity monad over a pure combinatory algebra,
\item[(2)] $\MCCont$ defined using the continuation monad over the same codes,
\item[(3)] $\MCContcc$ using the continuation monad over an extended set of codes  with control operators.
\end{itemize}

\paragraph{Evidenced frames}The initial question can now be made more precise: what is the relation between these monadic cores, and is it preserved on the associated induced semantics?  Intuitively, $\MCId$ and $\MCCont$ and the semantics they induce are tightly related, while $\MCCont$ corresponds to a fragment of $\MCContcc$, hence the specifications satisfied by the former should transport to the latter. This intuition is made precise in~\Cref{ex:continuation2,ex:continuation3}:
each monadic core induces a corresponding evidenced frame and then a tripos, and simulations between monadic cores induce corresponding simulations between the associated evidenced frames and morphisms between the associated triposes (see \Cref{fig:recap-strata}).
Specifically, the \emph{computationally dense} simulation from $\MCId$ to $\MCCont$ gives rise to a geometric simulation of the corresponding evidenced frames and a geometric
transformation between the associated triposes, while the  simulation from $\MCCont$ to $\MCContcc$ induces a regular simulation between the associated evidenced frames and a regular morphism of triposes.

\paragraph{Nuclei} Just like any Lawvere-Tierney topology on a topos induces a subtopos, from an evidenced frame $\ef$ and a nucleus one defines a new evidenced frame $\ef_j$~(\Cref{prop:appx-dermod-1}).
In the case of continuations, the geometric simulation from $\ef(\MCId)$ to $\ef(\MCCont)$ thus yields a nucleus $j$ such that $\ef(\MCId)_j$ is isomorphic to~$\ef(\MCCont)$~(\Cref{ex:continuation4}).

\section{Preliminaries: Reasoning Internally in a Topos}\label{sec:premin}

Realizability models are routinely built over bases other than $\Set$, such
as presheaf and sheaf toposes and realizability toposes. To provide a
unifying setting we develop the effectful frameworks of this paper
internally to an arbitrary elementary topos $\E$, which requires separating
the meta-theory from the topos's internal logic. This section fixes the notions we
use.

Every  statement is in the internal
(Mitchell--B\'enabou) language of $\E$~\cite{maclane1992sheaves}, whose
subobject classifier we write $\Omega$.
A \emph{type} is an object of $\E$,
a \emph{term in context} $x_1 : X_1, \dots, x_n : X_n \vdash \term : Y$ is a
morphism $X_1 \times \cdots \times X_n \to Y$, and a \emph{formula} is a
term of type $\Omega$.
We reason in generalized-element
(``set-theoretic'') notation, writing $x : X$ for a variable and arguing
with the intuitionistic connectives $\top, \bot, \sqcap, \sqcup,
\Rightarrow, \neg$, the quantifiers $\forall, \exists$, and the subobject
order $\sqsubseteq$ carried by the Heyting structure of $\Omega$. Such
reasoning is sound over any topos, so we identify
an internal term with the morphism it denotes.
For an internal function $\Gamma \vdash f : B^A$
and a generalized element $\Gamma \vdash a:A$, we write $\Gamma \vdash f(a) : B$ for
the morphism $\ev_{A,B} \circ \langle f , a \rangle$ where
$\ev : B^A \times A \to B$ is the evaluation morphism.
For internal predicates
$\varphi, \psi : \Omega^X$, $\varphi \sqsubseteq \psi$ means
$\forall x : X.\; \varphi(x) \Rightarrow \psi(x)$ internally, and for a
closed $\varphi$ we write $\vdashE{\E} \varphi$ when its characteristic
morphism is $\top$.

For a subobject $\iota : S \rightarrowtail A$ we write $\chi_S : A \to
\Omega$ for its characteristic morphism, $\{\, \vec{x} \mid \varphi(\vec{x})
\,\}$ for the subobject classified by $\varphi$, and $\{ \_ \}_A$ for the
singleton map. For a relation $R : A \times B \to \Omega$ and a predicate
$\varphi : A \to \Omega$, the \emph{direct image of $\varphi$ along $R$} is
$R^*\varphi \define \{\, y \mid \exists x : A.\; R(x,y) \sqcap \varphi(x)
\,\}$, and $\bigcup_A : \Omega^{\Omega^A} \to \Omega^A$, the multiplication
of the powerset monad, is $\bigcup_A \vec{\phi} \define \{\, x \mid \exists
\phi \in \vec{\phi}.\; x \in \phi \,\}$.

Taking $\E = \Set$ throughout loses nothing of the main development: a
proposition is an ordinary subset, a realizer an element, and every
construction reads as elementary set theory. Working over an arbitrary
topos makes the results model-independent and admits  richer
models.

The one construct we add to the internal language is the effectful
sequencing of a strong monad, used pervasively below. Fix a strong monad
$(M, \eta, \mu)$ on $\E$ with strength $\strength$. For $\Gamma \vdash \term:
X$ we write $\Gamma \vdash \Eret{\term} : M X$ for $\eta_X(\term)$, and for
$\Gamma \vdash \term[1] : M X_1$ and $\Gamma, x : X_1 \vdash \term[2] : M X_2$ the
bind $\Gamma \vdash \Elet{x}{\term[1]}{\term[2]} : M X_2$ is
$\mu_{X_2} \circ M(\term[2]) \circ \strength[\Gamma, X_1] \circ
\pair{\id[\Gamma]}{\term[1]}$. These satisfy the usual monad laws.
The exponential transpose
of the morphism $M (\ev_{A,B}) \circ \strength : B^A \times M(A) \to M(B)$
gives a map $ B^A \to M(B)^{M(A)}$. For $\Gamma \vdash f : B^A$,
we use this map to define a generalized element $\Gamma \vdash M f : M(B)^{M(A)}$.

\section{Monadic Cores and Simulations}\label{sec:mca-simulations}
Monadic cores were introduced in~\cite{MCA25} as the semantic objects of
effectful realizability.
A monadic core bundles a computational effect, a
space of realizers, and a way of observing effectful computations as meeting
specifications.
It can be read as an untyped language of evidence terms with an effectful operational semantics: the object of codes plays the role of
programs, application may exhibit effects, and the modality says when an
effectful computation satisfies a postcondition. The logical interface stays
fixed while the operational meaning of evidence changes with the effect.

Here we make cores the objects of a category by equipping them with
morphisms, \emph{simulations}. A simulation relates the realizers of a source
core to those of a target core, compatibly with effectful application, so
that target execution preserves every postcondition the source can express,
the semantic counterpart of an effect-aware implementation such as a
state-passing or list-based translation. What matters is not only that
programs translate but that the evidence witnessing logical reasoning
survives. For ordinary partial combinatory algebras a simulation specializes
to Longley's applicative morphism, and we isolate an algebraic fragment
induced by monad morphisms. A simulation is thus a structural criterion for
when a change of effect preserves the evidenced higher-order logic the source
core generates, turning a change of effect into a mathematical object rather
than a case-by-case argument.

\subsection{Monadic Cores}\label{subsec:monadic-cores}
A monadic core bundles four ingredients: an effect, given by a strong monad $M$, an observation policy, given by a \emph{modality} $\modality$, an untyped object of codes $\mca$ closed under an effectful application, and a \emph{separator} $\Separator$ marking which codes count as evidence.
We recall each component in turn, generalizing the $\Set$-based definition
of~\cite{MCA25} to an arbitrary topos $\E$, and assemble them in Definition~\ref{def:monadic-core}.
Throughout, the reader may keep in mind a single running example: \emph{demonic nondeterminism}, where a code applied to an argument returns a \emph{set} of possible outcomes, and a specification is satisfied only if \emph{all} outcomes satisfy it (\Cref{ex:demonic-core}).

\begin{definition}[$M$-Modality]
\label{def:modality}
Let $M : \E \to \E$ be a strong monad with unit $\eta$, multiplication $\mu$, and strength $\strength$.
An \textit{$M$-modality} is a family of morphisms $\modality[X] : \Omega^X \Longrightarrow \Omega^{M X}$,
satisfying for all objects $X$ and $Y$:

\begin{tabular}{lll}
     \textbf{\emph{Monotonicity:}} & $\vdashE{\E}
     \forall \varphi, \psi \!:\! \Omega^X. $ & $ \varphi \sqsubseteq \psi \; \Longrightarrow\;
       \modality \varphi \sqsubseteq \modality \psi$\\
     \textbf{\emph{Unit:}} & $\vdashE{\E} \forall \varphi \!:\! \Omega^X. $&$ \varphi \sqsubseteq \modality \varphi \circ \eta_{X}$\\
     \textbf{\emph{Bind:}} & $\vdashE{\E} \forall \varphi \!:\! \Omega^Y . \forall f \!:\! (M Y)^X. $&$ \modality \left(\modality \varphi \circ f\right) \sqsubseteq \modality \varphi \circ f^{*}$\\
     \textbf{\emph{Strength:}} & $\vdashE{\E} \forall \varphi_1 \!:\! \Omega^{X},
     \varphi_2 \!:\! \Omega^{Y}. $&$ \sqcap \circ (\varphi_1 \times \modality \varphi_2) \sqsubseteq \modality \left( \varphi_1 \times \varphi_2 \right) \circ \strength[X,Y]$
\end{tabular}

where
$\sqcap$ is the conjunction operator, $f^{*} \!:\! (M Y)^{(M X)}$ is the internal
Kleisli extension of $f$, and
$t_{A,B} \!:\! A \times M B \to M\left(A \times B\right)$ is the left strength of $M$.

The \textbf{\emph{bind}} law is equivalent to the combination of the following two laws:

\begin{tabular}{lll}
\textbf{\emph{Composition:}} & $\vdashE{\E} \forall \varphi \!:\! \Omega^X. $ & $ \modality \modality \varphi \sqsubseteq \modality \varphi \circ \mu_{X}$\\
\textbf{\emph{Lax Naturality:}} & $\vdashE{\E} \forall \varphi \!:\! \Omega^Y. \forall f \!:\! Y^X. $&$\modality\left(\varphi \circ f\right) \sqsubseteq \modality\varphi \circ M f$\\
\end{tabular}

\end{definition}
The intended reading is operational: while semantics such as realizability defined postconditions in terms of values or pure computations, modalities allow us to lift such semantics to effectful computations.
Given $\Gamma \vdash \term : M X$, and $\Gamma, x:X \vdash \varphi$, a modality allows us to consider the internal formula $\Gamma \vdash \after{x}{\term}{\varphi\paren{x}}$,
defined by the morphism $\interp{\after{x}{\term}{\varphi\paren{x}}} :=  \modality \varphi \circ \strength[\Gamma,X] \circ \pair{\id[\Gamma]}{\term}$,
which intuitively expresses that, after running $\term$, the resulting value satisfies $\varphi$.

For partiality with total correctness, $\after{x}{\term}{\varphi\paren{x}}$ says that the computation terminates with a value satisfying $\varphi$.  For example, for  nondeterminism it can express may or must satisfaction, and for step-indexed or modal models it can express satisfaction only after applying the chosen stability policy.
Concretely, monotonicity says the modality respects the logical ordering: a weaker postcondition
is easier to satisfy.
The unit axiom says that returning a value satisfying $\varphi$ already suffices,
so a pure return is at least as strong as an effectful computation.
The composition axiom makes the modality compatible with monadic sequencing:
a computation that produces a sub-computation satisfying $\varphi$ itself satisfies $\varphi$.
The strength axiom allows a prior assumption to be carried into the postcondition:
if $\psi$ holds in the current context and the computation satisfies $\varphi$,
then the computation satisfies $\psi \sqcap \varphi$.

\begin{definition}[Monadic Applicative Structure]
For a strong monad $M : \E \to \E$, a \emph{monadic applicative structure} (MAS) over $M$ is an object $\mca$ with an \emph{application} morphism
$ \App : \mca \times \mca \to M \mca  $.
\end{definition}

Given a MAS $\mca$, we add the following to the internal language:
$\interp{\eApp{E_1}{E_2}} := \App \circ \pair{\interp{E_1}}{\interp{E_2}}$.
Thus, applying one code to another may perform effects, characterized by the monad $M$.  This is the main point of passing from ordinary PCAs to MCAs.
Here, a \emph{pure} morphism means an effect-free one, $\mca^n \to \mca$, which can be lifted to a computation via $\eta$.
The $\App$-monomials are the effectful operations built from application, constants, identities, and sequencing.  They are the effectful analogue of the applicative terms generated by application in a PCA.

\begin{definition}[$\App$-Term of $\mca$] \label{def:appterm}
A term $\Gamma \vdash \term : M\mca$ in the internal language is called a $\App$-term~(of~$\mca$) if $\Gamma$ is of the form $\Gamma = x_1 : \mca , x_2 : \mca , \ldots , x_n : \mca$ and $\term$ is one of the following:
\begin{enumerate}
    \item $\Eret{x_k}$ where $x_k \in \Gamma$

    \item An application term $\eApp{x_i}{x_j}$ where $x_i,x_j \in \Gamma$
    \item Binding of $\mca$-terms. That is, $\Elet{x}{\term[1]}{\term[2]}$ where $\term[1]$ and $\term[2]$ are $\App$-terms
\end{enumerate}
\end{definition}

Note that the language of $\App$-terms does not have closed terms with no free variables, but it is sufficient for our purpose.

\begin{definition}[$\App$-Monomial]
A $\App$-monomial of arity $n$ is a morphism $f :\mca^{n} \to M\mca$ such that there exists an $\mca$-context $\Gamma = x_1 :
\mca , \ldots , x_n : \mca$ and a $\App$-term $\Gamma \vdash \term : M \mca$ such that $\interp{\term} = f$.
\end{definition}

\begin{definition}[$k$-ary Application]
The application operator $\App : \mca^2 \to M \mca$ is generalized to $\App^k : \mca^{k+1} \to M \mca$ by recursion on $k$:
\begin{align*}
    \App^0 &:= \eta \\
    \App^{k+1} &:= \mu_{\mca} \circ M \App \circ t^{L}_{A,A} \circ \App^{k} \circ \alpha^{k}_{\mca}
\end{align*}
where $t^{L}_{X,Y} : M X \times Y \to M \left(X \times Y\right)$ is the left strength, and $\alpha^k_{X} : X \times X^k \to X^k \times X$ is the $k$'th iteration of the associator of $\times$.

Using the bind added to the internal language of the topos
as described in Section~\ref{sec:premin}:
$$\App^{n} \circ \langle \mathsf c, \mathsf c_1 , \ldots , \mathsf c_n \rangle =
\Elets{\EletRib{x_1}{\mathsf c \cdot \mathsf c_1} \; ; \;
  \EletRib{x_2}{x_1 \cdot \mathsf c_2} \; ; \; \ldots  ;
  \EletRib{x_{n-1}}{x_{n - 2} \cdot \mathsf c_{n-1}}}{x_{n - 1} \cdot \mathsf c_n}
$$
\end{definition}

\begin{definition}[Monadic Combinatory Algebra]\label{def:mca}
A MAS $\mca$ over a monad $M : \E \to \E$ is said to be a \emph{monadic combinatory algebra (MCA)} when
for every $\App$-monomial $f$ of arity $n > 0$ there exists a pure global element  $e_f : \boldsymbol{1} \to \mca$
and, for every $k < n$, a pure morphism $e_f^k : \mca^k \to \mca$ such that the following diagrams commute.

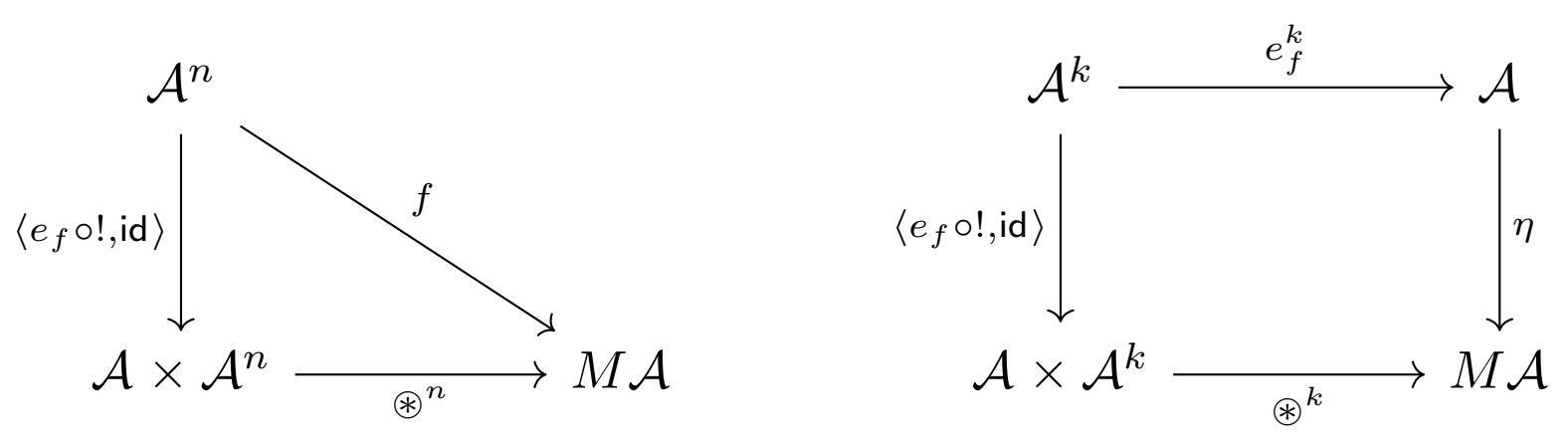
\[\begin{tikzcd}
	{\mca^{n}} && && {\mca^{k}} && {\mca} \\
	\\
	{\mca\times\mca^{n}} && {M \mca} && {\mca\times\mca^{k}} && {M \mca}
	\arrow["{\pair{e_f \circ !}{\id}}"', from=1-1, to=3-1]
	\arrow["f", from=1-1, to=3-3]
	\arrow["{\pair{e_f \circ !}{\id}}"', from=1-5, to=3-5]
    \arrow["{e_f^k}", from=1-5, to=1-7]
	\arrow["{\eta}", from=1-7, to=3-7]
	\arrow["{\circledast^n}"', from=3-1, to=3-3]
	\arrow["{\circledast^k}"', from=3-5, to=3-7]
\end{tikzcd}\]

\end{definition}

This is the combinatory completeness condition.  It says that every effectful operation with one distinguished input can be represented by a code.  In programming language terms, the untyped evidence language has enough abstraction: effectful computations built from application and sequencing can be internalized as programs.

When $\mca$ is an MCA, we add the following to the internal language:
$ \interp{\eLam{x_1, \ldots , x_n}{\term(x_1,\ldots,x_n)}
} := e_{\interp{\term}}$,
it satisfies $
\App^{n} \circ \langle
\interp{\eLam{x_1, \ldots , x_n}{\term(x_1,\ldots,x_n)}} ,
\mathsf{c}_1, \ldots,\mathsf{c}_n \rangle = \term(\mathsf{c}_1,
\ldots, \mathsf{c}_n)$.
As for partial abstraction,
$ \interp{\eLam{x_n}{\term(x_1,\ldots,x_n)}} :=
e_{\interp{\term}}^{n-1}$, it satisfies
$\interp{\eLam{x_n}{\term(x_1,\ldots,x_n)}} \cdot \mathsf{c} = \term(
x_1, \ldots, x_{n-1}, \mathsf{c})$.
The term $\interp{\eLam{x_n}{\term(x_1,\ldots,x_n)}}$
is a morphism $\mca^{n-1} \to \mca$ and its free
variables $x_1, \ldots, x_{n-1}$ can be substituted
by generalized elements using composition.

We also define the following encoding of pairs: $\cpair{\mathsf{c}_1}{\mathsf{c}_2} := \eLam{\mathsf{x}}{\Elet{\mathsf{y}}{\eApp{x}{\mathsf{c}_1}}{\eApp{y}{\mathsf{c}_2}}}$, along with the projections $\mathsf{e}_{\pi_1} := \eLam{\code}{\code \cdot (\eLam{\code[1],\code[2]}{\Eret{\code[1]}})}$ and $\mathsf{e}_{\pi_2} := \eLam{\code}{\code \cdot (\eLam{\code[1],\code[2]}{\Eret{\code[2]}})}$.
As noted in~\cite{MCA25},  the combinatory completeness condition above is equivalent to the existence of $\cS = \eLam{\code[f], \code[g], \code[a]}{\Elet{\code[1]}{\eApp{\code[f]}{\code[a]}}{\Elet{\code[2]}{\eApp{\code[g]}{\code[a]}}{\eApp{\code[1]}{\code[2]}}}}
$ and $\cK = \eLam{\code[1],\code[2]}{\Eret{\code[1]}}$ combinators and their partial applications\footnote{These partial applications
define morphisms of the topos. For instance,
the partial application $\cK_1$ of $\cK$ is a morphism
$\mca \to \mca$.}.

Some situations require a clear separation between codes used for programs and codes used for data, with only programs counting as evidence for validity. For example, we may wish to use Turing
machines for programs and representatives of real numbers for data, or to separate proof-like terms from the rest, as in classical
realizability. To that end, we define a special
separator subobject designating the codes to be used for evidence,
everything outside it being data.

\begin{definition}[Separator]\label{def:separator}
Given an MCA $\mca$ over $M$, a \emph{separator}~$\Separator$ is a consistent sub-MCA of $\mca$ closed by $\mca$-abstraction.
That is, a subobject $i_{\Separator} : \Separator \rightarrowtail \mca$ equipped with an application morphism $\App[\Separator] : \Separator \times \Separator \to M\Separator$ agreeing with the application morphism $\App[\mca]$ of $\mca$ in the sense of $\App[\mca] \circ (i_{\Separator} \times i_{\Separator}) = M i_{\Separator} \circ \App[\Separator]$, combinatory complete, and satisfying:

\begin{description}
\item[Consistent] $ \vdashE{\E}
\Entail{\code[f] : \Separator , \code[a] : \Separator}{ \after{\code[r]}{\eApp{\code[f]}{\code[a]}}{\bot} }{ \bot  }$

\item[Closure by $\mca$-abstraction]
For every
$\mca$-monomial $\mathsf{f} :
\mca^{n} \to M \mca$,
the global element $e_f : 1 \to \mca$
and the morphisms $e_f^k \circ i_{\Separator}^k : S^k \to \mca$
($1 \leq k < n$)
factor through $i_{\Separator}$.

    \end{description}
\end{definition}

An $M$-modality is called \emph{consistent} when it satisfies, for all $X$,
$ \Entail{m : M X}{ \after{x}{m}{\bot} }{\bot }$.
When the modality is consistent, we can take the top predicate as the separator $\Separator = \lambda x . \top$. When the top predicate is the separator, we call it the trivial separator.
In the examples, when a separator is not mentioned, it is implicitly assumed that the separator is trivial.
We abuse notation and sometimes write $\Separator(c)$ to denote $\chi_{\Separator}(c)$ for the characteristic morphism $\chi_{\Separator} : \mca \to \Omega$.
Similarly, we sometimes implicitly use  $i_{\Separator}$ and treat elements of $\Separator$ as elements of $\mca$.

\begin{definition}[Monadic Core]\label{def:monadic-core}
A monadic core in a topos $\E$ is a tuple  $\effshell = \left( M , \modality , \mca ,\Separator \right) $
where $M : \E \to \E$ is a strong monad, $\modality[X] : \Omega^X \to \Omega^{M X}$ is an $M$-modality,
 $\mca$ is a monadic combinatory algebra over $M$, and $\Separator$ is a separator.
\end{definition}

Ordinary PCAs are recovered as the partiality instance of monadic cores. Let $M$ be the partiality monad, so that $M\mca = \mca[\bot]$, and let $\modality$ be the total correctness modality
 $\after{x}{m}{\varphi\paren{x}} \; \equiv \; \exists x . m \downarrow x \sqcap \varphi\paren{x}$ .
Then a monadic applicative structure is an object of codes with a partial application operation.
The MCA completeness condition is then precisely the combinatory completeness condition for a PCA.
Other examples of monadic cores include, among others,  RCAs and SCAs with the demonic and angelic modalities for the powerset and powerset-state monad from~\cite{cohen2019effects, EF2021}, PPCAs with the demonic modality for the reader-partiality monad from~\cite{bauer2024countable}, and CPSCA with their corresponding modalities for the continuation monad~\cite{OlivaStreicher08,GardelleMiquey23} (see~\cite{MCA25}).
  We first recall the definition of a monadic core based on the continuation monad to capture Krivine realizability, and then we give an example of a monadic core that captures demonic nondeterministic computation.

\begin{example}[Continuation]
\label{ex:continuation}
Consider a total combinatory algebra $\mca$ with application~$\App$.
 The continuation monad extends $\mca$ by passing computations into explicit continuations. As was observed by Oliva and Streicher~\cite{OlivaStreicher08},
Krivine classical realizability can be recovered using a CPS translation, and therefore also using the continuation monad~\cite{MCA25}.
The monad is defined by
\[\Cont(X) \coloneq
        (X \to \Omega) \to \Omega \]
with $\eta^{\Cont}(x) \coloneq \lambda k.\, k(x)$ and
$\mu^{\Cont}(m) \coloneq \lambda k.m\, (\lambda v.v(k))$.
To capture Krivine's realizability, the modality has to account for an orthogonality relation, induced by a \emph{pole} $\pole$, which in our setting can be defined as a subobject of $\Omega$.
We say that $m:\Cont(X)$ is orthogonal to a continuation $k:X\to \Omega$,
whenever $m(k) \in \pole$, and for $x:X$ we similarly define $k\pole x \coloneq k(x)\in \pole$. As was observed in~\cite{MCA25}, the formula
$\after{x}{m}{\varphi(x)}$ expresses that $m$ is orthogonal to any continuation $k$, itself orthogonal to all $x$ satisfying $\varphi$.
Following this intuition, we define:
\[\modality[X]^\Cont(\varphi)(m) \coloneq
\forall k:X \to \Omega.(\forall x:\Omega.\varphi(x) \Rightarrow k \pole x)\Rightarrow m\pole k\]
This indeed defines a modality, inducing the expected interpretation of $\after{x}{m}{\varphi(x)}$.
Two monadic cores can then be defined: one based on the original combinatory algebra $\mca$ and any separator $\Separator$, yielding $\MCCont=(\Cont, \modality^\Cont,\mca,\Separator)$, where no new computational power is allowed (as we shall see in \Cref{ex:continuation3}).
But we can also benefit from the monad to extend $\A$ with codes for \texttt{call/cc} and continuations (see \cite{MCA25}), we write $\mcacc$ for the extended combinatory algebra,
$\Appcc$ for the application
and $\sepcc$ for the original separator $\Separator$ extended with \texttt{call/cc}. We write $\MCContcc$ for the corresponding monadic core, and we shall examine in \Cref{ex:continuation2,ex:continuation3} the nature of the simulations between the core induced by the identity monad on $\A$, $\MCCont$ and $\MCContcc$.
  \exqed
\end{example}
\begin{example}[Demonic nondeterminism]\label{ex:demonic-core}
Take $M = \mathcal{P}_{\mathrm{fin}}$, the finite powerset monad, so a computation $m : \mathcal{P}_{\mathrm{fin}}\mca$ is a finite set of possible outcomes.  Let $\mca$ be the codes generated by the $\cS$ and $\cK$ combinators (and their partial applications), together with one extra code $\mathsf{tag}$.
We write $\bar{n}$ for the Church encoding of $n \in \N$.
Let application $\App$ follow the usual $\cS$/$\cK$ rules (each returning a singleton set), along with the following nondeterministic clause for $\mathsf{tag}$:
\[
  \mathsf{tag} \cdot \mathsf{c} \;=\; \bigl\{ \cpair{\bar 0}{\mathsf{c}},\; \cpair{\bar 1}{\mathsf{c}} \bigr\},
\]
so $\mathsf{tag}$ is a binary choice returning both tagged copies of its argument at once.  Take the \emph{demonic} modality
\[
  \after{x}{m}{\varphi\paren{x}}
  \;\coloneqq\; \bigl(\exists x \in m.\,\top\bigr)\;\sqcap\;\bigl(\forall x \in m.\;\varphi\paren{x}\bigr),
\]
which deems a computation valid for $\varphi$ exactly when it has at least one outcome and \emph{all} of its outcomes satisfy $\varphi$.
Checking the modality axioms is routine: on a singleton, $\modality(\{p\})$ reduces to $p$ (unit), and enlarging $\varphi$ preserves both conjuncts (monotonicity).  For composition, $\modality(M\modality(m))$ says that $m$ is inhabited and every $t \in m$ is inhabited with all elements true, so $\mu(m) = \bigcup m$ is inhabited with all elements true, which is $\modality(\mu(m))$.  For strength, $p \sqcap \modality(m)$ pushes $p$ under the universal conjunct, giving $\modality\bigl(\{\, p \sqcap q \mid q \in m \,\}\bigr)$.
Moreover, $\modality$ is consistent, as $\after{x}{m}{\bot}$ asserts that an inhabited set consists of elements satisfying $\bot$, so we may take the trivial separator~$\Separator = \lambda x.\top$.
The tuple $\textbf{Tag-RCA} \coloneqq \bigl(\mathcal{P}_{\mathrm{fin}},\,\modality,\,\mca,\,\Separator\bigr)$ is then a monadic core.
\exqed
\end{example}

\subsection{General Simulations}

Given two monadic cores $\effshell[1,2] = \left( M_{1,2} , \modality[1,2] , \mca[1,2] , \Separator[1,2] \right)$, we want to be able to ``imitate'' computations in $\mca[1]$ with computations in $\mca[2]$ in a manner that is consistent with specifications, turning every specification in $\effshell[1]$ into a specification in $\effshell[2]$ over simulations.
To capture this idea, we use a binary relation $\Sim : \mca[1] \times \mca[2] \to \Omega$, where $\Sim\left(\mathsf{c},\mathsf{d}\right)$ is taken to mean that $\mathsf{d}$ is a simulation of $\mathsf{c}$.

For pure maps this would suffice, but source and target computations live in
different monads $M_1\mca[1]$ and $M_2\mca[2]$ and are not directly
comparable. What is comparable is the specifications they satisfy, so the
condition is postcondition-based: if a source computation $\mathsf{f}$ on
inputs $\mathsf{c}_1,\ldots,\mathsf{c}_n$ satisfies $\varphi$ and the
$\mathsf{d}_i$ simulate the $\mathsf{c}_i$, then $\mathsf{f}^{\Sim}$ must
satisfy the relational image of $\varphi$, producing a target result that
simulates some source result satisfying $\varphi$.

Our definition is the effectful analogue of Longley's applicative morphisms.
Recall Longley's definition of an applicative morphism for PCAs~\cite{Longley95}: a total relation $\Sim$ between the sets of codes $\mca[1]$ and $\mca[2]$, such that there exists a code $\mathsf{r}$ in $\mca[2]$, satisfying the following:
For all $\mathsf{c}_f , \mathsf{c}_a \in \mca[1]$ and $\mathsf{d}_f , \mathsf{d}_a \in \mca[2]$, if:
$
\Sim\left(\mathsf{c}_f,\mathsf{d}_f\right)\,\sqcap\,\Sim\left(\mathsf{c}_a,\mathsf{d}_a\right)\,\sqcap\,\mathsf{c}_f\cdot\mathsf{c}_a\downarrow\mathsf{c}_r
$
then:
$
\mathsf{r} \cdot \cpair{\mathsf{d}_f}{\mathsf{d}_a}\downarrow\mathsf{d}_r\sqcap\Sim\left(\mathsf{c}_r,\mathsf{d}_r\right)
$.
The code $\mathsf{r}$, the \emph{application translator}, sends target
implementations of a source function and argument to a target implementation
of their application.

To generalize it to arbitrary monadic cores, two changes are needed.
First, the definition must be read internally in the base topos.
Second, in the PCA case, the total correctness modality $\after{x}{m}{\varphi\paren{x}}
\;\equiv\; \exists x. \, m \downarrow x \sqcap \varphi\paren{x}$ absorbs the postcondition into
the convergence predicate, so the condition $c_f \cdot c_a \downarrow c_r$
already carries full information about $c_r$, in the sense that simulating for the convergence implies the simulation for every  postcondition.
For general modalities this absorption fails, and the postcondition $\varphi$
must be quantified over explicitly.

\begin{definition}[Simulation]
\label{def:simulation}
    Given two monadic cores
    $ \effshell[1,2] = \left( M_{1,2} , \modality[1,2] , \mca[1,2] , \Separator[1,2] \right) $
    in a topos $\E$, a simulation from $\effshell[1]$ to $\effshell[2]$ is a relation $\Sim : \mca[1] \times
    \mca[2] \to \Omega$ satisfying the following properties:
        \begin{description}
\item[(1) Evident totality]
        $
        \vdashE{\E}
        \forall \codeI : \Separator[1] . \exists \codef{t}_{\codeI} : \Separator[2] . \eAfter{ \codeII }{ \eApp{ \codef{t}_{\codeI}}{\id} }{ \Sim\paren{ \codeI , \codeII } }$

        \item[(2) Application translator] $\vdashE{\E}\ \exists \SRealizer\!:\!\Separator[2].\
  \forall \varphi\!:\!\mca[1]\!\to\!\Omega,\
  \mathsf c_f,\mathsf c_a\!:\!\mca[1],\
  \mathsf d_f,\mathsf d_a\!:\!\mca[2].$\\
  $\Sim(\mathsf c_f,\mathsf d_f) \sqcap \Sim(\mathsf c_a,\mathsf d_a) \sqcap
  \after{\mathsf c_r}{\mathsf c_f\cdot\mathsf c_a}{\varphi(\mathsf c_r)}
  \Rightarrow
  \after{\mathsf d_r}{\SRealizer\cdot\cpair{\mathsf d_f}{\mathsf d_a}}
       {\exists \mathsf c_r.\ \Sim(\mathsf c_r,\mathsf d_r) \sqcap \varphi(\mathsf c_r)}$
    \end{description}
\end{definition}

A relation~$\Sim$ is evidently total when, up to effectful computation,
it is total on codes in the separator. The witness of totality~$\codef{t}_{\codeI}$
is applied to~$\id := \eLam{\code}{ \Eret{\code}}$ and then computes a code~$\codeII$ simulating~$\codeI$. The choice of~$\id$ here is arbitrary,~$\mathsf{t}_{\codeI}$
can similarly be applied to any other element of the separator.
Thus, denoting $\mathtt{E}_{\text{tot}}\paren{d} \coloneqq\forall \codeI : \Separator[1].\,
        \exists \codef{t}_{\codeI} : \Separator[2].\,
        \eAfter{\codeII[r]}{\eApp{\codef{t}_{\codeI}}{\mathsf{d}}}{\Sim\paren{\codeI,\codeII[r]}}$, evident totality is equivalent to both $\vdashE{\E}\  \exists \mathsf{d} : \Separator[2].\mathtt{E}_{\text{tot}}\paren{d}$ and $\vdashE{\E}\  \forall \mathsf{d} : \mca[2].\mathtt{E}_{\text{tot}}\paren{d}$.

As in the pure case, $\SRealizer$ is the application translator.
What changes is that it now produces a \emph{computation} in $M_2\mca[2]$
rather than a code in $\mca[2]$, and the simulation condition is expressed
through the target modality $\modality[2]$ over every postcondition rather than through convergence alone.

A simulation is thus validated by the existence of two components: the evidence of totality, tracking the existence of simulating codes, and the application translator $\SRealizer$, tracking the application operation of the MCA. This is deliberately less strict than a strong monad morphism,
which fixes the translation of \emph{every} computation by preserving $\eta$ and $\mu$ on the nose.  We make this precise in \Cref{sec:algsim}, where every monad morphism is shown to induce an (algebraic) simulation.

By combinatory completeness, tracking the  application translator already fixes the translation of every effectful operation built from application and sequencing.  The next proposition characterizes condition (2) monomial-by-monomial.
A relation is a simulation exactly when it is evidently total and transports every source $\App$-monomial to a target one uniformly in the postcondition.

\begin{restatable}[\prooficon{equivsim}]{proposition}{equivsim}\label{prop:general-simulation}
Let $\effshell[1,2] = \left( M_{1,2} , \modality[1,2] , \mca[1,2] , \Separator[1,2] \right)$ be monadic cores in a topos $\E$, and let $\Sim : \mca[1] \times \mca[2] \to \Omega$ be an evidently total relation.
$\Sim $ satisfies the application-translator condition iff:%
    \begin{enumerate}

        \item[($2'$)] for every $n \geq 1$
        \vspace{-1em}

\[
\begin{array}{@{}l@{}}
\vdashE{\E}\ \forall \mathsf{c} : \Separator[1] \exists
\mathsf{c^{\gamma}} : \Separator[2]
\forall \varphi\!:\! \Omega^{\mca[1]},\
  \mathsf c_1 ,..., c_n \!:\!\mca[1],\
  \mathsf d_1 ,..., d_n \!:\!\mca[2]. \\[2pt]
  \hspace{-0.3cm}
   \bigsqcap_{i=1}^{n} \Sim(\mathsf c_i,\mathsf d_i)  \sqcap
  \after{\mathsf c_r}{
  \App^{n} \circ \langle \mathsf c, \mathsf c_1 , \ldots , \mathsf c_n \rangle}{\varphi(\mathsf c_r)}
  \Rightarrow
  \after{\mathsf d_r}{
  \App^{n} \circ \langle \mathsf c^{\Sim}, \mathsf d_1 , \ldots , \mathsf d_n \rangle
  }
       {\exists \mathsf c_r.\ \Sim(\mathsf c_r,\mathsf d_r) \sqcap \varphi(\mathsf c_r)}
\end{array}
\]

    \end{enumerate}
\end{restatable}

The per-monomial condition is deliberately uniform in the postcondition $\varphi$.
This uniformity  is what makes simulations logical rather than merely operational.
It says that target execution preserves all tests that the source logic can express, after translating those tests along the relation $\Sim$.

The conclusion of ($2'$) is a
direct image: condition ($2'$) says precisely that whenever $\mathsf{c}$
satisfies $\varphi$ on related source inputs, the translated computation
$\mathsf{c}^{\Sim}$ satisfies $\dimg{\Sim}\varphi$
on the related target
inputs.  This is the semantic analogue of a type translation: a source
specification is translated to the target by asking for an implementation
of some source realizer satisfying that specification.

For every monadic core, the identity relation on codes is a simulation, and the composite of two simulations is again a
simulation: totality of the composite follows from totality of each factor, and
the two application translators compose to a translator for the composite. The
simulations can moreover be compared through a preorder compatible with this composition. Thus, we obtain
a preorder-enriched categorical structure,
that is, a $2$-category where the hom-categories are preorders~\cite{johnsonyau2categories}.
\begin{definition}[Category $\MCcat$ over
$\E$]\label{def:MC-category}
Objects of $\MCcat$ are monadic cores over $\E$.
A morphism~$\gamma$ of~$\MCcat$ from~$\effshell[1]$
to $\effshell[2]$ is a simulation.
We define a preorder~$\gamma \leq \gamma'$ on morphisms from~$\effshell[1]$ to~$\effshell[2]$, it holds
when the following internal statement is valid
$\exists \mathsf{e} \! : \! \Separator[2] . \; \forall \mathsf{c} \! : \! \mca[1].\;
\forall \mathsf{d} \! : \! \mca[2] . \; \gamma(\mathsf{c},\mathsf{d}) \Longrightarrow \after{\mathsf{d_r}}{\mathsf{e} \cdot \mathsf{d}}{\gamma'(\mathsf{c},\mathsf{d_r})}$.
It gives the structure of a preorder-enriched category
to~$\MCcat$.
\end{definition}

Definition~\ref{def:simulation} is a conservative generalization of Longley's notion:
when the effect is partiality and the modality is total correctness, the two definitions coincide.
\begin{restatable}[\prooficon{PCAequiv}]{proposition}{PCAequiv}\label{prop:appx-mca-1}
When $M_1$ and $M_2$ are both the partiality monad, $\modality[1]$ and $\modality[2]$ are both the total correctness modality, with $\Separator[1]$ and $\Separator[2]$ both being trivial separators, a simulation from $\effshell[1]$ to $\effshell[2]$ is  a PCA applicative morphism.
\end{restatable}

\subsection{Examples and No-Go Theorem}\label{sec:MC-examples}

We illustrate simulations with three instances. A sanity check: total
combinatory algebras are exactly the cores with trivial effect and
observation, and their homomorphisms induce simulations, so the
effectful theory conservatively extends the pure one.
An
effectful transport: nondeterminism implemented by scheduling, related
by a simulation although no monad morphism exists.
And a limit: a no-go theorem separating
angelic from demonic observation,  under arithmetic preservation.

\subsubsection{Simulation Examples}

\begin{example}[Total Combinatory Algebras and Trivial Monadic Cores]
The simplest monadic cores have the identity monad, $M(X) = X$.
Taking $\modality = \id[\Omega]$ and
the trivial separator yields the \emph{trivial} monadic cores, and these are precisely the \emph{total combinatory algebras}.
The free trivial core on a set $X$ is the free combinatory algebra
$\mathrm{Free}_{CL}(X)$: combinatory terms over the generators
$\cS, \cK, X$ modulo combinatory equality.
Every function
$f : X \to \mca$ into the code object of a core
$(\mathbf{Id}, \modality, \mca, \Separator)$ over the identity monad
extends uniquely to a homomorphism
$\hat f : \mathrm{Free}_{CL}(X) \to \mca$, and the graph of $\hat f$
is a simulation
$(\mathbf{Id}, \id, \mathrm{Free}_{CL}(X), \top) \to
 (\mathbf{Id}, \modality, \mca, \Separator)$:
it is evidently total because it is strictly total, and $\SRealizer := \eLam{\code[f],\code[a]}{\code[f]
\cdot \code[a]}$ satisfies the application-translator condition
by the unit law.
\exqed
\end{example}

\begin{example}[Nondeterminism as scheduling]\label{ex:tag}

This example shows that a nondeterministic computation
can be simulated by one that \emph{reads a parameter} to resolve each choice.
The same code runs in two monadic cores,
one with set-valued nondeterminism and one with a \emph{scheduler} (i.e.~a
read-only parameter $i \in \mathbf{2}$ whose value resolves each
nondeterministic choice), and the identity relation on codes witnesses the
simulation.
What makes the example non-trivial is that the two cores carry the same computations but observe them through different lenses:
\emph{all outcomes at once} in the source versus
\emph{one outcome per scheduler input}
in the target.
The simulation certifies that, for demonic specifications, these two perspectives are logically indistinguishable.

For the source of the simulation, take
$MX = \mathcal{P}_{\mathrm{fin}}^{+}\left(X_{\bot}\right)$,
the finite nonempty powerset monad over the partiality monad,
so a computation $m : \mathcal{P}_{\mathrm{fin}}
^{+}\left(\mca[\bot]\right)$ is a finite set of possible.
The MCA is similar to the one defined in Example~\ref{ex:demonic-core}, with the same set of codes, and keeping the same notation for the behavior of $\mathsf{tag}$, but the computational behavior is different: For each element in the set of computations, rather than nonterminating computations being discarded, they are kept as an explicitly nonterminating branch.
The modality now becomes
\[
  \after{x}{m}{\varphi\paren{x}}
  \;\coloneqq\; \forall x \in m.\; x \downarrow \sqcap \varphi\paren{x},
\]
which deems a computation valid for $\varphi$ when \emph{all} of its outcomes terminate and satisfy $\varphi$.

Define the core $\textbf{Tag-PPCA}$
for demonic scheduling over the reader-partiality monad with parameter
set $\mathbf{2} = \{0, 1\}$: a computation is a function from
$\mathbf{2}$ to results, and, following~\cite{bauer2024countable}, we
write $(i \mid m) \downarrow \mathsf{r}$ when running $m$ with
scheduler input $i$ returns $\mathsf{r}$. Application
$\odot : \mca \times \mca \to \mathbf{2} \to \mca[\bot]$ follows the same $\cS$/$\cK$ rules (independently of
$i$), together with
\[
  (0 \mid \mathsf{tag} \odot \mathsf{c}) \downarrow \cpair{\bar{0}}{\mathsf{c}}
  \qquad
  (1 \mid \mathsf{tag} \odot \mathsf{c}) \downarrow \cpair{\bar{1}}{\mathsf{c}},
\]
and on the codes of $\mca$ it is total for each scheduler input, so we
may write $(i \mid m)\downarrow$ for the unique result. The modality
is again demonic, now over the scheduler:
\[
  \after{x}{m}{\varphi\paren{x}}
  \;\coloneqq\; \forall i \in \mathbf{2}.\;\exists x.\;
  (i \mid m) \downarrow x \sqcap \varphi\paren{x}.
\]
Here $\mathsf{tag}$ \emph{consults the scheduler}: $i$
selects which tagged copy to return, and the modality says that a
computation is good for $\varphi$ if the result satisfies $\varphi$ no
matter what the scheduler chooses.

The two cores share the code object $\mca$ and agree on $\mathsf{K}$ and $\mathsf{S}$,
but differ on how $\mathsf{tag}$ is observed.
Their relationship is captured by the following:
for every $\App$-monomial $\mathsf{f}$ and every $\mathsf{c}_1, \ldots, \mathsf{c}_n \in \mca$,
\begin{equation}\label{eq:tag-bridge}
  \bigl\{\,
  (0 \mid \mathsf{f}^{\odot}(\mathsf{c}_1, \ldots, \mathsf{c}_n)),\;
  (1 \mid \mathsf{f}^{\odot}(\mathsf{c}_1, \ldots, \mathsf{c}_n))
  \,\bigr\}
  \; \subseteq \;
  \mathsf{f}(\mathsf{c}_1, \ldots, \mathsf{c}_n),
\end{equation}
where $\mathsf{f}^{\odot}$ is the same syntactic term as $\mathsf{f}$ evaluated
in $\effshell[2]$.
That is, the set of source outcomes includes the image of the target scheduler function.
This holds at the level of $\mathsf{tag}$ by definition, and extends to
all $\App$-monomials by induction on their structure, using the fact that
$\mathsf{K}$ and $\mathsf{S}$ behave the same in both cores.

Define a simulation from $\textbf{Tag-RCA}$ to  $\textbf{Tag-PPCA}$ by
$\Sim(\mathsf{c}, \mathsf{d}) \;\coloneqq\; (\mathsf{c} \equiv \mathsf{d})$.
It is total since every code is related to itself,  so it is
evidently total by taking $\mathsf{t}_{\mathsf{c}} := \cK_1(\mathsf{c})$.
For each $\App$-monomial $\mathsf{f} : \mca^n \to \mathcal{P}_{\mathrm{fin}}(\mca)$, define
$\mathsf{f}^{\Sim} : \mca^n \to M_{\mathbf{2}}(\mca)$ to be the same term evaluated in $\textbf{Tag-PPCA}$.
To show that it is indeed a simulation, fix $\mathsf{c}_1, \ldots, \mathsf{c}_n \in \mca$, a postcondition
$\varphi : \mca \to \Omega$, and suppose $\Sim(\mathsf{c}_i, \mathsf{d}_i)$ for all $i$ (so
$\mathsf{c}_i \equiv \mathsf{d}_i$) and that the source modality
holds.
Let $i \in \mathbf{2}$.
By~\eqref{eq:tag-bridge}, $(i \mid \mathsf{f}^{\Sim}(\mathsf{c}_1, \ldots, \mathsf{c}_n))  \in \mathsf{f}(\mathsf{c}_1, \ldots, \mathsf{c}_n)$.
Write $r := (i \mid \mathsf{f}^{\Sim}(\mathsf{c}_1, \ldots, \mathsf{c}_n))$, then by assumption, $r \downarrow$, $\varphi(\mathsf{r})$ holds, and $\Sim(\mathsf{r}, \mathsf{r})$ holds by reflexivity.
Thus, the target modality holds.
At the level of the application translator, the
realizer is application itself,
$\SRealizer \cdot \cpair{\mathsf{d}_f}{\mathsf{d}_a} = \mathsf{d}_f
\odot \mathsf{d}_a$.
We will see in~\Cref{sec:algsim} that this simulation
is nevertheless not \emph{algebraic}: the two cores are logically
related, but no structural monad morphism witnesses the relation.
\exqed
\end{example}

\subsubsection{No-Go Theorem}
\label{sec:nogo}

The transport correspondence guarantees that a simulation carries realized
entailments from a source core to a target core. It is therefore an obstruction
to the \emph{existence} of a simulation that the two monadic cores disagree on a principle the simulation would be obliged to preserve.
We prove that angelic observation cannot be simulated by demonic observation.
The separation is elementary as it takes place at the level of the modalities alone, and the proof uses nothing beyond numerals and a single choice code.
In particular, to rule out trivial simulations between almost any two cores, such as $\Sim(\mathsf{c},\mathsf{d}) \coloneqq \top$, we require arithmetic-preservation, a minimal assumption demanding that numerals simulate only themselves.

\begin{definition}[Arithmetic-preserving   simulation]
Assume $\E$ has a natural-number object and
let $\effshell[1], \effshell[2]$ be two monadic cores over
$\E$, whose shared MCA carries
numeral codes $\bar{n}$, and write
$\varphi_n \coloneqq ({-} \equiv \bar{n})$ for the canonical numeral
predicates.
A simulation $\Sim : \effshell[1] \to \effshell[2]$ is
\emph{arithmetic-preserving} if
$\Sim(\bar{n}, \mathsf{d}) \iff \mathsf{d} \equiv \bar{n}$ for every
$n$, equivalently, $\Sim^{*}(\varphi_n) = \varphi_n$ as predicates.
\end{definition}
Let $\mathrm{Flip}_{\exists}$ and $\mathrm{Flip}_{\forall}$ denote two
monadic cores over the powerset monad $M(X) = \Omega^{X}$ that share
the same MCA that contains a code $\mathsf{flip}$ with
$\mathsf{flip} \cdot \mathsf{c} = \{\bar{0},\bar{1}\}$ for every
$\mathsf{c}$. Each monadic core is equipped with its own modality:
$\mathrm{Flip}_{\exists}$ uses the \emph{angelic} modality
$\after{x}{m}{\varphi\paren{x}} \coloneqq \exists x \in m.\;
\varphi(x)$, and $\mathrm{Flip}_{\forall}$ the \emph{demonic} one,
$\after{x}{m}{\varphi\paren{x}} \coloneqq (\exists x \in m . \top)
\sqcap (\forall x \in m.\; \varphi(x))$.

\begin{restatable}[\prooficon{maymustnogo}]{proposition}{maymustnogo}
\label{prop:may-must-nogo}
There is no arithmetic-preserving simulation
$\Sim : \mathrm{Flip}_{\exists} \to \mathrm{Flip}_{\forall}$.
\end{restatable}
The proposition holds because a simulation must serve every
postcondition with one fixed translator, while the two modalities
price postconditions differently.
Angelically, $\mathsf{flip}$
validates the incompatible tests $\varphi_0$ and $\varphi_1$ at
once, since each asks only for one suitable outcome. A simulation
would have to reproduce both validations demonically through a single
outcome set $R_0$, produced by the application translator before the
test is chosen, and arithmetic-preservation leaves the demonic side
no slack: all of $R_0$ must equal $\bar{0}$ for one test and all of
it $\bar{1}$ for the other, while remaining inhabited.
In the reverse direction the identity relation
$\Sim(\mathsf{c},\mathsf{d}) \coloneqq (\mathsf{c} \equiv \mathsf{d})$ defines an
arithmetic-preserving simulation $\mathrm{Flip}_{\forall} \to \mathrm{Flip}_{\exists}$.
Consequently every entailment realized in $\mathrm{Flip}_{\forall}$ is also realized in
$\mathrm{Flip}_{\exists}$.

\subsection{Monad Morphisms as Algebraic Simulations}
\label{sec:algsim}

A simulation relates two monadic cores, and a core is built around a monad, so it is natural to ask how the standard notion relating
monads, the \emph{monad morphism}, sits inside our framework. The answer sharpens a theme of the paper: simulations are laxer.
A strong monad morphism preserves  unit, join, and strength on the nose,
and
sees only part of a core: it says nothing about the modality or the codes. When it does agree with both,
it induces a particularly tractable simulation, the
\emph{algebraic} one: the global simulation obligation, quantifying over all
monomials and all postconditions, reduces to local equations on the generating data.
The payoff is practical: given a source core, a strong monad morphism
$\Phi : M_1 \Rightarrow M_2$, and a compatible $M_2$-modality, one can canonically equip $M_2$ with an application so that $\Phi$
transports the whole source logic, as the examples below illustrate.

First, observe that from any monad morphism $\Phi : M_1 \Rightarrow M_2$ and any $M_2$-modality, an $M_1$-modality can be defined.
\begin{restatable}[\prooficon{monadmodality}]{lemma}{monadmodality}
\label{lemma:appx-algsim-1}
Given a strong monad morphism $\Phi : M_1 \Rightarrow M_2$, i.e.~a monad morphism that preserves unit, join, and strength, and an $M_2$-modality $\modality[X] : \Omega^X \to \Omega^{M_2 X}$,
$\modality^{\Phi}$ defined by
$ \modality[X]^{\Phi} := \Omega^{\Phi_{X}} \circ \modality[X] $ is an $M_1$-modality.
\end{restatable}

While a monad morphism relates monads over the same base, simulations
also relate cores with different code objects. Pairing $\Phi$ with a
code map $h : \mca[1] \to \mca[2]$ yields the algebraic notion:
\begin{definition}[Algebraic Simulation]
Given monadic cores
$\effshell[1,2] = ( M_{1,2} , \modality[1,2] , \mca[1,2], \Separator[1,2])$
in $\E$, an \emph{algebraic simulation} from $\effshell[1]$ to
$\effshell[2]$ is a pair $(\Phi , h )$ of a strong monad morphism
$\Phi : M_1 \Rightarrow M_2$ and a morphism $h : \mca[1] \to \mca[2]$
of $\E$ such that:
\begin{description}
\item[Separator preservation]\label{algsim:sep} There exists a morphism $h^s : \Separator[1] \to \Separator[2]$ s.t $h \circ i_{\Separator[1]} = i_{\Separator[2]} \circ h^s$ where $i[\Separator[1,2]] : \Separator[1,2] \rightarrowtail \mca[1,2]$ are the inclusion monomorphisms of the separators in the MCAs
\item[Modality compatibility]\label{algsim:mon} \quad$\modality[1] \sqsubseteq \modality[2]^{\Phi}$, i.e., for all $X$ and $\psi : \Omega^X$:
 $ \modality[1]\psi \;\sqsubseteq\; \modality[2]^{\Phi}(\psi)
  \;=\; \modality[2]\psi \circ \Phi_{X}$
\item[Application compatibility]\label{algsim:app} \quad $\vdashE{\E} \exists \mathsf{r} : \Separator[2]. \mathsf{r}_{\App[2]} \circ h \times h = \Phi h \circ \App[1]  $,  where
$\mathsf{r}_{\App[2]}(x,y) \coloneq \App[2](\mathsf{r},\cpair x y)$
and
$ \Phi h = \Phi_{\mca[2]} \circ M_1 h = M_2 h \circ \Phi_{\mca[1]} $.
\end{description}

\end{definition}

Separator preservation says the code translation morphism $h$ restricts to the separators.
Modality compatibility says the source observation discipline is no stronger than the pullback of the target's along $\Phi$: whatever
$\modality[1]$ certifies remains certified after translation.
Application compatibility says that the realizer $\mathsf{r}$, applied to the $h$-images of two codes, computes the same $M_2$-computation as
running the source application and translating the result through $\Phi h$ (the two composites defining $\Phi h$ agree by naturality of
$\Phi$). Thus $\mathsf{r}$ plays the role of the application translator, and the two conditions make
$\Sim(\mathsf{c}, \mathsf{d}) \coloneqq h(\mathsf{c}) \equiv \mathsf{d}$
a simulation, the extension from binary application to all
$\App$-monomials following by induction on their structure.

\begin{restatable}[\prooficon{algebraicsim}]{proposition}{algebraicsim}
\label{proposition:algebraicsim}
    An algebraic simulation is a simulation.
\end{restatable}

This is the sense in which algebraic simulations are \emph{algebraic}:
like an algebra homomorphism, the pair $(\Phi, h)$ is data on generators, and the induction behind \Cref{proposition:algebraicsim}
extends it to every derived operation. Two local equations thereby discharge an obligation that quantifies over all monomials and all postconditions.

Observe that algebraic simulations are strictly stronger than general
simulations: the simulation of \Cref{ex:tag} is not algebraic,
because no monad morphism exists from the powerset monad to the
reader-partiality monad of that example. Indeed, suppose
$\Phi$ were one and let $A = \{a_0, a_1\}$. Unit preservation forces
$\Phi_A(\{a_j\}) = (i \mapsto a_j)$, and naturality in the
transposition $\sigma : a_0 \leftrightarrow a_1$ forces
$\Phi_A(\{a_0,a_1\})$, a function $\mathbf{2} \to \{a_0,a_1\}$, to be
fixed by postcomposition with $\sigma$, which no such function is.
There is no natural way to turn a set of outcomes into a scheduler:
any assignment must pick an ordering that the set does not carry.

We illustrate algebraic simulations with three canonical examples, in
each case extending the effect while leaving codes and separator
unchanged, so that the source logic transports at no cost.

\begin{example}[Continuation]
\label{ex:continuation2}
Even though the monadic core described in \Cref{ex:continuation} is meant to capture Krivine realizability, as long as the set of codes is not extended, adding continuations in that way is still benign: the canonical morphism $\iota := \eta_{\mca} $ embeds a source computation $c$ as $\lambda k.k(c)$.
Formally, given a choice of a separator~$\Separator$ on~$\mca$,
we can consider the trivial monadic core~$\effshell[\textrm{Id}]\coloneq (\mathrm{Id},\id[\Omega], (\mca,  \App) , \Separator)$.
Then we get the following algebraic simulations:
\begin{center}
\scalebox{0.8}{
\begin{tikzpicture}[arrow/.style={thick,->,>=latex}]]
\node (mca) {$\MCId$};
\node [right=2cm of mca] (mcacont) {$\MCCont = (\Cont, \modality^\Cont, (\mca, \eta_{\mca} \circ \App),
     \Separator)$};
\node [right=2cm of mcacont] (mcacontcc) {$\MCContcc = (\Cont, \modality^\Cont, (\mcacc,
        \Appcc),
     \sepcc)$};
\draw[arrow] (mca) -- (mcacont) node[anchor=south,inner sep=2pt,midway] {$(\iota,\id[\mca])$};
\draw[arrow] (mcacont) -- (mcacontcc) node[anchor=south,inner sep=2pt,midway] {$(\id,\mathrm{inj})$};
 \end{tikzpicture}
}
\end{center}
where the first simulation relates $\A$ and its image through a CPS translation, and is actually stronger (see \Cref{ex:continuation3}).
 While the second simulation extend the target to include \texttt{call/cc} using the canonical injection $\mathrm{inj}:\mca\to\mcacc$, inducing thereby a classical EF.\exqed
\end{example}

\begin{example}[Writer]
\label{ex:writer}
The writer monad extends $M$ with logging: each computation additionally produces an
entry in a monoid $W$.
Adding this capability is benign because the lifted modality $\modality[W]$ discards the log entirely before evaluating the proposition.

Let $\effshell[1] = (M, \modality, (\mca, \App), \Separator)$ be a monadic core and
$(W, \cdot, \varepsilon)$ a monoid.  Define:
\begin{itemize}
  \item \emph{Writer monad}: $M_W(X) \coloneq M(W \times X)$, with
        $\eta^{M_W}(x) \coloneq \eta^M(\varepsilon, x)$, where sequencing is given by
        $\Elet{(w,x)}{t}{\Elet{(w',y)}{f(x)}{\eta^M(w \cdot w', y)}}$.
  \item \emph{Lifted modality}: $\modality[W]\psi \coloneq \modality\psi \circ M\pi_2$
        for any $X$ and $\psi : X \to \Omega)$.
  \item \emph{Canonical morphism}: $\iota^W_X(m) \coloneq M(\langle \varepsilon,
        \mathrm{id}_X \rangle)(m)$, giving $\iota^W : M \Rightarrow M_W$.
\end{itemize}
Then $(\iota^W, \mathrm{id}_{\mca})$ is an algebraic simulation
from $\effshell[1]$ to $\effshell[W]\coloneq (M_W, \modality[W], (\mca, \iota^W_{\mca} \circ \App), \Sep{})$.
\exqed
\end{example}

\begin{example}[Reader]
\label{ex:reader}
The reader monad extends $M$ with an environment: each computation may consult a
parameter $e \in E$.
Adding this dependency is benign because the canonical morphism
$\Phi^{E}$ embeds source computations as environment-independent, and the modality
$\modality[E]$ takes the meet over all environments, so any proposition realized in
$\effshell[1]$ remains realized no matter what $e$ holds.

Let $\effshell[1] = (M, \modality, (\mca, \App), \Separator)$ be a monadic core and $E$ a
non-empty set .  Define:
\begin{itemize}
  \item \emph{Reader monad}: $M_{E}(X) \coloneq E \to M(X)$, with
        $\eta^{M_{E}}(x) \coloneq \lambda e.\, \eta^{M}(x)$, where sequencing is given by
        $\lambda e.\, \Elet{x}{t(e)}{f(x)(e)}$.
  \item \emph{Relativised modality}: $\modality[E]\psi\left(t\right) \coloneq \bigwedge_{e \in E}
        \modality \psi (t(e))$ for $X$, $\psi : X \to \Omega$, and $t : E \to M(X)$.
  \item \emph{Canonical morphism}: $\iota^{E}_{X}(m) \coloneq \lambda e.\, m$,
        giving $\iota^{E} : M \Rightarrow M_{E}$.
\end{itemize}
Then $(\iota^{E}, \mathrm{id}_{\mca})$ is an algebraic simulation
from $\effshell[1]$ to $\effshell[E]\coloneq (M_{E}, \modality[E], (\mca, \iota^{E}_{\mca} \circ \App), \Sep{})$.%
\exqed
\end{example}

Among the features provided by monad morphisms, in situations such as the reader and writer monads build for a monad $M$, any monad morphism $\Phi: M^1 \Rightarrow M^2$ induces monad morphisms from the reader and writer monads build over $M^1$ to its twin over $M^2$. This also applies to the algebraic simulations between monadic cores, which can also be lifted to algebraic simulations between the corresponding reader/writer monadic cores.
\begin{restatable}[\prooficon{liftingalgsim}]{proposition}{liftingalgsim}
\label{proposition:liftingalgsim}
Let $(\Phi,h)$ be an algebraic simulation from $(M^1, \modality[1], \mca[1], \Separator[1])$ to $(M^2, \modality[2], \mca[2], \Separator[2])$. Then,
writing $\Phi^W$ for $\Phi^W_X \coloneq \Phi_{W\times X}$ and
$\Phi^E$ for $\Phi^E_X(m)(e) \coloneq \Phi_{X}(m(e))$, we have
that $(\Phi^\ddagger,h)$ is an algebraic simulation from $\effshell[\ddagger]^1$ to $\effshell[\ddagger]^2$ for $\ddagger\in\{W,E\}$.

\end{restatable}

\section{Evidenced Frames and Logical Simulations}\label{sec:ef-sims-main}

Monadic cores describe effectful computation. Evidenced frames (EFs) describe the
logic generated by such computation~\cite{EF2021}.
They abstract away from the particular
shape of the effect and retain the proof-relevant structure needed for
reasoning: propositions, evidence, and a relation saying when an evidence term
transforms realizers of one proposition into realizers of another.

Moving to this layer is what makes transport a statement about \emph{logics}
rather than codes. A simulation of monadic cores relates source and target
codes so that target execution preserves every source postcondition.
At the
level of EFs  this becomes a translation of propositions and
evidence preserving realized entailments.
The two are tightly linked: every monadic core simulation induces an EF
simulation (soundness), and for regular simulations the converse holds
(completeness).
Moreover, both correspond to morphisms of the induced triposes.

We set up evidenced frames over an arbitrary topos (\Cref{sec:ef}) and their
simulations (\Cref{sec:ef-sims}), then characterize the latter on both sides:
they correspond to simulations of monadic cores
(\Cref{sec:monadic-and-ef-simulations}) and to transformations of the induced
triposes (\Cref{sec:ef-simulations-to-models}).

\subsection{Evidenced Frames}\label{sec:ef}

An evidenced frame, which we recast here over an arbitrary base
topos $\E$, is a Heyting-style logic equipped with explicit evidence
witnessing entailments, together with the standard evidence combinators for
composition, pairing, abstraction, and quantification.
The judgment
$
  \phi_1 \xle{e} \phi_2
$
should be read as: the evidence $e$ transforms realizers of $\phi_1$ into realizers
of $\phi_2$.  In other words, $\phi_1$ is an assumption or
precondition, $\phi_2$ is a guarantee or postcondition, and $e$ is the program
evidence witnessing the transformation.

\begin{definition}[Evidenced Frame]
\label{def:evidenced-frame-topos}
An \emph{evidenced frame} in $\E$ is a triple
$\ef^\E \coloneqq( \Phi, E,  \mbox{$\cdot \xle{\cdot} \cdot$} )$, where:
\begin{itemize}[leftmargin=0.8cm]
\item $\Phi\in \E$ is the \emph{object of propositions}
\item $E\in \E$ is the \emph{object of evidence}
\item $\mbox{$\cdot \xle{\cdot} \cdot$}$ is a \emph{ternary evidence relation}, i.e. a subobject      $\holds\rightarrowtail \Phi \times E \times \Phi$
      , written $\phi_1 \xle{e} \phi_2$
\end{itemize}
 with a proposition $\top : 1\to\Phi$, a \emph{conjunction} operator $\wedge : \Phi\times\Phi\to\Phi$, and a \emph{universal implication} operator $\imp : \Phi\times\Omega^{\Phi}\to\Phi$ satisfying the properties in~\Cref{fig:ef}.
\begin{figure*}[t]
\[
\renewcommand{\arraystretch}{1.35}
\begin{array}{@{}l@{\;\;}l@{}}
\text{\textsc{(refl)}}  & \vdashE{\E} \exists \eid \!: \! E. \; \forall \phi\!:\!\Phi.\; \phi \xle{\eid} \phi \\
\text{\textsc{(trans)}} & \vdashE{\E} \exists \mathrm{tr} \!:\! E^{E \times E }. \,
\forall \phi_1,\phi_2,\phi_3\!:\!\Phi.\;\forall e,e'\!:\!E.\;
      \big(\phi_1 \xle{e} \phi_2 \sqcap \phi_2 \xle{e'} \phi_3\big) \Rightarrow
      \phi_1 \xle{\mathrm{tr}(e,e')} \phi_3 \\[3pt]
\text{\textsc{(top)}}   & \vdashE{\E} \exists \etrue \! : \! E. \; \forall \phi\!:\!\Phi.\; \phi \xle{\etrue} \top \\[3pt]
\text{\textsc{(fst)}}   &  \vdashE{\E}
\exists \efst \! : \! E. \;
\forall \phi_1,\phi_2\!:\!\Phi.\; \phi_1 \wedge \phi_2 \xle{\efst} \phi_1 \\
\text{\textsc{(snd)}}   & \vdashE{\E}
\exists \esnd \! : \! E. \;
\forall \phi_1,\phi_2\!:\!\Phi.\; \phi_1 \wedge \phi_2 \xle{\esnd} \phi_2\\
\text{\textsc{(pair)}}  & \vdashE{\E}
\exists \mathrm{pr} \!:\!  E^{E \times E}. \;
\forall \phi,\phi_1,\phi_2\!:\!\Phi.\;\forall e_1,e_2\!:\!E.\;
      \big(\phi \xle{e_1} \phi_1 \sqcap \phi \xle{e_2} \phi_2\big)
      \Rightarrow \phi \xle{\mathrm{pr}(e_1,e_2)} \phi_1 \wedge \phi_2 \\[3pt]
\text{\textsc{(curry)}} & \vdashE{\E}
\exists \elambda{} \! : \! E^E. \;
\forall \phi_1, \phi_2, \vec{\phi}, e.\; (\forall \phi \in \vec{\phi}.\; \phi_1 \wedge \phi_2 \xle{e} \phi) \implies \phi_1 \xle{\elambda{e}} \phi_2 \imp \vec{\phi} \\
\text{\textsc{(eval)}}  & \vdashE{\E}
\exists \eeval:E . \;
\forall \phi_1, \vec{\phi}, \phi \in \vec{\phi}.\; (\phi_1 \imp \vec{\phi}) \wedge \phi_1 \xle{\eeval} \phi
\end{array}
\]
\vspace{-0.3cm}
\caption{Axioms of an evidenced frame}
\vspace{-0.4cm}
\label{fig:ef}
\end{figure*}

\end{definition}

We write $\ecomp{e}{e'}$ for $\mathrm{tr}(e,e')$ and $\epair{e}{e'}$ for $\mathrm{pr}(e,e')$.
Standard connectives are definable as:
$$
\begin{array}{@{}l@{\qquad\qquad}l@{}}
\begin{array}[t]{@{}l@{\;\;\define\;\;}l@{}}
\phi_1 \arrowE \phi_2 & \phi_1 \imp \{\phi_2\} \\
\prod \vec{\phi}       & \top \imp \vec{\phi}\\
\bot                   & \prod \Phi
\end{array}
&
\begin{array}[t]{@{}l@{\;\;\define\;\;}l@{}}
\phi_1 \vee \phi_2 & \prod \{ ((\phi_1 \imp \phi') \wedge (\phi_2 \imp \phi')) \imp \phi' \mid \phi' \in \Phi \}\\
\coprod \vec{\phi}  & \prod \{ (\prod \{ \phi \imp \phi' \mid \phi \in \vec{\phi}\}) \imp \phi' \mid \phi' \in \Phi \}
\end{array}
\end{array}
$$

This impredicative interface is what lets the EF support higher-order reasoning
uniformly over arbitrary bases.
When $\E=\Set$, the definition reduces to
the usual $\Set$-based notion of evidenced frame~\cite{EF2021}.
When generalized over an arbitrary topos, the definition states propositions externally, as morphisms of the topos, whereas the existence of evidence is stated internally.
This key difference allows us to
construct externally a Heyting prealgebra from an internal evidenced frame. For every object $A : \E$, the set $\Hom_{\E}(A,\Phi)$
can be endowed
with the structure of a Heyting prealgebra.
The Heyting operations (meet, implication...) on~$\Hom_{\E}(A,\Phi)$
are then constructed externally using the morphisms that define propositions of the EF:
for instance, for propositions $\phi_1,\phi_2 : A \to \Phi$ ,
their meet is defined to be $\wedge \circ \langle \phi_1 , \phi_2 \rangle$.
The preorder on this set is then computed in the internal logic of the topos:
the ordering $ \phi \sqsubseteq \psi$ holds if and only if $\vdashE{\E}
\exists e\! :\: E.\;
\forall a\! :\! A. \; \phi(a) \xle{e} \psi(a)$.

\subsection{Simulations of Evidenced Frames}
\label{sec:ef-sims}

Since an evidenced frame is a proof-relevant entailment judgment, not merely an
ordering on propositions, a simulation between evidenced frames must preserve the
evidence witnessing entailments, not only the truth of propositions.
Following~\cite{tarantino2026}, we define \emph{lax} and \emph{regular
simulations} of EFs.
EFs equipped with lax (resp. regular) simulations form a preorder-enriched category
$\eflax$~(resp. $\efreg$).

\begin{definition}[Lax and regular simulations of EF]\label{def:ef-sims}
    A \textit{lax} EF \textit{simulation} from~$\ef_1 = ( \Phi_1, E_1, \cdot \xle{\cdot}_1 \cdot )$ to
$\ef_2 = ( \Phi_2, E_2, \cdot \xle{\cdot}_2 \cdot )$ is
a map~$F : \Phi_1 \to \Phi_2$ in $\E$
satisfying:
\begin{description}
\item[Monotonicity]
$\vdashE{\E} \forall e_1\!:\!E_1.\; \exists e_2\!:\!E_2.\; \forall \phi,
\phi'\!:\!\Phi_1.\;
\phi \xle{e_1}_1 \phi' \Longrightarrow F(\phi) \xle{e_2}_2 F(\phi')$
\item[Top preservation] $\vdashE{\E} \exists e_2\!:\!E_2.\; \top_2 \xle{e_2}_2 F(\top_1)$
\item[Implication distributivity]
$\vdashE{\E} \exists r \!:\! E_2. \; \forall \phi,\phi'\!:\!\Phi_1.\;
F(\phi\supset_1 \phi')\ \xrightarrow{\,r\,}_2\ (F(\phi) \supset_2 F(\phi'))
$
\end{description}
It is \textit{regular} if it additionally satisfies: \begin{description}%
\item[Regularity]
$\vdashE{\E} \exists e_\exists \!:\! E_2. \; \forall \vec{\phi} \!:\! \Omega^{\Phi_1} .\; F(\coprod_1 \vec{\phi} ) \; \xle{e_\exists}_2\; \coprod_2\, \left\{ F(\phi) \mid \phi \in \vec{\phi} \right\}$.
\end{description}
\end{definition}

The definitions of monadic cores simulation and EF simulation are
parallel.
Evident totality corresponds to top preservation
with monotonicity, and the evidence $r$ of implication distributivity is the
logical counterpart of the application translator, making the translation
compatible with application and hence with higher-order reasoning.

For every evidenced frame $\ef$, the identity map is a lax simulation
$\ef \to \ef$, and the composite of two lax (regular) simulations is a lax (regular) simulation. Thus, we obtain a categorical structure.

\begin{definition}[Category \eflax{} and $\efreg$ over
$\E$]\label{def:ef-simulation}
Objects of \eflax{} and $\efreg$ are evidenced frames over $\E$.
Morphisms of~$\eflax{}$ ($\efreg$) are lax (regular) simulations, respectively.
For
parallel simulations $F, G : \ef_1 \to \ef_2$ we set $F \leq G$ when
$\vdashE{\E}\ \exists e_2\!:\!E_2.\ \forall \phi\!:\!\Phi_1.\ F(\phi) \xle{e_2}_2 G(\phi)$.
This gives the structure of preorder-enriched categories
to~$\eflax{}$ and $\efreg$.
\end{definition}
The preorder $  F \leq G$
records when one logical translation is uniformly refined by
another.

\begin{example}[Booleanization]  \label{ex:double-neg}
    In~\cite{Vanoosten16}, the booleanization of a relative realizability tripos is
    defined. It adapts to the framework of EFs as follow.
    Let $\ef \triangleq (\Phi, E, \cdot \xle{\cdot}{\cdot})$
    be an evidenced frame and fix a proposition $P : 1 \to \Phi$.
We introduce the following notation:
$$\negP \phi \define \phi \arrowE P
 \qquad \qquad \coprodP \vec{\phi} \define \negP
\prod \{ \negP \phi \mid \phi \in \vec{\phi} \}.$$

The \emph{booleanization} of $\ef$
with respect to $P$ is defined to be $\EFC = (\Phi, E, \phi_1 \xleC{e} \phi_2)$
where $\phi_1 \xleC{e} \phi_2
\define (\negP \phi_1 \wedge \phi_2) \xle{e} P$.
To see that this is indeed an evidenced frame, take
$\topC  \define  P$, $\phi_1 \andC \phi_2  \define
     \negP ( \negP \phi_1 \wedge \negP \phi_2)$, and
         $\phi_1 \arrowC \vec{\phi} \define
     \negP \phi_1 \wedge \coprodP \vec{\phi}$.
     Comparing to~$\ef$ the order in~$\EFC$ is reversed:
     if $\phi_1 \xle{e} \phi_2$ then $\phi_2 \xleC{\epair{\efst}{\esnd;e};\eeval} \phi_1$. However, the other direction does not need to hold:
     $\phi_2$ entails $\phi_1$ in~$\EFC$ does not imply that $\phi_1$
     entails $\phi_2$ in~$\ef$.
     The EF properties of~$\EFC$
     are proved by doing similar computations inside~$\ef$. For instance,
     $(\negP \phi \wedge P) \xle{\esnd} P$ gives that
     $\phi \xleC{\esnd} P$, thus showing that~$P$ acts
     as the top proposition.

There is a regular simulation $L_P : \ef \to \EFC$ and a lax simulation
$R_P : \EFC \to \ef$, both given by $\phi \mapsto \negP \phi$. In general $R_P$ is
not regular: in the locale of opens of~$\R$ with $P = \emptyset$ the EF is a mere
order (one evidence), where $U_1 \xleC{} U_2$ is $U_2 \subseteq \neg \neg U_1$ with
$\neg U \define \IntR{\ComplR{U}}$ and join
$\coprod \vec{U} \define \IntR{\bigcap \{ \neg \neg U \mid U \in \vec{U} \}}$.
But
$\neg \coprod \{ U \mid 0 \in U \} \nsubseteq \bigcup \{ \neg U \mid 0 \in U \}$, so
$R_P$ fails regularity.

A similar construction underlies the CPS implementation of classical
realizability~\cite{OlivaStreicher08}, later recast algebraically~\cite{Miquey20}.
Since negation is self-adjoint~\cite{mellies17},
~$L_P$ is a geometric simulation as stated
in Example~\ref{ex:double-negation2}.
\exqed
\end{example}

\subsection{From Monadic Cores to Evidenced Frames}\label{sec:MC-EF}

The section shows that every monadic core generates an evidenced frame, and every simulation
of monadic cores generates a simulation of the evidenced frames. Moreover, this
passage is functorial, and on simulations the
operational and the logical presentation carry exactly the same information.

\subsubsection{The evidenced frame of a monadic core}\label{sec:MCtoEF}

We now recall how a monadic core gives rise to an evidenced frame~\cite{MCA25}, generalizing the construction where both structures are over a  topos $\E$.

\begin{restatable}[\prooficon{effromcore}]{theorem}{effromcore}
\label{thm:ef-from-monadic-core}
    For
    $\effshell = (M,\modality,\mca,\Separator)$ in $\E$, define \mbox{$\ef(\effshell) \define (\Omega^{\mca},
    \Separator, \cdot \xle{\cdot}_{\scriptscriptstyle{\effshell}} \cdot)$} where $$  \phi \xle{\mathsf{e}}_{\scriptscriptstyle{\effshell}} \predB
  \define
  \forall \mathsf{c} \!:\! \mca.\;
  \phi(\mathsf{c})
  \Rightarrow
  \after{\mathsf{c}_r}{\mathsf{e} \cdot \mathsf{c}}
        {\predB(\mathsf{c}_r)},$$
    with $\top(c) \define \top_{\Omega}$,
$ (\phi \wedge \predB)(\mathsf{c})
  \define \after{\mathsf{c_r}}{\mathsf{e_{\pi_1}} \cdot \mathsf{c}}{\phi(\mathsf{c_r})}
  \sqcap \after{\mathsf{c_r}}{\mathsf{e_{\pi_2}} \cdot \mathsf{c}}{\predB(\mathsf{c_r})}$, and
$
  (\phi \imp \vec{\predB})(\code[f])
  \define
  \forall
  \predB \in \vec{\predB}
  . \;
  \forall \mathsf{c}\!:\!\mca.\;
  \phi(\mathsf{c})
  \Rightarrow
  \after{\mathsf{c}_r}{\code[f]\cdot \mathsf{c}}
        {\predB(\mathsf{c}_r)}
$.
    Then, $\ef(\effshell)$ is an evidenced frame.
\end{restatable}

A proposition $\phi : \Phi_{\scriptscriptstyle\effshell}$ assigns to each code
$\mathsf{c}:\mca$ a truth value $\phi(\mathsf{c})$,  read
as `$\mathsf{c}$ realizes $\phi$'.
The object of evidence is the separator $\Separator$.
An evidence code witnesses an entailment
$
  \phi \xle{\mathsf{e}} \predB
$
when, for every realizer $\mathsf{c}$ of $\phi$, applying $\mathsf{e}$ to
$\mathsf{c}$ produces, as observed by the modality, a realizer of $\predB$.

The evidence combinators required by Definition~\ref{def:evidenced-frame-topos} are  represented by codes in $\Separator$:
combinatory completeness of the MCA
provides them, and the separator axioms ensure they are themselves evidence. The
modality laws then make these codes validate the corresponding EF axioms once
their computations are observed through $\modality$.
Thus a monadic core gives both an operational semantics for evidence and the
logical structure generated by that semantics.

Before continuing, we show that in~$\ef(\effshell)$ the existential
$\coprod \vec{\phi}$ (defined after Definition~\ref{def:evidenced-frame-topos})
is computed as a union. This is standard for partial combinatory algebras, and we
show it persists for monadic cores, despite entailment being mediated by
$\modality$.
\begin{restatable}[\prooficon{unioncoprod}]{lemma}{unioncoprod}
\label{lemma:union-coprod}
Let $\effshell = (M, \modality, \mca, \Separator)$ be a monadic core. In its
evidenced frame $\ef(\effshell)$, the existential $\coprod$
coincides with the union $\bigcup$, that is, the following hold:
$$
\begin{array}{@{}l@{\qquad\quad}l@{}}
 \vdashE{\E} \exists \mathsf{e}_1 \!:\! \Separator.\,
  \forall \vec{\phi} \!:\! \Omega^{\Omega^{\mca}}.\,
  \bigcup \vec{\phi} \xle{\mathsf{e}_1}_{\scriptscriptstyle\effshell} \coprod \vec{\phi}
&
 \vdashE{\E} \exists \mathsf{e}_2 \!:\! \Separator.\,
  \forall \vec{\phi} \!:\! \Omega^{\Omega^{\mca}}.\,
  \coprod \vec{\phi} \xle{\mathsf{e}_2}_{\scriptscriptstyle\effshell} \bigcup \vec{\phi}
\end{array}$$
\end{restatable}

\subsubsection{The Correspondence of Simulations}
\label{sec:monadic-and-ef-simulations}

We now relate the operational and logical levels.  Relational image sends every simulation of
monadic cores to a regular EF simulation (soundness),  a singleton construction recovers a simulation of cores from every regular EF
simulation (completeness), and the two constructions are mutually inverse. In categorical terms, $\ef(-)$ is a strict 2-fully-faithful functor\footnote{A strict
2-functor~\cite{johnsonyau2categories} between preorder-enriched
categories is a functor that is order preserving on hom-preorders. It
is 2-fully-faithful when it induces an equivalence of hom-preorders.} from monadic cores to evidenced frames.

A simulation $\Sim : \effshell[1] \to \effshell[2]$ is a relation on
codes, while propositions of $\ef(\effshell[i])$ are predicates on
codes, so $\Sim$ acts on propositions by direct image: writing $\Sim(\mathsf{c}) \define
\{\, \mathsf{d} \mid \Sim(\mathsf{c},\mathsf{d}) \,\}$ for the image of
a single code, the translated proposition is
\[
  \Sim^{*}(\phi)
  \;\define\;
  \bigcup_{\mathsf{c} \in \phi} \Sim(\mathsf{c})
  \;=\;
  \{\, \mathsf{d} \mid \exists \mathsf{c} : \mca[1].\;
       \Sim(\mathsf{c},\mathsf{d}) \sqcap \phi(\mathsf{c}) \,\}.
\]
Thus a target code realizes the translated proposition when it
implements some source code realizing the original one: the semantic
analogue of a type translation.

\begin{restatable}[{Soundness \prooficon{efsoundness}}]{theorem}{efsoundness}
\label{thm:mca-simulation-induces-ef-simulation}
If $\Sim : \effshell[1] \to \effshell[2]$ is a simulation of monadic
cores, then $\Sim^{*} : \ef(\effshell[1]) \to \ef(\effshell[2])$ is a
regular EF simulation.
\end{restatable}

Conversely, an EF simulation between generated frames determines a
relation on the underlying codes: a target code simulates a source code
when it realizes the translation of the singleton proposition selecting
that code.

\begin{restatable}[{Completeness \prooficon{efcompleteness}}]{theorem}{efcompleteness}
\label{thm:ef-simulation-induces-mca-simulation}
If $F : \ef(\effshell[1]) \to \ef(\effshell[2])$ is a regular EF
simulation, then
$\Sim[F](\mathsf{c},\mathsf{d}) \define F(\{\mathsf{c}\})(\mathsf{d})$
is a simulation of monadic cores
$\Sim[F] : \effshell[1] \to \effshell[2]$.
\end{restatable}

In the completeness proof, the simulation~$F$ needs to be regular.
Because of Lemma~\ref{lemma:union-coprod}, $F(\phi)$ can then be computed (up to evidences)
from the images of the elements of~$\phi$
as~$\bigcup \{ F(c) \mid c \in \phi\}$ and it is the key point
of the proof.
The two constructions invert each other, strictly in one direction and
up to the evidence-preorder in the other.

\begin{restatable}[\prooficon{mcefcorrespondence}]{theorem}{mcefcorrespondence}
\label{thm:MCEFfullyfaithful}\label{prop:MCisom}
The assignment $\ef(-)$ extends to a strict 2-functor
$\MCcat \to \efreg$ which is 2-fully-faithful:
on hom-preorders, $\Sim \mapsto \Sim^{*}$ and $F \mapsto \Sim[F]$ are
mutually inverse, with
$  \Sim[\Sim^{*}] \;=\; \Sim$
and
$  (\Sim[F])^{*} \;\simeq\; F$,
the latter an equivalence in the preorder of $\efreg$.
\end{restatable}

\subsection{From Evidenced Frames to Triposes}\label{sec:ef-simulations-to-models}

This section shows that every evidenced frame generates a realizability model, a tripos of
uniform families of propositions~\cite{HJP_1980,Vanoosten_08}, and every EF simulation generates a
transformation of the triposes. As in \Cref{sec:MC-EF}, the passage is functorial, and on simulations the logical and the
model-level presentation carry exactly the same information.

\subsubsection{The Tripos of an Evidenced Frame}
\label{sec:ef-tripos}
The precise meaning of tripos varies across settings. Following the
$\Set$-based treatment in~\cite{EF2021}, we take a
tripos over the base topos $\E$ to be a
pseudofunctor\footnote{A pseudofunctor is a relaxation of a functor
wherein identities and composition need only be preserved up to
isomorphism.} $\T : \E\op \to \pHA$ valued in Heyting prealgebras,
with the structure below. Triposes over an arbitrary cartesian closed
category are standard in the
literature~\cite{Pittsthesis,Vanoosten_08}, and so are triposes
constructed from internal partial combinatory algebras in a
topos~\cite{Pittsthesis}.

\begin{definition}[$\E$-based Tripos]
\label{def:tripos-over-E}
An \emph{$\E$-based tripos} is a pseudofunctor~$\T : \E\op \to \pHA$ with:
\begin{description}
 \item[Quantifiers] For each morphism $s : \Gamma \to \Gamma'$ in $\E$,
 ${s^*}$ has both a left and a right adjoint $\coprod_{s}/\prod_{s}$ in $\PreOrd$, that is:
 $\coprod_{s}$ and $\prod_{s}$ are order-preserving maps,
 $ \phi \leq s^*(\psi)  \Leftrightarrow  \coprod_{s} (\phi) \leq \psi$ and $s^*(\phi)\leq \psi  \Leftrightarrow  \phi \leq \prod_{s} (\psi)$.

 \item[Compatibility] Given any pullback square $u \circ r = s \circ v$ in $\E$,
we have $s^* \circ \coprod_u \le \coprod_v \mathop{\circ} r^*$ and $\prod_v \mathop{\circ} {r^*} \le s^* \circ \prod_u$.

 \item[Generic predicate] An object~$\Omega' \in \E$, a predicate~$\holds \in \T(\Omega')$,
 and, for any object~$\Gamma \in \E$ and any predicate~$\phi \in \T(\Gamma)$,
 a morphism~\mbox{$\chi_\phi : \Gamma \to \Omega'$} such that $\chi_\phi^*(\holds) \simeq \phi$.

 \item[Reflective surjective substitutions] Given any \emph{regular epimorphism}~$s : \Gamma \to \Gamma'$ in $\E$, one has
   $s^* \phi \leq s^* \phi' \ \implies\ \phi \leq \phi'$.
\end{description}
\end{definition}

\begin{definition}[$\UFam$ over $\E$]
\label{def:ufam-body}
Given an EF $\ef = ( \Phi,E,\xle{}{})$ over $\E$, we define a
pseudofunctor $\UFam(\ef):\E^{\mathrm{op}}\to\pHA$ by:
\begin{itemize}[leftmargin=1.1em,itemsep=0.4em]
\item $\UFam(\ef)(\Gamma) := \Gamma \to \Phi$ (predicates in context),
\item the preorder on $\UFam(\ef)(\Gamma)$ is defined as follows:
for $\phi , \psi: \Gamma \to \Phi$,
$\phi\leq \psi$ holds if and only if
      $\vdashE{\E} \exists e\! : \: E.
      \forall\gamma \!: \! \Gamma.\ \phi(\gamma)\xle{e}\psi(\gamma)$,
\item substitution is precomposition,
\item quantifiers along $s:\Gamma\to\Gamma'$ are obtained using the
      EF's impredicative $\coprod,\prod:\Omega^\Phi\to\Phi$,
\item a generic predicate is given by $\Phi$ itself.
\end{itemize}
\end{definition}
\begin{restatable}[\prooficon{eftotripos}]{theorem}{eftotripos}\label{thm:ef-to-tripos}
For every evidenced frame $\ef$ over $\E$,
$  \UFam(\ef)$
is an $\E$-based tripos.  Moreover, $\UFam(\ef)$ is consistent iff
$\ef^\E$ is consistent, witnessed in the terminal context.
\end{restatable}

\subsubsection{The Correspondence of Transformations}
\label{sec:ef-tripos-correspondence}

We now connect EF simulations and tripos transformations.
A lax (resp. regular) EF simulation induces a
cartesian (resp. regular) transformation
between the triposes generated by
the two evidenced frames.
On each context, the induced translation is simply
postcomposition with the map on propositions.
As for arrow algebras~\cite{tarantino2026}, this passage is 2-fully-faithful (\Cref{prop:UFamisom}).
We adapt that development to
evidenced frames over an arbitrary base topos~$\E$, first fixing the
notion of transformation, following~\cite{Vanoosten_08}.

\begin{definition}[Cartesian and regular transformations of triposes]
    A cartesian transformation of triposes from~$\T_1$ to~$\T_2$ is an $\E$-indexed collection of functions
\mbox{$F_\Gamma: \T_1(\Gamma) \to \T_2(\Gamma)$} (for all $\Gamma\in \E$) such that:
\begin{description}
\item[Naturality] $F$ is a pseudonatural transformation
between $\T_1$ and $\T_2$ considered as pseudofunctors
valued in
preorders\footnote{It means that for $X : \E$,
$F_X : \T_1(X) \to \T_2(X)$ needs only to preserve the order.};
\item[Meet preservation] $F$ preserves finite meets up to isomorphism.
\end{description}
A regular transformation also satisfies the following extra condition:
\begin{description}
    \item[Regularity]
 $\forall s : \Gamma \to \Gamma'.\; \forall \gamma \in \T_1(\Gamma).\; F_{\Gamma'}(\coprod_{s} \gamma) \;\leq\; \coprod_s\, F_\Gamma(\gamma)$.
\end{description}
\end{definition}

\begin{definition}[Categories {$\tripcart$} and {$\tripreg$}]
Objects of $\tripcart$ and $\tripreg$ are $\E$-based triposes.
Morphisms of~$\tripcart$ ($\tripreg$) are cartesian (regular) transformations, respectively.
We define a preorder \mbox{$F \leq G$} on morphisms
from~$\T_1$ to~$\T_2$
(in $\tripcart$ and $\tripreg$), it holds when
for all $\Gamma : \E$ and for all $\gamma \in \T_1(\Gamma)$
$ F_\Gamma (\gamma) \leq_{\T_2(\Gamma)} G_\Gamma (\gamma)$.
This makes
$\tripcart$ and $\tripreg$ preorder-enriched categories.
\end{definition}

\begin{restatable}[\prooficon{efsimtomodel}]{theorem}{efsimtomodel}\label{thm:ef-simulation-induces-model-morphism}
Let
 $ F:\ef_1\to\ef_2$
be a lax EF simulation.
Then $F$ induces a cartesian transformation of the
associated realizability models
 $ \UFam(F):\UFam(\ef_1)\to\UFam(\ef_2)$, whose action in context
$\Gamma$ is postcomposition:
$\UFam(F)_\Gamma(\phi) \define F\circ \phi$ for
$\phi:\Gamma\to\Phi_1$. Moreover, if $F$ is regular then so is
$\UFam(F)$.
\end{restatable}

The operation $\UFam(\_)$ is in fact an isomorphism of preorders
between the hom-sets of~$\eflax$ and~$\tripcart$. The next lemma
is used to construct its inverse.

\begin{restatable}[\prooficon{triptrans}]{lemma}{triptrans}\label{lemma:triptrans}
    Let $\tau : \UFam(\ef_1) \to \UFam(\ef_2)$ be a pseudonatural transformation. For all $X$ and~$\chi : \UFam(\ef_1)(X)$,
    $\tau(X)(\chi) \simeq \tau(\Phi_1)(\id) \circ \chi$.
\end{restatable}

It shows that a transformation $\tau$ between
effectful triposes is fully characterized
by $\tau(\Phi_1)(\id) : \Phi_1 \to \Phi_2$.
Therefore, we write~$F_{\tau} \define \tau_{\Phi_1}(\id)$ and
we show that $F_{\tau}$ is a lax (resp. regular) simulation of~$\ef$ in the case that
$\tau$ is cartesian (resp. regular).

\begin{restatable}[\prooficon{cartesiantolax}]{proposition}{cartesiantolax}\label{prop:appx-ef-1}
    If $\tau : \UFam(\ef_1) \to \UFam(\ef_2)$ is cartesian
    transformation then $F_{\tau}\define \tau_{\Phi_1}(\id)$ is a lax simulation.
    Moreover, if it is regular, $F_{\tau}$ is also a
    regular simulation.
\end{restatable}

The two constructions invert each other, strictly in one direction and
up to the pseudonatural isomorphism of \Cref{lemma:triptrans} in the
other.
Thus, EF simulations are  the logical presentation of transformations of
realizability models, completing the second step of the ladder.

\begin{restatable}[{Correspondence \prooficon{ufamcorrespondence}}]{theorem}{ufamcorrespondence}
\label{prop:UFamisom}
The assignment $\UFam(-)$ extends to a strict 2-functor
$\eflax \to \tripcart$ which is 2-fully-faithful: on hom-preorders,
$F \mapsto \UFam(F)$ and $\tau \mapsto F_{\tau}$ are mutually inverse,
with
 $ F_{\UFam(F)} \;=\; F$ and $
  \UFam(F_{\tau}) \;\simeq\; \tau$ .
Moreover, $\UFam(-)$ restricts to a 2-fully-faithful functor
$\efreg \to \tripreg$.
\end{restatable}

\subsection{Geometric Morphisms
and Computationally Dense Simulations}
\label{sec:cd-simulations}

A simulation transports reasoning in one direction. To reason about what an
implementation \emph{preserves}, one must send a source specification into the
target, reason there, and pull the conclusion back, witnessed by evidence. The
classical answer is an adjunction: a \emph{geometric} morphism, a translation $F$
with a right adjoint $H$, is the standard notion of logical transport between
toposes, and in pure realizability it is what computationally dense applicative
morphisms induce on the triposes~\cite{HOFSTRA_VANOOSTEN_2003,Vanoosten_08}.

We define geometric simulations at the level of evidenced frames and triposes,
show the correspondence of \Cref{sec:ef-simulations-to-models} restricts to them,
and  identify their operational counterpart: computational density, which
produces geometric simulations over any base, with a converse exactly over $\Set$.

\begin{definition}[Geometric simulation of EF]
\label{def:computational-density}
Let
  $F:\ef_1\to\ef_2$
be an EF simulation. We say that~$F$ is \emph{geometric} if there exists
a lax simulation $H:\Phi_2 \to \Phi_1$ such that
$$\vdashE{\E} \exists e_\varepsilon \!:\!E_2. \;
\forall \phi \!: \! \Phi_2.\; F(H_\Phi(\phi)) \xle{e_\varepsilon} \phi \qquad
\qquad \vdashE{\E} \exists e_\eta \!:\! E_1. \;
\forall \phi \!: \! \Phi_1.\; \phi \xle{e_\eta} H_\Phi(F(\phi))
$$
Using the preorder
between lax simulations, it can be
rephrased as $F \circ H \leq \id$ and $\id \leq H \circ F$. We write
$F \dashv H$ and say that $H$ is a right adjoint of $F$.
Evidenced frames and geometric simulations
form a subcategory~\efcd{} of~\efreg{}.
\end{definition}

    In the case that $F \dashv H$, $F$ is always regular.
    Moreover, in the previous definition, it is enough for
    $H$ to be monotone (in the sense of Definition~\ref{def:ef-sims}):
    as a right adjoint, it will imply that it respects
    finite meets
    (and therefore that it is a lax simulation).

\begin{example}[Booleanization]
\label{ex:double-negation2}
The regular simulation $L_P$ from example~\ref{ex:double-neg}
    is in fact a geometric simulation: it admits $R_P$
    as right adjoint.
\exqed
\end{example}

\begin{definition}[Geometric transformation of triposes]
\label{def:geomtrictransformation}
Let
  $F:\T_1\to\T_2$
be a tripos transformation. We say that $F$ is \emph{geometric} if there exists $H:\T_2\to\T_1$ such that $F \circ H \leq \id$ and $\id \leq H \circ F$. We write
$F \dashv H$ and say that $H$ is a right adjoint of $F$.
Triposes and geometric transformations
form a subcategory~\tripgeom{} of~\tripreg{}.
\end{definition}

The correspondence between EF simulations and tripos transformations
respects adjoints on both sides, so it restricts to the geometric
subcategories: geometric simulations are exactly the frame-level
presentation of geometric transformations of realizability models.

\begin{restatable}[{Geometric EF simulations induce logical transport \prooficon{geomtransport}}]{theorem}{geomtransport}\label{thm:dense-ef-simulation-logical-transport}
The functor~$\UFam$
restricts to a strict 2-fully-faithful functor
from~$\efcd$
to $\tripgeom$.
\end{restatable}

We now descend to the operational level and ask which simulations of
monadic cores induce geometric EF simulations.
In the $\Set$-based setting with partial combinatory algebras, this is the
notion of computational density~\cite{HOFSTRA_VANOOSTEN_2003,Vanoosten_08} and
it
captures when the target does not provide new ways to compute on data from the source.
Its generalization to monadic cores is the following.

\begin{definition}[Computationally Dense Simulation]
A simulation $\Sim \! :\! \effshell[1] \to \effshell[2]$ is said to be \emph{computationally dense} if there exists a code
$\evCD \! :\! 1 \to \Separator[2]$ such that
$$ \vdashE{\E}
\forall b \!:\! \Separator[2].\; \exists \mathsf{a} \! :\! \Separator[1]. \;
\forall \mathsf{a'} \!:\! \mca[1] .\; \forall \phi_2 \! :\! \Omega^{\mca[2]}.
\; \left(\Sim^* (\{\mathsf{a'}\}) \xle{\mathsf{b}} \phi_2  \Longrightarrow
\after{\mathsf{c_a}}{\mathsf{a} \cdot \mathsf{a'}}{\Sim^*(\{\mathsf{c_a}\}) \xle{\evCD} \phi_2} \right).
$$
\end{definition}

Note that $\evCD$ is required to exist as a global element, not
merely internally. This is what
orients the two propositions below: a global $\evCD$ is exactly what
the proof of \Cref{prop:appx-ef-2} (1) consumes, whereas an internally
stated density is what a geometric simulation can produce.

\begin{example} [Continuation]
\label{ex:continuation3}
    The  simulation in Example~\ref{ex:continuation2} from~$\MCId$
    to~$\MCCont$
    is computationally
    dense. Define $\mathsf{b}_{\epsilon} \define \id$
    and for $\mathsf{b} \in \mca$ , pick
    $\mathsf{a} \define \mathsf{b}$. Let $a'  \in \mca$ and $\phi : \Omega^{\mca}$,
    then computational density unfolds to:
$
\{\mathsf{a'}\} \xle{\mathsf{b}}_{\mathrm{Cont}} \phi  \Longrightarrow
\{\mathsf{b} \cdot \mathsf{a'}\} \xle{\id}_{\mathrm{Cont}} \phi,
$
which follows from the definitions of~$\modality^\Cont$ and of
$\App^{\mathrm{Cont}} \define \eta \circ \App$.
However, the simulation
from~$\MCCont$ to $\MCContcc$ is not computationally dense: \texttt{call/cc} in the
target cannot be simulated in the source.
\exqed
\end{example}

\begin{restatable}[\prooficon{cdgeometric}]{proposition}{cdgeometric}\label{prop:appx-ef-2}
The following hold:
\begin{enumerate}
    \item
    If $\Sim : \effshell[1] \to \effshell[2]$
    is a computationally dense simulation,
    the induced regular EF simulation~$\Sim^{\ast}$
    is geometric.
    \item In~$\Set$, if
$
  F:\ef(\effshell[1])\to\ef(\effshell[2])
$ is a geometric EF simulation,
the induced simulation
$
  \Sim[F]:\effshell[1]\to\effshell[2]
$ is computationally dense.
    \end{enumerate}
\end{restatable}

\definecolor{recapOut}{HTML}{FDF4E6}
\definecolor{recapMid}{HTML}{F7E3C2}
\definecolor{recapIn}{HTML}{EFCE9C}
\definecolor{recapStroke}{HTML}{C7A56A}%
\definecolor{recapLab}{HTML}{5A4A2E}
\definecolor{recapWorld}{HTML}{A99F8C}
\definecolor{recapArr}{HTML}{12726E}

\begin{wrapfigure}{r}{0.42\linewidth}
  \vspace{-0.7cm}
  \centering
  \resizebox{0.82\linewidth}{!}{%
\begin{tikzpicture}[
    x=.7cm, y=.7cm,
    ring/.style ={draw=recapStroke, line width=.5pt},
    slab/.style ={text=recapLab, font=\footnotesize, inner sep=0pt},
    wlab/.style ={text=recapWorld, font=\footnotesize\scshape,
                  anchor=east, align=right},
    cor/.style  ={recapArr, line width=.6pt,
                  {Stealth[length=1.5mm]}-{Stealth[length=1.5mm]}},
    scor/.style ={recapArr, line width=.6pt, -{Stealth[length=1.5mm]}},
    oil/.style  ={text=recapArr, font=\tiny\itshape, anchor=west}
  ]

    \draw[ring, fill=recapMid] (-0.75,0) ellipse [x radius=1.55, y radius=.92];
    \draw[ring, fill=recapIn]  (-1.33,0) ellipse [x radius=0.85, y radius=.62];
    \node[slab] at (-1.33,0) {CD};
    \node[slab] at ( 0.16,0) {App.};
    \node[wlab] at (-2.60,0) {Monadic\\cores};

    \draw[ring, fill=recapOut] ( 0.00,-3.1) ellipse [x radius=2.3,  y radius=1.4];
    \draw[ring, fill=recapMid] (-0.63,-3.1) ellipse [x radius=1.55, y radius=.92];
    \draw[ring, fill=recapIn]  (-1.21,-3.1) ellipse [x radius=0.85, y radius=.62];
    \node[slab] at (-1.21,-3.1) {Geo.};
    \node[slab] at ( 0.28,-3.1) {Reg.};
    \node[slab] at ( 1.61,-3.1) {Lax};
    \node[wlab] at (-2.60,-3.1) {Evidenced\\frames};

    \draw[ring, fill=recapOut] ( 0.00,-6.2) ellipse [x radius=2.3,  y radius=1.4];
    \draw[ring, fill=recapMid] (-0.63,-6.2) ellipse [x radius=1.55, y radius=.92];
    \draw[ring, fill=recapIn]  (-1.21,-6.2) ellipse [x radius=0.85, y radius=.62];
    \node[slab] at (-1.21,-6.2) {Geo.};
    \node[slab] at ( 0.28,-6.2) {Reg.};
    \node[slab] at ( 1.61,-6.2) {Cart.};
    \node[wlab] at (-2.60,-6.2) {Triposes};

    \draw[cor]  ( 0.22,-0.73) -- ( 0.22,-2.35);
    \draw[scor] (-1.39,-0.62) -- (-1.39,-2.48);
    \draw[scor, dashed] (-1.15,-2.48) -- (-1.15,-0.62);
    \node[oil] at (-1.3,-1.56) {(in Set)};

    \draw[cor] (-1.21,-3.72) -- (-1.21,-5.58);
    \draw[cor] ( 0.28,-3.87) -- ( 0.28,-5.43);
    \draw[cor] ( 1.61,-4.10) -- ( 1.61,-5.20);
  \end{tikzpicture}
    }%
  \label{fig:recap-strata}
  \vspace{-0.4cm}
\end{wrapfigure}

The asymmetry in the above proposition has a precise source. The proof of
\Cref{prop:appx-ef-2} (2) extracts $\evCD$ from the counit of
$F \dashv H$, but over an arbitrary base the adjunction provides that
counit only internally, as
$\vdashE{\E} \exists e_\varepsilon : E_2.\ \forall \phi : \Phi_2.\
F(H(\phi)) \xle{e_\varepsilon} \phi$, and internal existence does not
yield a global element $\evCD : 1 \to \Separator[2]$. Over $\Set$ the
two coincide, which is exactly why the converse holds there.

The correspondences established in this section are summarized in the figure on the right,
which connects the different types of
simulations across the three levels of the framework.

\section{Transport-derived Modalities}
\label{sec:transport-modalities}

A simulation transports the source logic into the target, but even a geometric
transport rarely preserves it on the nose. The discrepancy is itself structured:
every geometric simulation, with translation $F:\ef_{src}\to\ef_{tgt}$ and
abstraction $H:\ef_{tgt}\to\ef_{src}$, induces a closure
$HF:\ef_{src}\to\ef_{src}$ on the source logic, an \emph{$\ef$-nucleus}, and the
transport factors through it:
\[
  \ef_{src} \;\xrightarrow{\ \id\ }\; \ef_{src}^{\,HF}
           \;\xrightarrow{\ F\ }\; \ef_{tgt}.%
\]
Here $\ef_{src}^{\,HF}$ is the source logic closed under the induced modality.
The first map inserts the abstraction barrier and the second transports only
invariant-closed reasoning. Thus the barrier a translation enforces is not
postulated but \emph{computed} from the translation itself.

Closure operators on Heyting algebras and locales are
standard~\cite{johnstone2002sketches,maclane1992sheaves}, and our
$\ef$-nuclei are their arrow-algebra
analogue~\cite{tarantino2026}. Our contribution is at the
level of evidence: the return and bind of a nucleus are explicit codes, so
closing a specification is itself a computation and $\ef_j$ is again an
evidenced frame on the \emph{same} propositions and evidence, and the
modality is \emph{derived} from a transport rather than postulated as in
classical realizability~\cite{Krivine03}.

We define $\ef$-nuclei and  prove a factorization theorem (\Cref{sub:nuclei}),
and then instantiate it at sheafification (\Cref{sub:sheaf}), where the derived route
meets its ceiling and a built-in construction takes over.

\subsection{Nuclei and Factorization}
\label{sub:nuclei}

An $\ef$-nucleus is a closure on a program logic, presented in the same
evidence language as the underlying evidenced frame. It is a strong monad
on the frame: a \emph{return} turning a proposition into its closure, and
a \emph{bind} propagating entailments through the closure.

\begin{definition}[nucleus]
Let $\ef = (\Phi, E , \cdot \xle{\cdot} \cdot)$ be an evidenced frame.
    A map $j : \Phi \to \Phi$ is an $\ef$-nucleus if
    \begin{description}
        \item[Return]
         $\vdashE{\E} \exists \eR \!:\! E. \; \forall \phi \! :\! \Phi. \; \phi \xle{\eR} j (\phi)$
        \item[Bind] $\ \vdashE{\E} \exists
        \eBind \!:\! E. \;
        \forall \phi_1, \phi_2 \! : \Phi. \;
        \phi_1 \arrowE j(\phi_2) \xle{\eBind} j(\phi_1) \arrowE j(\phi_2)$
    \end{description}
\end{definition}
The bind in the definition implies, up to evidence, its externalized
forms: any realized entailment $\phi_1\xle{e}\phi_2$ can be transposed to
$\top \xle{e'}\phi_1\arrowE \phi_2$ and back. We use this transposition
freely below, since it lets us reason about $j$ either as an operator on
propositions or as an operator on entailments.
Returns and binds make $j$ a closure.%

A nucleus determines a new evidenced frame on the \emph{same} propositions
and evidences, in which entailment tests its conclusion after closure.

\begin{restatable}[{EF from nucleus \prooficon{prop:nucleustoef}}]{proposition}{prop:nucleustoef}
\label{prop:appx-dermod-1}
   Given an evidenced frame~$\ef = (\Phi, E , \cdot \xle{\cdot} \cdot)$  and a nucleus $j:\ef \to \ef$,
   define
     $\efj \define (\Phi, E, \cdot \xlej{\cdot} \cdot)$
where $$\phi_1 \xlej{e} \phi_2 \define  \phi_1  \xle{e} j(\phi_2),$$
with
$
     \topj  \define  \top ,
     \phi_1 \andj \phi_2 \define  \phi_1 \wedge \phi_2 , ~~~\text{and}~~~
     \phi_1 \arrowj \vec{\phi}  \define  \phi_1 \arrowE \{ j (\phi) \mid
\phi \in \vec{\phi}\} $.
    Then, $\efj$ is an
    evidenced frame.
\end{restatable}

Adding a nucleus does not change the raw evidence language, it merely changes
what counts as a stable conclusion. The passage $\ef\to\ef_j$ is itself a
geometric transport.

\begin{restatable}[\prooficon{prop:nucleusadj}]{proposition}{prop:nucleusadj}
\label{prop:appx-dermod-2}
     If $j$ is a nucleus,
    $id : \ef \to \efj$
    is a geometric simulation with $j$ as right adjoint.
\end{restatable}

The above proposition shows that every nucleus arises
from a geometric simulation. The factorization theorem says the converse: \emph{every} geometric simulation induces a nucleus, and
the transport factors through the modal frame it induces.

\begin{restatable}[{\prooficon{prop:adjnucleus}}]{proposition}{prop:adjnucleus}
\label{prop:appx-dermod-4}
    Let $F : \ef_1 \to \ef_2$
be a geometric simulation and $H$ a right adjoint to~$F$.
Then $HF : \ef_1 \to \ef_1$ is a nucleus.
\end{restatable}

As an operator on the underlying preorder, $HF$ is just the monad of the adjunction
$F\dashv H$, and its idempotency up to evidence is the standard closure attached to
a reflective adjunction. The content of the proposition is therefore not the closure
but its \emph{realizers}: the bind (equivalently, the strength of the monad) is an
explicit evidence term built from the laxness of $H$, the counit of $F\dashv H$, and
$\eeval$. Thus $HF$ is a nucleus in the \emph{evidence language} of $\ef_1$, not
merely on its preorder, which is what makes $\ef_1^{HF}$ an evidenced frame and
\Cref{prop:factor} proof-relevant.

\begin{example}[Booleanization]
    The geometric simulation  $L_P$ from Example~\ref{ex:double-negation2}
induces the nucleus~$\phi \mapsto (\phi \arrowE P) \arrowE P$ on~$\ef$.
\exqed\end{example}

\begin{example}[Continuation]
\label{ex:continuation4}
    The computationally dense simulation from~$\MCId$ to~$\MCCont$ of Example~\ref{ex:continuation3}
    induces a geometric morphism between the corresponding EFs and therefore
    also a nucleus~$j$, given at the level of EFs by the right adjoint,
    $\phi \mapsto
    \{\mathsf{a} \mid \{\mathsf{a}\} \xle{\id}_{\Cont} \phi\}
    $ (see Proof of~\Cref{prop:appx-ef-2}).
    In~$\ef(\MCCont)$, classical logic is not realized as it misses
    computational apparatus to handle continuation (\texttt{call/cc} and continuation
    constants).
    The second step, as mentioned in Example~\ref{ex:continuation2},
    is to add these codes to~$\MCCont$: it gives the monadic core~$\MCContcc$
    and an applicative simulation from~$\MCId$ to~$\MCContcc$.
    At the level of evidenced frames, this morphism lifts to a regular
    simulation that factors through~$\ef(\MCId)_j$.
    We see~$\ef(\MCId)_j$ as the first step (or the direct image) of the CPS
    translation.
    At this stage, no computational content is added but the ambient modality
    has changed,
    enabling the definition of codes implementing
    control operators.
    \exqed
\end{example}

\begin{restatable}[Factorization {\prooficon{propfactor}}]{proposition}{propfactor}
\label{prop:factor}
    Let $F : \ef_1 \to \ef_2$
be a geometric simulation with $H$ as adjoint,
and let $\ef_{HF}$ be the evidenced frame constructed
from the nucleus $HF$. Then $F : \ef_{HF} \to \ef_2$
is a geometric simulation (with $H$ as adjoint).
\end{restatable}

Intuitively, the composite $HF$ measures the effect of translating a source specification
to the target and then abstracting it back to the source. The factorization
through $EF_1^{HF}$ says that the adjoint transport induced by $F \dashv H$
does not see arbitrary source specifications directly, but sees them through
their $HF$-closure.

\subsection{Built-in Modality}
\label{sub:sheaf}

The factorization theorem produces a modality from a transport, and
\Cref{sub:nuclei} turns any such nucleus into a modal program logic by closing
the \emph{postconditions} of the source logic.  This \emph{derived} route is
canonical, it comes for free from a transport, but it has a ceiling: a closure of
conclusions can only ever act on conclusions.  A modality that must constrain the
operational \emph{run} itself, for instance one requiring termination only
locally, lies beyond it.  We therefore give an alternative
construction%
, the \emph{built-in} modal evidenced frame, which boxes the whole
operational judgement rather than only the postcondition.

\label{subsub:modal-machinery}

\begin{definition}[Modality as a nucleus on $\Sub_\E(-)$]
\label{def:modality-nucleus}
A \emph{modality} is a family of closure operators
$$  \Box_\Gamma:\Sub_\E(\Gamma)\to \Sub_\E(\Gamma)
\qquad(\Gamma\in \E)$$
such that for every $h:\Delta\to \Gamma$ and $A,B\mono \Gamma$ the following conditions hold:
\begin{description}
\item[ \textit{Extensive}] $A\sqsubseteq \Box_\Gamma A$.
\item[ \textit{Idempotent}]  $\Box_\Gamma(\Box_\Gamma A)=\Box_\Gamma A$.
\item[ \textit{Monotone}]  $A\sqsubseteq B\Longrightarrow \Box_\Gamma A\sqsubseteq \Box_\Gamma B$.
\item[ \textit{Pullback stability}]  $h^*(\Box_\Gamma A)=\Box_\Delta(h^*A)$.
\item[ \textit{Finite meets and top}] $\Box_\Gamma(A\sqcap B)=\Box_\Gamma A\sqcap \Box_\Gamma B$
      and $\Box_\Gamma(\top)=\top$.
\end{description}
\end{definition}
Every such $\Box$ is induced by a Lawvere--Tierney topology $j:\Omega\to\Omega$
\cite{johnstone2002sketches,maclane1992sheaves}: $\Box_\Gamma$ acts on
$m:S\rightarrowtail\Gamma$ by postcomposing its classifier $\chi_m:\Gamma\to\Omega$
with $j$, and conversely applying $\Box_{\Omega}$ to the generic subobject
$\top:1\rightarrowtail\Omega$ recovers such a $j$ by pullback stability.

We fix a \emph{computational system} $\mathbf{C}$ in $\E$, that is,  an abstract effectful
operational semantics internal to the topos. It provides an object of states
$\Sigma$ with a ``possible future'' preorder $\sigma\le\sigma'$, an object of codes
$C$, a state-threading application $e_f\cdot e_a \downarrow^{\sigma}_{\ \sigma'} c_r$
(reduce $e_f\cdot e_a$ from $\sigma$, reaching $\sigma'$ with result $c_r$), and a
termination predicate $e_f\cdot e_a \Downarrow_\sigma$. These satisfy the
state-passing laws of~\cite{EF2021}: application is reflexive and
compositional, reduction only passes to futures (preservation), and termination
yields a reduct (progress).
Moreover, the calculus is \emph{functionally complete},
so every metalevel expression is realized by a code. A \emph{separator}
$S\rightarrowtail C$ singles out the admissible evidence codes: a functionally
complete subobject, closed under reduction and satisfying progress.
Intuitively, this is an operational semantics with a mutable store,
axiomatized flexibly to accommodate any reduction discipline, deterministic
or not, terminating or divergent. The full axiomatization can be found in~\cite{EF2021}.

We write $\Phi_0\subseteq \Sub_\E(\Sigma\times C)$ for the object of future-stable
predicates, i.e., subobjects $\phi\rightarrowtail \Sigma\times C$ satisfying
$(\sigma\le \sigma'\ \wedge\ \phi(\sigma,c))\ \Longrightarrow\ \phi(\sigma',c)$.
We define the \emph{angelic} operational judgement:
\[
(e_f\cdot e_a \Downarrow_\sigma \phi)\ \define\
e_f\cdot e_a \Downarrow_\sigma\ \wedge\
\exists \sigma',c'.\ \big(e_f\cdot e_a \downarrow^{\sigma}_{\ \sigma'} c'\big)\wedge \phi(\sigma',c'),
\]
that is, the computation terminates at $\sigma$ and \emph{some} reduct satisfies $\phi$.

\begin{restatable}[{Modal EF \prooficon{modalefsound}}]{theorem}{modalefsound}
\label{thm:modal-ef-sound}
Given a computational system $\mathbf{C}$, a separator $S\rightarrowtail C$, and a
modality $\Box$, define
$\efboxm\ \define\ \langle \Phi_0,\ S,\ \cdot\xle{\cdot}_\Box\cdot\rangle$,
where
$$\phi_1 \xle{e}_\Box \phi_2 \define
\forall \sigma,c.\ \phi_1(\sigma,c)\ \Longrightarrow\
\Box_{\Sigma\times C}\big(\, e\cdot c\ \Downarrow_{\sigma}\ \phi_2 \,\big),$$
with $\top(\sigma,c) \coloneqq \top_\Omega$,
$(\phi_1\wedge\phi_2)(\sigma,p) \coloneqq
\exists c_1,c_2.\ p=\cpair{c_1}{c_2} \wedge \phi_1(\sigma,c_1)\wedge \phi_2(\sigma,c_2)$,
and $(\phi_1\imp\vec\phi)(\sigma,f) \coloneqq
\forall\sigma'\ge\sigma,\,c,\,\phi\in_{\Phi_0}\vec\phi.\ \phi_1(\sigma',c)\Rightarrow
\Box\big(f\cdot c\Downarrow_{\sigma'}\phi\big)$.
Then $\efboxm$ is an evidenced frame over $\E$.
\end{restatable}

Note that for the demonic operational judgement
$e_f\cdot e_a \Downarrow^{\D}_\sigma \phi \define e_f\cdot e_a\Downarrow_\sigma
\wedge \forall\sigma',c'.\,(e_f\cdot e_a\downarrow^\sigma_{\sigma'}c')\Rightarrow
\phi(\sigma',c')$, \Cref{thm:modal-ef-sound} additionally requires $\Box$ to
commute with the universal along the reduction relation,
$\forall_\pi\Box\sqsubseteq\Box\forall_\pi$ (for transitivity).
In particular, for
\emph{deterministic} systems the reduct is unique, so the demonic and angelic readings coincide, so
\Cref{thm:modal-ef-sound} is unconditional.

\begin{example}[Sheafification]
\label{ex:sheaf}
This example compares the derived and built-in modalities over the one base topos
in the paper that is not $\Set$: a \emph{presheaf} topos with a Grothendieck
topology as observation policy, where the paper's topos-level generality is
exercised. The two constructions \emph{coincide} for stage-independent computation
and \emph{come apart} once computation is local, where only the built-in
construction survives.

Take $\E=\Psh(\C)$ for a small site $(\C,J)$ and let $\Box_j$ be the modality of
its Lawvere--Tierney topology: $\Box_j\phi(U)=\{a\mid\chi_\phi(a)\in J(U)\}$
with $\chi_\phi(a)=\{f:V\to U\mid a|_f\in\phi(V)\}$, the internal
presheaf-topos form of sheafification.
We use $\C=\{G,L,R\}$ with only
$\ell:L\to G$, $r:R\to G$ (and identities) and $J(G)=\{\{\ell,r\},\text{max}\}$,
every covering sieve inhabited.
Then $\Box_j$ is the identity at $L,R$ and
$\Box_j\phi(G)=\{a\mid a|_\ell\in\phi(L)\wedge a|_r\in\phi(R)\}$.

The derived construction of \Cref{sub:nuclei} imports $\Box$ into the logic as the
postcondition closure $j_\Box:\phi\mapsto\Box_{\Sigma\times C}\phi$: when this
is an $\ef$-nucleus, \Cref{prop:appx-dermod-1} gives the frame $\efboxmj$ with
\emph{postcondition-boxing} entailment
$\phi_1\xle{e}_{j_\Box}\phi_2 \iff \forall\sigma,c.\,\phi_1(\sigma,c)\Rightarrow
e\cdot c\Downarrow_\sigma\Box\phi_2$, closing only what the program delivers. Its
return is the identity code, but the crux is the bind.

Fix a PCA $\mca$ in $\Set$ and consider the constant presheaf
$\Delta(\underline{\mca})$. The \emph{constant system} has $\Sigma=1$, codes and
separator $\mathbf C=S=\Delta(\underline{\mca})$, and stage-independent (hence
deterministic) PCA application as reduction; predicates are sub-presheaves
$\phi\rightarrowtail\Delta(\underline{\mca})$. Because application is
stage-independent, $e\cdot c$ has the same value at every stage, so boxing the run
and boxing the postcondition agree:
$\Box_j(e\cdot c\Downarrow\phi)=(e\cdot c\Downarrow\Box_j\phi)$ as subobjects
of $1$. Thus $j_\Box:\phi\mapsto\Box_j\phi$ is an $\ef$-nucleus on $\efD$, and
the derived route yields the modal frame $\efboxmj$.

Fix distinct $a_0,a_1\in\mca$ and let
$\phi(G)\define\{a_0\}$, $\phi(L)=\phi(R)\define\{a_0,a_1\}$.
Here $a_1$
is a \emph{potential global element}, present at every stage of the cover but absent
at $G$, and $\Box_j\phi(G)=\{a_0,a_1\}$.  With $e_1\define\mathbf k\,a_1$, the
entailment $\top\xle{e_1}\phi$ \emph{fails} in $\efD$ (as $e_1\cdot c=a_1\notin
\phi(G)$) but \emph{holds} in $\efboxmj$ since $e_1$ is correct locally on the
cover, and the modality accepts this local correctness.  For
stage-independent computation the base modality acts purely through
postconditions, and the derived route captures it.

To see the ceiling of the derived route, take codes and separator global, $\mathbf C=S=\Delta(\N)$, but let reduction
at stage $U$ be Kleene application relative to a partial oracle $\omega_U$, with
$\omega_G\define\emptyset$, $\omega_L\define\lambda n.0$,
$\omega_R\define\lambda n.1$.
Convergent computations persist under oracle
extension, so reduction and termination are sub-presheaves, and the combinators
(never querying the oracle) give functional completeness uniformly.
Let $q\in\N$
query the oracle at its argument: $q\cdot c$ diverges at $G$ but returns $0$ at
$L$ and $1$ at $R$, so termination is genuinely stage-dependent.
Take, for example, $\mathsf{bool}\define\underline{\{0,1\}}$.  Then $\top\xle{q}\mathsf{bool}$
\emph{fails} in $\efD$ and $\top\xle{q}_{j_{\Box}} \mathsf{bool}$ also fails: both demand the bare
conjunct $q\cdot c\Downarrow$ at $G$, which is false.

Here $j_\Box$ is not an $\ef$-nucleus at all: it has no bind. The bind would require
$(e\cdot(-)\Downarrow\Box_j\psi)$ to be $\Box_j$-closed, i.e.\ the inclusion
$\Box_j(e\cdot c\Downarrow\psi)\sqsubseteq(e\cdot c\Downarrow\Box_j\psi)$, which
fails when termination is stage-dependent.
With $e=q$, $\psi=\mathsf{bool}$ at $G$, the left side is $\top$ (the run converges
on the cover) while the right is $\bot$ (no convergence at $G$).  A closure of
conclusions cannot reach the termination conjunct, which lives in the run, not the
postcondition.

To capture such a modality we insert $\Box$ \emph{inside} the operational
judgement rather than after it via the modal EF construction of Theorem~\ref{thm:modal-ef-sound}.
In the oracle system, $\top\xle{q}\mathsf{bool}$ now \emph{holds} in $\efboxm$: at
$L,R$ the plain judgement holds, and at $G$ it holds on the cover $\{\ell,r\}$,
hence under $\Box_j$.  The built-in construction reaches exactly the local
termination the derived route could not.
\exqed
\end{example}

\section{Related Work}
\label{sec:related}

\paragraph{Morphisms of realizability models.}

Applicative morphisms were first introduced in~\cite{Longley95}, relating PCAs and establishing the connection to $\Delta\Gamma$-functors between assemblies.
This concept was later extended to typed PCAs~\cite{Longley1999}
and to C-structures in~\cite{longley2014computability}.
Sub-PCAs were developed in~\cite{awodey1999elementary,awodey2002local,birkedal1999developing}, following the ideas of relative realizability developed in~\cite{Kleene1965}. A notion of morphism for relative realizability appears in~\cite{BirkedalVanOosten2002}.
Applicative morphisms between PCAs are the special case of simulations between monadic cores with the partiality monad, the total correctness modality, and the trivial separator.
Furthermore, sub-PCAs are the special case when the separator is non-trivial.
On the algebraic side of realizability, morphisms were studied
in the frameworks of arrow algebras~\cite{Berg2023ArrowA, tarantino2026} and implicative algebras~\cite{Miquel20}.
We use in this paper evidenced frames, which are closer to uniform preorders~\cite{frey2014}, and as for arrow algebras, the correspondence we obtain between evidenced frames and triposes gives a 2-fully-faithful functor.
To the best of our knowledge, the most general development of the concept of simulations was done in~\cite{cockett2010categorical}, which relates computability structures by a relational, lax notion of tracking. Our simulations share this relational and lax character, and the relation on codes underlying one can be read as a simulation of that kind.
A precise formal comparison is left to future work.

\paragraph{Monads, effects and modalities}

Monads model computational effects~\cite{moggi1989computationallamdba,moggi1991notions},
a view completed by the algebra--monad--effect correspondence for algebraic
effects~\cite{plotkin2001semantics}.
Modalities over a monad, introduced as modal
operators~\cite{moggi1991notions} and developed in evaluation
logic~\cite{pitts1991evaluation,moggi1995semanticsEL}, specify how effectful outcomes
are observed. This is the role of the $M$-modality in a monadic core. Their fibrational
form is the \emph{lifting} of a monad~\cite{Katsumata2005,goubault2008logical,Katsumata2013,katsumata2018codensity,aguirre2020weakest,KammarKatsumata22},
where a monad in the fibers binds specifications to computations, and modalities are
precisely such liftings over the subobject classifier~\cite{aguirre2020weakest}.
Related uses include quantitative logics for behavioural
equivalence~\cite{voorneveld2019quantitative}, preorders on
monads~\cite{KatsumataSato2013}, and the effect observations of Dijkstra
monads~\cite{ahman2017dijkstra,maillard2019dijkstra}.
All of these lift or index
structure over a \emph{single, fixed} monad, within one semantics. Our simulations
instead relate two cores, over possibly different monads, transporting the entire
realizability logic together with its evidence from one to the other.
Similarly, sound and complete
program logics for algebraic effects index admissible observations by modalities within
a fixed semantics~\cite{MatacheStaton2019}, whereas our observation policies are
additionally transported along simulations.
A precise account of simulations as
morphisms in a category of monad liftings is a natural formal home for this picture,
which we leave to future work.

\paragraph{Modal and step-indexed program logics.}

 Modalities are also central to step-indexed reasoning: the later modality of guarded recursion arises semantically in the topos of
trees~\cite{Birkedal2012}, and step-indexed modalities underpin
Iris~\cite{Jung2018} and its logical-relations
methodology~\cite{Timany2024}. There the modality is part of the
ambient logic, built into the model from the start, and it is used to
reason about programs of a fixed effectful language.
Whether the later modality can be exhibited
as the nucleus of a suitable simulation, for instance into a step-counting monadic core, is a question we leave for future work.

\paragraph{Dense transport and nuclei on realizability models.}
At the level of realizability models, nuclei are induced by dense transport~\cite{HOFSTRA_VANOOSTEN_2003}. They induce nuclei and
geometric transformations on
arrow algebras~\cite{tarantino2026} and
triposes~\cite{Vanoosten_08}.
The situation is similar in our setting: computationally
dense simulations of monadic cores induce geometric simulations
and nuclei of evidenced frames and triposes.

\section{Conclusion}\label{sec:conc}

This paper makes a change of effect a first-class mathematical
object: a simulation between effectful realizability cores.
Simulations of monadic cores, regular morphisms of the induced
evidenced frames, and morphisms of the generated triposes determine
one another through $2$-fully-faithful assignments, so transport may
be performed at whichever level is most convenient, with nothing lost
and, by completeness, nothing gained: the logics are related exactly
as the cores are.
Two features distinguish this account from a soundness theorem for a
particular translation. First, transport carries its own
justification: evidence is a program, so a proof crosses the gap by
computation, producing an explicit realizer in the target rather than
a promise that one exists. Second, the framework does not just
perform transport, it prices it: the round trip of a computationally
dense simulation computes a nucleus through which the transport
factors, with double negation arising as an instance.
Modalities built into a logic by hand can then be compared with
transport-derived ones, and the sheafification  case study shows the two routes coincide
in some settings and provably diverge in others.
The formal account of simulation also makes impossibility provable, allowing us to identify cases for which no simulation exists.
Thus, while a change of effect has long been the point at which reasoning is discarded and rebuilt from scratch, with the formal account of simulations as evidenced morphisms it becomes an object of
mathematical study in its own right, one that can be composed,
factored through modalities, and shown, in some cases, not to exist.

Several directions remain open.
The development in the paper is semantic, but it has a worked
syntactic counterpart: a concrete type system whose derivations
denote simulations and whose type-directed translation produces the transported realizers. Its full presentation is deferred to a future paper.
Since the framework is internal to an arbitrary topos, it can be
instantiated over a realizability topos and iterated, connecting to
Pitts' iteration theorem~\cite{Pittsthesis} and raising the question
of how simulations compose across iterations.
Effects are also routinely combined, yet broad classes of monads
admit no distributive law, and clarifying when simulations are
compatible with monad composition would extend transport to combined
effects.
Since the algebraic fragment factors through monad morphisms, a
natural question is its relation to algebraic effects and handlers.
We would also like to recover standard
relational methods as transport, exhibiting step-indexed and biorthogonal logical-relations proofs of compiler correctness as simulations.

\clearpage

\bibliographystyle{plainurl}
\bibliography{ref}

\clearpage
\appendix
\section{Appendix}
\label{app:proofs}

This appendix contains proofs of the main results in the paper.

\subsection{Proofs from Section~\ref{sec:mca-simulations}}

\proofappendix[prop:general-simulation]{equivsim}
\begin{proof}
Anything that satisfies the condition in the formula above also satisfies the application-translator condition in \Cref{def:simulation}, by considering $\mathsf{c} := \eLam{x,y}{x \cdot y}$
and taking $\SRealizer$ to be $
\eLam{x}{
\Elets{\EletRib{x_1}{\mathsf{e}_{\pi_1} \cdot x} \; ; \; \EletRib{x_2}{\mathsf{e}_{\pi_2} \cdot x}
 \; ; \; \EletRib{x_3}{\mathsf{c}^{\Sim} \cdot x_1}
} {x_3 \cdot x_2 }}$ .

In the other direction, assume the application-translator condition
along with evident totality and let $n$ be an integer.
Let $\mathsf{c} : \Separator$ and consider the code
$\mathsf{f} \define \eLam{x_1 , \ldots, x_n}{
\App^{n} \circ \langle
\mathsf{c} ,x_1 ,\ldots , x_n \rangle}$.
Write~$\mathsf{t}_{\mathsf{f}}$ as the evidence of totality for
$\mathsf{f}$. We show that $\mathsf{c}^{\Sim}$ can be taken
as the code
\[\mathsf{c}^{\Sim} \coloneq
\eLam{x_1, \ldots, x_n}{
\Elets{
  \EletRib{\codeII[\mathsf{f}]}{\mathsf{t}_{\mathsf{f}} \cdot \id} \;;\;
   \EletRib{x_1'}{\SRealizer \cdot \cpair{\codeII[\mathsf{f}]}{x_1}} \; ;\;
  \EletRib{x_2'}{\SRealizer \cdot \cpair{x_1'}{x_2}} \;;\; \ldots
}{\SRealizer \cdot \cpair{x_{n-1}'}{x_n}}}\]

Let $\mathsf{c_1},\ldots,\mathsf{c_n} : \mca[1], \mathsf{d_1},\ldots,
\mathsf{d_n} : \mca[2]$, $\phi : \mca[1] \to \Omega$.
First, as a lemma, show by induction, through repeated use of the
unit law of the modality, that:
    $$\after{\codeI[r]}{
     \App^{n} \circ \langle \mathsf c, \mathsf c_1 , \ldots , \mathsf c_n \rangle
    }{ \varphi\paren{\codeI[r]} }
    \sqsubseteq{} \underbrace{\after{\codeI[1]'}{\mathsf{f} \cdot \codeI[1]}{ \after{\codeI[2]'}{\codeI[1]' \cdot \codeI[2]}{ \ldots \after{\codeI[r]}{\codeI[n-1]' \cdot \codeI[n]}{ \varphi\paren{\codeI[r]} } \ldots } }}_{\eqqcolon\, \Psi}
$$

We further abbreviate the transported postcondition
$\theta\paren{\codeII[r]} \coloneq \exists \codeI[r] .
\Sim\paren{\codeI[r], \codeII[r]} \sqcap \varphi\paren{\codeI[r]}$.
Now, assuming the premise:
\[\bigsqcap_{i=1}^{n} \Sim\paren{\codeI[i],\codeII[i]} \sqcap \after{\codeI[r]}{  \App^{n} \circ \langle \mathsf c, \mathsf c_1 , \ldots , \mathsf c_n \rangle }{ \varphi\paren{\codeI[r]} }\]
combined with evident totality for $\mathsf{f}$, we get:
\[\after{\codeII[\mathsf{f}]}{\mathsf{t}_{\mathsf{f}} \cdot \id}{ \Sim\paren{\mathsf{f} , \codeII[\mathsf{f}]} } \sqcap \bigsqcap_{i=1}^{n} \Sim\paren{\codeI[i],\codeII[i]} \sqcap \after{\codeI[r]}{
 \App^{n} \circ \langle \mathsf c, \mathsf c_1 , \ldots , \mathsf c_n \rangle
}{ \varphi\paren{\codeI[r]} } \]
Using strength, we push all the conjuncts into the modality of the
computation of $\codeII[\mathsf{f}]$, giving us:
\[\after{\codeII[\mathsf{f}]}{\mathsf{t}_{\mathsf{f}} \cdot \id}{ \Sim\paren{\mathsf{f} , \codeII[\mathsf{f}]} \sqcap \bigsqcap_{i=1}^{n} \Sim\paren{\codeI[i],\codeII[i]} \sqcap \after{\codeI[r]}{
 \App^{n} \circ \langle \mathsf c, \mathsf c_1 , \ldots , \mathsf c_n \rangle}{ \varphi\paren{\codeI[r]} } }\]
Using the lemma, it gives us:
\[\after{\codeII[\mathsf{f}]}{\mathsf{t}_{\mathsf{f}} \cdot \id}{ \Sim\paren{\mathsf{f} , \codeII[\mathsf{f}]} \sqcap \bigsqcap_{i=1}^{n} \Sim\paren{\codeI[i],\codeII[i]} \sqcap \Psi }\]
Now, by repeatedly using the assumption of the
application-translator condition followed by monotonicity over
strength, using the simulation of the result of the previous step as
the simulation of the operator part of the next step, while removing
the existential quantifier resulting from the previous step when its
bound variable disappears from the next step, we get
$\after{\codeII[\mathsf{f}]}{\mathsf{t}_{\mathsf{f}} \cdot \id}{\Theta}$
for
\[\Theta \coloneq \after{\codeII[1]'}{\SRealizer \cdot \cpair{\codeII[\mathsf{f}]}{\codeII[1]}}{\after{\codeII[2]'}{\SRealizer \cdot \cpair{\codeII[1]'}{\codeII[2]}}{ \ldots \after{\codeII[r]}{\SRealizer \cdot \cpair{\codeII[n-1]'}{\codeII[n]}}{ \theta\paren{\codeII[r]} } \ldots } }\]
By repeated use of the composition law of the modality, we
obtain $\after{\codeII[r]}{\, \mathsf{d}^{\Sim} }{ \theta\paren{\codeII[r]} }$
for the expression
\[\mathsf{d}^{\Sim} \coloneq \Elets{
  \EletRib{\codeII[\mathsf{f}]}{\mathsf{t}_{\mathsf{f}} \cdot \id} \;;\;
   \EletRib{\codeII[1]'}{\SRealizer \cdot \cpair{\codeII[\mathsf{f}]}{\codeII[1]}} \;;\;
  \EletRib{\codeII[2]'}{\SRealizer \cdot \cpair{\codeII[1]'}{\codeII[2]}} \;;\; \ldots}{\SRealizer \cdot \cpair{\codeII[n-1]'}{\codeII[n]}}\]

Finally, note that $\mathsf{d}^{\Sim} =
\App^{n} \circ \langle \mathsf{c}^{\Sim} ,
 \mathsf d_1 ,
\ldots \mathsf d_n \rangle$.

\end{proof}

\proofappendix[prop:appx-mca-1]{PCAequiv}
\begin{proof}
One direction is trivial, because we can just take $\varphi\left(x,\mathsf{c}_r\right) = x \equiv \mathsf{c}_r$.
In the other direction, assume the definition of a PCA applicative morphism, we get:
$$\begin{array}{rl}
\Sim\left(\mathsf{c}_{f},\mathsf{d}_{f}\right) \sqcap \Sim\left(\mathsf{c}_{a},\mathsf{d}_{a}\right) \sqcap \TotCorrectIgnore{\mathsf{c}_r}{\mathsf{c}_{f} \cdot \mathsf{c}_{a}}{\varphi\left(x, \mathsf{c}_{r}\right) }

\Longrightarrow & \TotCorrect{\mathsf{d}_{r}}{ \SRealizer \cdot \cpair{\mathsf{d}_{f}}{\mathsf{d}_{a}} }{ \Sim\left(\mathsf{c}_{r},\mathsf{d}_{r}\right) }
\end{array}$$
By conjoining both sides with $\varphi\left(x,\mathsf{c}_r\right)$ for some arbitrary $\varphi$ we get:
$$\begin{array}{l}
\Sim\left(\mathsf{c}_{f},\mathsf{d}_{f}\right) \sqcap \Sim\left(\mathsf{c}_{a},\mathsf{d}_{a}\right) \sqcap \TotCorrect{\mathsf{c}_r}{\mathsf{c}_{f} \cdot \mathsf{c}_{a}}{\varphi\left(x, \mathsf{c}_{r}\right)} \\[6pt]
\quad \Longrightarrow\; \TotCorrect{\mathsf{d}_{r}}{ \SRealizer \cdot \cpair{\mathsf{d}_{f}}{\mathsf{d}_{a}} }{ \Sim\left(\mathsf{c}_{r},\mathsf{d}_{r}\right) \sqcap \varphi\left(x, \mathsf{c}_{r}\right)} \\[6pt]
\qquad \Longrightarrow\; \TotCorrect{\mathsf{d}_{r}}{ \SRealizer \cdot \cpair{\mathsf{d}_{f}}{\mathsf{d}_{a}} }{ \exists \mathsf{c}_{r} . \Sim\left(\mathsf{c}_{r},\mathsf{d}_{r}\right) \sqcap \varphi\left(x, \mathsf{c}_{r}\right)}
\end{array}$$
Which is simply:
$$
\Sim\left(\mathsf{c}_{f},\mathsf{d}_{f}\right) \sqcap \Sim\left(\mathsf{c}_{a},\mathsf{d}_{a}\right) \sqcap \after{\mathsf{c}_r}{\mathsf{c}_{f} \cdot \mathsf{c}_{a}}{\varphi\left(x, \mathsf{c}_{r}\right) }
\Longrightarrow  \after{\mathsf{d}_{r}}{ \SRealizer \cdot \cpair{\mathsf{d}_{f}}{\mathsf{d}_{a}} }{  \exists \mathsf{c}_{r} . \Sim\left(\mathsf{c}_{r},\mathsf{d}_{r}\right) \sqcap \varphi\left(x, \mathsf{c}_{r}\right) }
$$
\qedhere
\end{proof}

\proofappendix[prop:may-must-nogo]{maymustnogo}
\begin{proof}
Suppose $\Sim$ is such a simulation.
For each $n \in \mathbb{N}$,
$
  \Sim^{*}(\varphi_n)(\mathsf{d})
  = \exists \mathsf{c}.\; \Sim(\mathsf{c},\mathsf{d}) \sqcap (\mathsf{c} \equiv \bar{n})
  = \Sim(\bar{n},\mathsf{d}),
$
so arithmetic-preservation gives
\begin{equation}\label{eq:sim-arith}
  \Sim(\bar{n},\mathsf{d}) \;\iff\; \mathsf{d} \equiv \bar{n}
  \qquad\text{for all } n \in \mathbb{N},\; \mathsf{d} \in \mca.
\end{equation}

Note that $\Sim$ is total, in this case: for each $\mathsf{c}$,
evident totality provides a code whose run the target modality
certifies, and the demonic modality of $\mathrm{Flip}_{\forall}$
certifies only inhabited outcome sets all of whose elements simulate
$\mathsf{c}$, so in particular some $\mathsf{d}$ with
$\Sim(\mathsf{c},\mathsf{d})$ exists. Pick $\mathsf{d}_{\mathsf{flip}}$
with $\Sim(\mathsf{flip}, \mathsf{d}_{\mathsf{flip}})$, and set
$R_0 \coloneqq \SRealizer \cdot
\cpair{\mathsf{d}_{\mathsf{flip}}}{\bar{0}}$, with $\SRealizer$ the
application translator of \Cref{def:simulation}. The argument code
$\bar{0}$ is chosen only because $\Sim(\bar{0},\bar{0})$ is
available by~\eqref{eq:sim-arith}; any code with a known simulant
would do, and $R_0$ does not depend on $n$.

Using $\Sim(\mathsf{flip}, \mathsf{d}_{\mathsf{flip}})$ and
$\Sim(\bar{0}, \bar{0})$, the simulation condition for
postcondition $\varphi_n$ gives, for each $n \in \{0,1\}$:
\begin{equation}\label{eq:sim-cond}
  \underbrace{
  \exists \mathsf{c}_r \in \mathsf{flip} \cdot \bar{0} \; . \; \varphi_n(\mathsf{c}_r)
  }_{\mathrm{Flip}_{\exists} \text{ modality}}
  \;\;\Longrightarrow\;\;
  \underbrace{
  \left(\exists \mathsf{d}_r \in R_0 \; . \; \top
  \right) \sqcap
  \left(
  \forall \mathsf{d}_r \in R_0 \; . \;
  \exists \mathsf{c}_r .\;\Sim(\mathsf{c}_r,\mathsf{d}_r) \sqcap \varphi_n(\mathsf{c}_r)
  \right)
    }_{\mathrm{Flip}_{\forall} \text{ modality}}.
\end{equation}
Since $\mathsf{flip} \cdot \bar{0} = \{\bar{0},\bar{1}\}$, the
premise holds for both $n = 0$ and $n = 1$. The right-hand side
requires $R_0$ inhabited and, using~\eqref{eq:sim-arith}, every
$\mathsf{d}_r \in R_0$ to satisfy $\mathsf{d}_r \equiv \bar{n}$, that
is, $R_0 \subseteq \{\bar{n}\}$. From $n = 0$ and $n = 1$ we get
$R_0 \subseteq \{\bar{0}\} \cap \{\bar{1}\} = \emptyset$,
contradicting the inhabitation of $R_0$.
\end{proof}

\proofappendix[lemma:appx-algsim-1]{monadmodality}
\begin{proof}
It is clear that $\modality^{\Phi}:\Omega^X \to \Omega^{M_1 X}$,
moreover $\modality^{\Phi} (\phi) = \Omega^{\Phi_{X}} (\modality[X] (\phi))$
written $(\modality[X] (\phi)) \circ \Phi_X$ in the internal language.
It remains to check that the several properties hold, which is straightforward from the definition of $\Phi$ and the fact that $\modality$ is a $M_2$ modality.
\begin{enumerate}
\item \emph{Monotonicity:} let $\varphi,\psi:X\to\Omega$ be such that $\varphi \sqsubseteq \psi$. By monotonicity of $\modality$ we get $\modality\varphi \sqsubseteq \modality \psi$. Using the naturality of $\Phi$, we obtain:
\[\modality^\Phi \varphi = \modality \varphi \circ \Phi_X \sqsubseteq \modality \psi \circ \Phi_X = \modality^\Phi \psi\]

\item \emph{Unit:} we have $\modality^\Phi \varphi \circ \eta^1_{\Omega} = \modality \varphi \circ \Phi_\Omega \circ \eta^1_{\Omega} = \modality \varphi \circ \eta^2_{\Omega} \sqsupseteq \id[\Omega] $.
\item \emph{Bind:}
we have
\begin{align*}
\modality^\Phi \left(\modality^\Phi \varphi \circ f\right) &= \modality \left(\modality \varphi \circ \Phi \circ f\right) \circ \Phi\\
&\sqsubseteq \modality \phi \circ \left( \Phi \circ f \right)^{*} \circ \Phi \\
&= \modality \varphi \circ \Phi \circ f^{*} \\
&= \modality^{\Phi} \varphi \circ f^{*}
\end{align*}
\item \emph{Strength:}\quad
\begin{align*}
\sqcap \circ (\varphi_1 \times \modality^{\Phi} \varphi_2) &= \sqcap \circ (\varphi_1 \times (\modality \varphi_2 \circ \Phi))\\
&= \sqcap \circ (\varphi_1 \times \modality \varphi_2) \circ (\id \times \Phi) \\
&\sqsubseteq \modality \left( \varphi_1 \times \varphi_2 \right) \circ \strength \circ (\id \times \Phi)\\
&= \modality \left( \varphi_1 \times \varphi_2 \right) \circ \Phi \circ \strength\\
&= \modality^{\Phi} \left( \varphi_1 \times \varphi_2 \right) \circ \strength
\end{align*}
\end{enumerate}
\end{proof}

\proofappendix[proposition:algebraicsim]{algebraicsim}
\begin{proof}
Let $(\Phi,h)$ be an algebraic simulation from $\left( M_{1} , \modality[1] , \mca[1],
\Separator[1] \right)$ to $\left( M_{2} , \modality[2] ,
\mca[2], \Separator[2] \right)$. We define $\Sim$ as the
subobject of $\mca[1] \times \mca[2]$ given by the pullback of
$h$ and $\id[{\mca[2]}]$, that is, $\Sim\left(\mathsf{c},
\mathsf{d}\right) := h\left(\mathsf{c}\right) \equiv
\mathsf{d}$.
We need to check the two items of the definition of simulations.
\begin{enumerate}
        \item We need to prove that $\Sim$ is evidently total, i.e. that
        \[\vdashE{\E}
        \forall \codeI : \Separator[1] . \exists \codef{t}_{\codeI} : \Separator[2] . \eAfter{ \codeII[r] }{ \eApp{ \codef{t}_{\codeI}}{\id} }{ \Sim\paren{ \codeI , \codeII[r] } }\]
        To that end, and for any $c:\mca[1]$, let $t_c = \Lambda \_.\Eret h(c)$. Writing $\psi(x) := h(c) \equiv x$, we then have  $\eAfter{ \codeII[r] }{ \eApp{ \codef{t}_{\codeI}}{\id} }{ \psi(\codeII[r]) } = (\modality[2]\circ M\psi \circ \eta_{\mca[2]})(h(c)) = (\modality[2]\circ \eta_{\Omega}\circ \psi)(h(c)) $ by naturality of~$\eta$. The result then follows directly from the unit rule of the modality (see \Cref{def:modality}) and the fact that $\varphi(h(c)) = h(c) \equiv h(c)$ holds.
        \item For the last item, we can actually directly use the code $\mathsf{r}:\Separator[2]$ from the third item.
        Let $\varphi \in \mca[1] \to \Omega$ and $c_f,c_a:\mathcal A_1$. From the definition, we get:
\begin{align*}
    &\after{\mathsf{c}_r}{\mathsf{c}_f \cdot \mathsf{c}_a}{\varphi\left(\mathsf{c}_r\right)}\\
    \sqsubseteq& \after{\mathsf{c}_r}{\Phi\left(\mathsf{c}_f \cdot \mathsf{c}_a\right)}{\varphi\left(\mathsf{c}_r\right)}\\
    \sqsubseteq& \after{\mathsf{c}_r}{\Phi\left(\mathsf{c}_f \cdot \mathsf{c}_a\right)}{\exists \mathsf{c}_r' . h\left(\mathsf{c}_r'\right) \equiv  h\left(\mathsf{c}_r\right) \sqcap \varphi\left(\mathsf{c}_r\right)}\\
    \sqsubseteq&\after{\mathsf{c}_r}{\Phi\left(\mathsf{c}_f \cdot \mathsf{c}_a\right)}{
    \after{\mathsf{d}_r}{\Eret{h\left(\mathsf{c}_r\right)}}{\exists \mathsf{c}_r' . h\left(\mathsf{c}_r'\right) \equiv \mathsf{d}_r \sqcap \varphi\left(\mathsf{c}_r\right)}}\\
    \sqsubseteq&\after{\mathsf{d}_r}{ \Elet{\mathsf{c}_r}{\Phi\left(\mathsf{c}_f \cdot \mathsf{c}_a\right)}{\Eret{h\left(\mathsf{c}_r\right)}} }{\exists \mathsf{c}_r' . h\left(\mathsf{c}_r'\right) \equiv \mathsf{d}_r \sqcap \varphi\left(\mathsf{c}_r\right)}\\
    =& \after{\mathsf{d}_r}{ \mathsf{r} \cdot \cpair{h\left(\mathsf{c}_f\right)}{h\left(\mathsf{c}_a\right)} }{\exists \mathsf{c}_r . h\left(\mathsf{c}_r\right) \equiv \mathsf{d}_r \sqcap \varphi\left(\mathsf{c}_r\right)}
\end{align*}

For any $d_f,d_a:\mathcal A_2$, we then get:
\begin{align*}
    &\Sim\left(\mathsf{c}_f,\mathsf{d}_f\right) \sqcap \Sim\left(\mathsf{c}_a,\mathsf{d}_a\right) \sqcap \after{\mathsf{c}_r}{\mathsf{c}_f \cdot \mathsf{c}_a}{\varphi\left(\mathsf{c}_r\right)}\\
    =~&h\left(\mathsf{c}_f\right)\equiv\mathsf{d}_f \sqcap h\left(\mathsf{c}_a\right)\equiv\mathsf{d}_a \sqcap \after{\mathsf{c}_r}{\mathsf{c}_f \cdot \mathsf{c}_a}{\varphi\left(\mathsf{c}_r\right)}\\
    \sqsubseteq~&h\left(\mathsf{c}_f\right)\equiv \mathsf{d}_f \sqcap h\left(\mathsf{c}_a\right)\equiv \mathsf{d}_a \sqcap \after{\mathsf{d}_r}{ \mathsf{r} \cdot \cpair{h\left(\mathsf{c}_f\right)}{h\left(\mathsf{c}_a\right)} }{\exists \mathsf{c}_r . h\left(\mathsf{c}_r\right) \equiv \mathsf{d}_r \sqcap \varphi\left(\mathsf{c}_r\right)}\\
    \sqsubseteq~&\after{\mathsf{d}_r}{ \mathsf{r} \cdot \cpair{\mathsf{d}_f}{\mathsf{d}_a} }{\exists \mathsf{c}_r . h\left(\mathsf{c}_r\right) \equiv \mathsf{d}_r \sqcap \varphi\left(\mathsf{c}_r\right)}\\
    =~&\after{\mathsf{d}_r}{ \mathsf{r} \cdot \cpair{\mathsf{d}_f}{\mathsf{d}_a} }{\exists \mathsf{c}_r . \Sim\left(\mathsf{c}_r , \mathsf{d}_r\right) \sqcap \varphi\left(\mathsf{c}_r\right)}
\end{align*}
\end{enumerate}
\end{proof}

\proofappendix[proposition:liftingalgsim]{liftingalgsim}
\begin{proof}
In both cases we need to show the modality and application compatibily
\begin{enumerate}
    \item For the writer monad, we have:
    \begin{itemize}
    \item Modality compatibility, using that this property holds for $\modality[1]$ and $\modality[2]$ and the naturality of $\Phi$:
    \begin{align*}
    \modality[W]^2 \varphi \circ \Phi_X^W &= \modality[2] \varphi \circ M_2(\pi_2)\circ \Phi_{W\times X}\\
    & = \modality[2] \varphi \circ \Phi_{X}\circ M_1(\pi_2) \\
    & \geq \modality[1] \varphi \circ M_1(\pi_2) = \modality[W]^1 \varphi
    \end{align*}
    \item Application compatibility. First recall that $\iota^{W,i}_X(m) \coloneq M^i(\langle \varepsilon,\mathrm{id}_X \rangle)(m)$.
    In particular, we have that
    \[\Phi_{W\times \mca[2]} \circ \iota^{W,1}_{\mca[2]}
    = \Phi_{W\times \mca[2]} \circ M^1(\langle \varepsilon,\id[{\mca[2]}]\rangle)
    = M^2(\langle \varepsilon,\id[{\mca[2]}] \rangle) \circ \Phi_{\mca[2]}
    = \iota^{W,2}_{\mca[2]} \circ \Phi_{\mca[2]} \\
    \]
    Besides, given $\mathsf{r}:\Separator[2]$, since $\mathsf{r}_{\App[2]} (x,y) =
    \App[2](\mathsf{r},\cpair{x}{x})$,
    \[\mathsf{r}_{\App[2]^W} (x,y) = \App[2]^W(\mathsf{r},\cpair{x}{y})
    = \iota^{W,2}_{\mca[2]} \circ \App[2] (\mathsf{r},\cpair{x}{y})
    = \iota^{W,2}_{\mca[2]} \circ \mathsf{r}_{\App[2]}(x,y)
    \]
    Therefore, we can derive:
    \begin{align*}
    \Phi^W h\circ \App[1]^W
    & = \Phi^W_{\mca[2]} \circ M_W^1 h \circ \iota^W_{\mca[1]} \circ \App[1] \\
    & = \Phi^W_{\mca[2]} \circ \iota^W_{\mca[2]} \circ M^1 h \circ \App[1] \\
    & = \iota^W_{\mca[2]} \circ \Phi_{\mca[2]} \circ  M^1 h \circ \App[1] \\
    & = \iota^W_{\mca[2]} \circ r_{\App[2]} \circ h\times h\\
    & = r_{\App[2]^W} \circ h\times h\\
    \end{align*}
    Therefore the same applicator works.
    \end{itemize}
    \item The proof for the reader monad is analogous, the modality compatibility follows from:
    \[
    (\modality[E]^2 \varphi \circ \Phi_X^E)(t) = \bigwedge_{e \in E} \modality[2] \varphi (\Phi_{X}(t(e)))
    \geq \bigwedge_{e \in E} \modality[1] \varphi (t(e)) = \modality[E]^1 \varphi (t)\\
    \]
    The applicator compatibility is similar to the case for the writer monad, using  the fact that we again have the commutation:
    \[(\Phi^E_{\mca[2]} \circ \iota^{E,1}_{\mca[2]})(t)
    = \Phi^E_{\mca[2]} (\lambda e.t)
    = \lambda e.\Phi_{\mca[2]}(t(e))
    = (\iota^{E,2}_{\mca[2]} \circ \Phi_{\mca[2]})(t)
    \]
\end{enumerate}
\end{proof}

\subsection{Proofs from Section~\ref{sec:ef-sims-main}}

\proofappendix[thm:ef-from-monadic-core]{effromcore}
\begin{proof}
This is the construction of~\cite[Thm.~16]{MCA25}, read in the internal language
of~$\E$ rather than in~$\Set$. The propositions $\Omega^{\mca}$, the evidence
$\Separator$, the entailment relation, and all combinators are unchanged.
The only difference is that propositions are stated externally as morphisms
$\mca\to\Omega$ while evidence exists internally. Each axiom is an internal
existence statement discharged, as there, from combinatory completeness and the
modality laws, so the proof transfers verbatim.
\end{proof}

\proofappendix[lemma:union-coprod]{unioncoprod}
\begin{proof}
The required realizers are obtained through combinatory completeness
    and correspond to the terms
    $t_1 \define \eLam{x, w, y}{\Elet{z}{y \cdot \mathsf{w}}{z \cdot x}}$ and
    $t_2 \define \eLam{x}{\Elet{z}{x \cdot \mathsf{c}}{z \cdot (\eLam{w, y}{\Eret{y}})}}$
    where $\mathsf{c} \!: \! \mca$ is arbitrary (it is used as the argument of a term
    that realizes $\top \supset \vec{\phi}$).

\end{proof}

\proofappendix[thm:mca-simulation-induces-ef-simulation]{efsoundness}
\begin{proof}
Given a simulation of monadic cores $\Sim : \mca[1] \times \mca[2] \to \Omega$, we
define~$\Sim^{*}\left(\varphi\right)\left(d\right) := \exists c . \Sim\left(c,d\right) \sqcap\varphi\left(c\right)$.
We show that $\Sim^{*}$ gives a regular EF simulation between the EFs derived from the source and target monadic cores, by going over each of the properties in Definition~\ref{def:ef-sims}.
\begin{enumerate}

    \item
    \underline{Monotonicity:}
    Assume $\phi_1 \xle{\mathsf{e}_1}_1 \phi_1'$, that is, for all $\mathsf{c}:\mca[1]$, $\phi_1\left(\mathsf{c}\right) \sqsubseteq \after{\mathsf{c}_r}{\mathsf{e}_1 \cdot \mathsf{c}}{\phi_1'\left(\mathsf{c_r}\right)}$. Take  $\mathsf{c} : \mca[1]$ and $\mathsf{d} : \mca[2]$, we get: $\Sim\left(\mathsf{c}, \mathsf{d}\right) \sqcap \phi_1\left(\mathsf{c}\right) \sqsubseteq \Sim\left(\mathsf{c}, \mathsf{d}\right) \sqcap \after{\mathsf{c}_r}{\mathsf{e}_1 \cdot \mathsf{c}}{\phi_1'\left(\mathsf{c_r}\right)} $.
    From totality applied to the code~$\mathsf{e}_1$
    which is in the separator, there is a code $\mathsf{t_e}$ such that
    $
    \vdashE{\E} \after{\mathsf{d}_f}{\mathsf{t_e} \cdot \id}{\Sim\left(\mathsf{e}_1, \mathsf{d}_f\right)}$.
    It implies that
    \[\Sim\left(\mathsf{c}, \mathsf{d}\right) \sqcap \phi_1\left(\mathsf{c}\right) \sqsubseteq
    \after{\mathsf{d}_f}{\mathsf{t_e} \cdot \id}{\Sim\left(\mathsf{e}_1, \mathsf{d}_f\right)}
    \sqcap
    \Sim\left(\mathsf{c}, \mathsf{d}\right)
    \sqcap
    \after{\mathsf{c}_r}{\mathsf{e}_1 \cdot \mathsf{c}}{\phi_1'\left(\mathsf{c_r}\right)}\]

    Using the strength of the modality, we push all the conjuncts inside the first modality:
    \[\Sim\left(\mathsf{c}, \mathsf{d}\right) \sqcap \phi_1\left(\mathsf{c}\right) \sqsubseteq
    \after{\mathsf{d}_f}{\mathsf{t_e} \cdot \id}{\Sim\left(\mathsf{e}_1, \mathsf{d}_f\right) \sqcap
    \Sim\left(\mathsf{c}, \mathsf{d}\right)
    \sqcap
    \after{\mathsf{c}_r}{\mathsf{e}_1 \cdot \mathsf{c}}{\phi_1'\left(\mathsf{c_r}\right)}}\]

    On the right hand side, inside the external modality, is precisely the form of the left hand side of the simulation condition for $\Sim$, so by monotonicity of the modality:
    \[ \Sim\left(\mathsf{c}, \mathsf{d}\right) \sqcap \phi_1\left(\mathsf{c}\right)
    \sqsubseteq
    \after{\mathsf{d}_f}{\mathsf{t_e} \cdot \id}{\after{\mathsf{d}_r}{\mathsf{r} \cdot \epair{\mathsf{d_f} }{\mathsf{d}}}{ \exists \mathsf{c_r} . \Sim\left(\mathsf{c_r},\mathsf{d_r}\right) \sqcap \phi_1'\left(\mathsf{c}_r\right) }} \]

    Using the properties of the modality, we combine the computations on the right hand side:
    \[ \Sim\left(\mathsf{c}, \mathsf{d}\right) \sqcap \phi_1\left(\mathsf{c}\right)
    \sqsubseteq
    \after{\mathsf{d}_r}{\Elet{\mathsf{d}_f}{\mathsf{t_e} \cdot \id }{\mathsf{r} \cdot \epair{\mathsf{d_f}}{\mathsf{d}}}}{ \exists \mathsf{c_r} . \Sim\left(\mathsf{c_r},\mathsf{d_r}\right) \sqcap \phi_1'\left(\mathsf{c}_r\right) }\]

    Writing $\mathsf{e'}$ for $\eLam{x} \Elet{\mathsf{d}_f}{\mathsf{t_e} \cdot \id }{\mathsf{r} \cdot \epair{\mathsf{d_f} }{x}}$, it rewrites as
    \[ \Sim\left(\mathsf{c}, \mathsf{d}\right) \sqcap \phi_1\left(\mathsf{c}\right)
    \sqsubseteq
    \after{\mathsf{d}_r}{\mathsf{e'} \cdot \mathsf{d}}{ \exists \mathsf{c_r} . \Sim\left(\mathsf{c_r},\mathsf{d_r}\right) \sqcap \phi_1'\left(\mathsf{c_r}\right) }\]

    Since it holds for an arbitrary $\mathsf{c}$, we existentially quantify it on the left to get:
    \[ \exists \mathsf{c} . \Sim\left(\mathsf{c}, \mathsf{d_r}\right) \sqcap \phi_1\left(\mathsf{c}\right)
    \sqsubseteq
    \after{\mathsf{d}_r}{\mathsf{e'} \cdot \mathsf{d}}{ \exists \mathsf{c_r} . \Sim\left(\mathsf{c_r},\mathsf{d}\right) \sqcap \phi_1'\left(\mathsf{c}_r\right) } \]

    By folding $\Sim^{*}$, it gives:
    \[
    \vdashE{\E}
    \Sim^{*}\left(\phi_1\right)\left(\mathsf{d}\right) \Rightarrow \after{\mathsf{d}_r}{ \mathsf{e'} \cdot \mathsf{d} }{ \Sim^{*}\left(\phi_1'\right)\left(\mathsf{d}_r\right) } \]

    which means
    $\Sim^{*}(\phi_1) \xle{\mathsf{e'} }_2 \Sim^{*}(\phi_1')$.
    Finally,
    since $\mathsf{t_e},\id$ and $\mathsf{r}$ are in $\Separator[2]$, and since $\Separator[2]$ is combinatory complete, the evidence $\mathsf{e'}$ also is
    in~$\Separator[2]$.

    \item \underline{Top preservation:}
    From evident totality,
    we get a code
    code~$\mathsf{t} \in \Separator[2]$
    such that $\{\id\} \xle{\mathsf{t}} \gamma^*(\{\id\})$.
    But $\top \xle{\eLam{x}{\Eret{\id}}} \{\id\}$
    and $\gamma^*(\{\id\}) \sqsubseteq \gamma^*(\top)$.
    Therefore, top preservation follows by composition.

    \item \underline{Implication distributivity:}
    As a first step,  the second condition of
    a monadic core simulation implies
    $$
    \Sim^{*}\left(\phi \supset \psi\right)\left(\mathsf{d}_f\right) \sqcap \Sim^{*}\left(\phi\right)\left(\mathsf{d}_a\right)
    \sqsubseteq
    \after{\mathsf{d}_r}{\mathsf{r} \cdot \epair{\mathsf{d}_f}{\mathsf{d}_a}}{\Sim^{*}\left(\psi\right)\left(\mathsf{d}_r\right)}
    $$
     and it gives:
    \[ \Sim^{*}\left(\phi \supset \psi\right)\left(\mathsf{d}_f\right) \sqsubseteq \Sim^{*}\left(\phi\right)\left(\mathsf{d}_a\right) \Rightarrow \after{\mathsf{d}_r}{ \left(\eLam{y}  \mathsf{r} \cdot \epair{\mathsf{d}_f}{y} \right) \cdot \mathsf{d}_a}{\Sim^{*}\left(\psi\right)\left(\mathsf{d}_r\right)} \]

    Since this is true for $\mathsf{d}_a$, we universally quantify $\mathsf{d}_a$ on the right:
    \[ \Sim^{*}\left(\phi \supset \psi\right)\left(\mathsf{d}_f\right) \sqsubseteq \forall \mathsf{d}_a . \;\Sim^{*}\left(\phi\right)\left(\mathsf{d}_a\right) \Rightarrow \after{\mathsf{d}_r}{\left(\lambda y . \mathsf{r} \cdot \epair{\mathsf{d}_f}{y} \right) \cdot \mathsf{d}_a}{\Sim^{*}\left(\psi\right)\left(\mathsf{d}_r\right)} \]

    Using the property of the modalities, it gives
    \[
    \Sim^{*}\left(\phi \supset \psi\right)\left(\mathsf{d}_f\right)
    \sqsubseteq
    \after{\mathsf{d}_t}{\Eret{\lambda y . \mathsf{r} \cdot \epair{\mathsf{d}_f}{y} }}{\forall \mathsf{d}_a . \Sim^{*}\left(\phi\right)\left(\mathsf{d}_a\right) \Rightarrow \after{\mathsf{d}_r}{\mathsf{d}_t \cdot \mathsf{d}_a}{\Sim^{*}\left(\psi\right)\left(\mathsf{d}_r\right)}}
    \]

    We write $\lambda\left(\mathsf{r}\right) :=
    \eLam{ x ,y} \mathsf{r} \cdot \cpair{x}{y}$, it satisfies $\lambda\left(\mathsf{r}\right) \cdot \mathsf{d}_f = \Eret{\eLam{y}{\mathsf{r} \cdot \cpair{\mathsf{d}_f}{y}}} $.
   It implies
    \[
    \Sim^{*}\left(\phi \supset \psi\right)\left(\mathsf{d}_f\right)
    \sqsubseteq
    \after{\mathsf{d}_t}{\lambda\left(\mathsf{r}\right) \cdot \mathsf{d}_f}{\forall \mathsf{d}_a . \Sim^{*}\left(\phi\right)\left(\mathsf{d}_a\right) \Rightarrow \after{\mathsf{d}_r}{\mathsf{d}_t \cdot \mathsf{d}_a}{\Sim^{*}\left(\psi\right)\left(\mathsf{d}_r\right)}}
    \]

    By folding $\Sim^*\left(\phi\right) \supset \Sim^*\left(\psi\right)$, that gives us:
    \[ \Sim^{*}\left(\phi \supset \psi\right)\left(\mathsf{d}_f\right)
    \sqsubseteq
    \after{\mathsf{d}_t}{\lambda\left(\mathsf{r}\right) \cdot \mathsf{d}_f}{\left(\Sim^{*}\left(\phi\right) \supset \Sim^{*}\left(\psi\right)\right)\left(\mathsf{d}_t\right)} \]
    So using $\lambda\left(\mathsf{r}\right)$ as evidence, it means:
    \[ \Sim^{*}\left(\phi \supset \psi\right) \xle{\lambda\left(\mathsf{r}\right)}_2 \Sim^{*}\left(\phi\right) \supset \Sim^{*}\left(\psi\right) \]
    Since $\mathsf{r}$ is in $\Separator$ and $\Separator$ is combinatory complete, the evidence $\lambda\left(\mathsf{r}\right)$ is also in $\Separator$.
    \item \underline{Regularity:}
    By definition, $\Sim^{*}$  preserves unions.
    From Lemma~\ref{lemma:union-coprod}, $\coprod$ is evidently equivalent to $\cup$, so we get that $\Sim^{*}$ preserves $\coprod$ up to evidence.

\end{enumerate}
\end{proof}

\proofappendix[thm:ef-simulation-induces-mca-simulation]{efcompleteness}
\begin{proof}
We show that $\Sim[F]$
defined by $\Sim[F](c,d) \define F(\{c\})(d)$
is an applicative simulation. First, because $F$
is regular and because $\coprod$ can be computed as an union (Lemma~\ref{lemma:union-coprod}),
$F(\phi)$ is equivalent (up to evidence)
to the predicate
$\left\lceil \Sim[F] \right\rceil^{\Omega}(\phi)$
defined by $\left\lceil \Sim[F] \right\rceil^{\Omega}(\phi) \define
\exists c . \; \Sim[F]\left(c,d\right) \sqcap \phi\left(c\right)$.

Because $F$ distributes with implication, there exists a code~$\mathsf{r}$ such that
for all $\mathsf{d} : \mca[2]$:
$$ \after{\mathsf{d_1}}{\mathsf{e}_{\pi_1} \cdot \mathsf{d}}{{\left\lceil \Sim[F] \right\rceil}^{\Omega}(\psi \supset \phi)(\mathsf{d_1})}
\sqcap
\after{\mathsf{d_2}}{e_{\pi_2} \cdot \mathsf{d}}{{\left\lceil \Sim[F] \right\rceil}^{\Omega}(\psi)(\mathsf{d_2})}
\sqsubseteq
\after{\mathsf{d}_r}{\mathsf{r} \cdot \mathsf{d}}{{\left\lceil \Sim[F] \right\rceil}^{\Omega}(\phi)(\mathsf{d}_r)}
$$
So, in particular, for $\mathsf{d}_f , \mathsf{d}_a : \mca[2]$, we replace $\mathsf{d}$ with $\epair{\mathsf{d}_f}{\mathsf{d}_a}$ and get:
\begin{align*}
    \after{\mathsf{d_1}}{\mathsf{e}_{\pi_1} \cdot \epair{\mathsf{d}_f}{\mathsf{d}_a}}{{\left\lceil \Sim[F] \right\rceil}^{\Omega}(\psi \supset \phi)(\mathsf{d_1})}
\sqcap &
\after{\mathsf{d_2}}{e_{\pi_2} \cdot  \epair{\mathsf{d}_f}{\mathsf{d}_a}}{{\left\lceil \Sim[F] \right\rceil}^{\Omega}(\psi)(\mathsf{d_2})} \\
  &  \qquad  \sqsubseteq
\after{\mathsf{d}_r}{\mathsf{r} \cdot \epair{\mathsf{d}_f}{\mathsf{d}_a}}{{\left\lceil \Sim[F] \right\rceil}^{\Omega}(\phi)(\mathsf{d}_r)}
\end{align*}
From this, using the properties of the modality, pairing, and projections, we get:
\[
{\left\lceil \Sim[F] \right\rceil}^{\Omega}(\psi \supset \phi)(\mathsf{d}_f)
\sqcap
{\left\lceil \Sim[F] \right\rceil}^{\Omega}(\psi)(\mathsf{d}_a)
\sqsubseteq
\after{\mathsf{d}_r}{\mathsf{r} \cdot \epair{\mathsf{d}_f}{\mathsf{d}_a}}{{\left\lceil \Sim[F] \right\rceil}^{\Omega}(\phi)(\mathsf{d}_r)}
\]
Unfolding further, it means:
$$
\exists \mathsf{c}_f. \Sim[F](\mathsf{c}_f,\mathsf{d}_f) \sqcap (\psi \supset \phi)(\mathsf{c}_f)
\sqcap
\exists \mathsf{c}_a . \Sim[F](\mathsf{c}_a,\mathsf{d}_a) \sqcap
\psi(\mathsf{c}_a)
\sqsubseteq
\after{\mathsf{d}_r}{\mathsf{r} \cdot \epair{\mathsf{d}_f}{\mathsf{d}_a}}{\exists \mathsf{c}_r . \Sim[F](\mathsf{c}_r,\mathsf{d}_r) \sqcap \phi(\mathsf{c}_r)}
$$
That is, for arbitrary $\mathsf{c}_f , \mathsf{c}_a : \mca[1]$, and unfolding $\psi \supset \phi$, we get:
\begin{align*}
\Sim[F](\mathsf{c}_f,\mathsf{d}_f) \sqcap (\forall \mathsf{c} . \psi(\mathsf{c}) \Rightarrow \after{\mathsf{c}_r}{\mathsf{c}_f \cdot \mathsf{c}}{\phi(\mathsf{c}_r)})
\sqcap&
\Sim[F](\mathsf{c}_a,\mathsf{d}_a) \sqcap
\psi(\mathsf{c}_a)
\\
\sqsubseteq &
\after{\mathsf{d}_r}{\mathsf{r} \cdot \epair{\mathsf{d}_f}{\mathsf{d}_a}}{\exists \mathsf{c}_r . \Sim[F](\mathsf{c}_r,\mathsf{d}_r) \sqcap \phi(\mathsf{c}_r)}
\end{align*}

Now, by taking $\psi$ to be the singleton predicate of $\mathsf{c}_a$, we get:
\begin{align*}
\Sim[F](\mathsf{c}_f,\mathsf{d}_f) \sqcap (\forall \mathsf{c} . \mathsf{c}_a \equiv \mathsf{c} \Rightarrow \after{\mathsf{c}_r}{\mathsf{c}_f \cdot \mathsf{c}}{\phi(\mathsf{c}_r)})
\sqcap
\Sim[F](&\mathsf{c}_a,\mathsf{d}_a) \sqcap
\mathsf{c}_a \equiv \mathsf{c}_a
\\
\sqsubseteq &
\after{\mathsf{d}_r}{\mathsf{r} \cdot \epair{\mathsf{d}_f}{\mathsf{d}_a}}{\exists \mathsf{c}_r . \Sim[F](\mathsf{c}_r,\mathsf{d}_r) \sqcap \phi(\mathsf{c}_r)}
\end{align*}
By removing the trivial components and reorganizing the conjuncts, we now get:
\[ \Sim[F](\mathsf{c}_f,\mathsf{d}_f)
\sqcap
\Sim[F](\mathsf{c}_a,\mathsf{d}_a)
\sqcap \after{\mathsf{c}_r}{\mathsf{c}_f \cdot \mathsf{c}_a}{\phi(\mathsf{c}_r)}
\sqsubseteq
\after{\mathsf{d}_r}{\mathsf{r} \cdot \epair{\mathsf{d}_f}{\mathsf{d}_a}}{\exists \mathsf{c}_r . \Sim[F](\mathsf{c}_r,\mathsf{d}_r) \sqcap \phi(\mathsf{c}_r)}
\]
which is precisely the simulation condition for $\Sim[F]$.

We show that $\Sim[F]$ is evidently total. Let $\mathsf{c} \in \Separator[1]$. We have $\top \xle{\lambda \_. \mathsf{c}} \{\mathsf{c}\}$ and by monotonicity there exists
$\mathsf{e_2} \in \Separator[2]$ such that
$F(\top) \xle{\mathsf{e_2}} F(\{\mathsf{c}\})$. Because $F$
preserves $\top$, there exists $\mathsf{e_1} \in \Separator[2]$
such that $\top \xle{\mathsf{e_1}} F(\top)$.
By composition of the EF, $\top \xle{\mathsf{e_1} ; \mathsf{e_2}}
F(\{c\})$. In particular,
$\after{\mathsf{d}}{\left(\mathsf{e_1} ; \mathsf{e_2}\right) \cdot \id}{\Sim[F](\mathsf{c} , \mathsf{d})}$.

\end{proof}

\proofappendix[prop:MCisom]{mcefcorrespondence}
\begin{proof}
First, notice that the maps $\_^*$ and $\gamma_{\_}$ are order preserving.
We show that they are inverses (up to~$\leq$).
\begin{enumerate}
\item Let $\Sim:\effshell[1]\to\effshell[2]$. Then we have $\Sim^*(\phi)(d) = \exists c.\Sim(c,d)\land \phi(c)$,
and therefore $\Sim[\Sim^*](c,d) = \Sim^*(\{c\})(d) =  \exists c'.\Sim(c',d)\land c' \in \{c\}$. Therefore,
as subobjects of $\mca[1] \times \mca[2]$, $\Sim[\Sim^*] = \Sim$.
\item Let $ F:\ef(\effshell[1])\to\ef(\effshell[2])$.
Then, we have $\Sim[F](c,d) \define F(\{c\})(d)$
and $\Sim[F]^*(\phi) =
\underset{c \in \phi}{\bigcup} F(\{c\})$.
Because $F$ is regular, lemma~\ref{lemma:union-coprod} entails
that $F(\phi)$ is equivalent (up to
uniform evidence) to~$\Sim[F]^*(\phi)$.
\end{enumerate}
\end{proof}

\proofappendix[thm:ef-to-tripos]{eftotripos}
\begin{proof}
This is the evidenced-frame-to-tripos of~\cite{EF2021}, read fiberwise in the
internal logic of~$\E$.
In each context $\Gamma$ the fiber
$\Gamma\to\Phi$ carries the Heyting prealgebra structure of~\Cref{def:evidenced-frame-topos} applied pointwise, the generic predicate is
$\Phi$, substitution is precomposition, and the quantifiers $\coprod_s,\prod_s$
along $s:\Gamma\to\Gamma'$ are the impredicative $\coprod,\prod$ of the EF on the
fibers of $s$.
These and their adjunctions to $s^{*}$ are verified exactly as
in~\cite{EF2021}, now internally.
Only two clauses change form.
\emph{Compatibility}  is required for pullback squares in~$\E$, and
since reindexing is precomposition and $\coprod,\prod$ are computed fiberwise, it
reduces to pullback pasting in~$\E$.
\emph{Reflective surjective substitution} is
stated for regular epimorphisms $s$: a realizer of $s^{*}\phi\sqsubseteq
s^{*}\phi'$ is uniform over $\Gamma$, and $s$ being a regular epi, hence
internally surjective, it is uniform over $\Gamma'$, giving $\phi\sqsubseteq\phi'$.
Finally $\UFam(\ef)(1)=\Hom_\E(1,\Phi)$ is the prealgebra of global propositions
of~$\ef$ under EF entailment, so $\top\sqsubseteq\bot$ in the terminal fiber iff
$\top\xle{e}\bot$ has evidence in~$\ef$.
Thus, $\UFam(\ef)$ is consistent iff
$\ef$ is.
\end{proof}

\proofappendix[thm:ef-simulation-induces-model-morphism]{efsimtomodel}
\begin{proof}
We show that for each~$\Gamma$,
$\UFam(F)_{\Gamma}$ preserves the order and finite meets.
\begin{enumerate}[leftmargin=0.5cm]
    \item
    Since $F$ preserves entailment,
    $\UFam(F)_{\Gamma}$ is an order-preserving function
    between  $\UFam(\ef_1)(\Gamma)$
    and $\UFam(\ef_2)(\Gamma)$.
    \item
    Let $e$ such that $\top \xle{e} F(\top)$ given
    by the definition of a lax morphism.
    For all $\phi \in \Phi_2, \phi \xle{(\etrue ; e)} F(\top)$.
    Because we exhibit a uniform evidence for the
    previous entailment,~$\UFam(F)_{\Gamma}$ preserves~$\top$.
    \item We show that $\UFam(F)_{\Gamma}$ preserves binary meet.
    Let $\phi,\psi \in \Phi_1$.
        We have that
        $F(\phi \wedge \psi) \xle{\epair{\widehat \efst}{\widehat \esnd}}
        F(\phi) \wedge F(\psi)$
            for some evidence~$\widehat \efst,\widehat \esnd$
            obtained
            from~$\efst,\esnd$. We also have
            $$F(\phi) \wedge F(\psi)
            \xle{(\epair{\widehat{\lambda(\eid)};r)}{\eid'};\eeval')}{F(\phi \wedge \psi)}$$
            for some evidence~$\widehat{\lambda(\eid)}$
            obtained
            from~$(\lambda(\eid))$.
Therefore, $\UFam(F)_{\Gamma}$ preserves binary meet.
\end{enumerate}
Finally,  regularity gives
compatibility with existential joins, and hence with the existential
quantifiers of $\UFam$.
\end{proof}

\proofappendix[lemma:triptrans]{triptrans}
\begin{proof}
It is a consequence of the naturality of~$\tau$.
    Namely, $\tau(X)(\chi^* (\id)) \simeq  \chi^* (\tau(\Phi_1) (\id))$,
    remembering that $\chi^* f = f \circ \chi$.
\end{proof}

\proofappendix[prop:appx-ef-1]{cartesiantolax}
\begin{proof}
We show that $F_{\tau} : \Phi_1 \to \Phi_2$ satisfies the three conditions of a lax simulation.
    \begin{enumerate}
        \item \underline{Monotonicity:} Let $e\in E_1.$
        Consider the subobject~$G_e$ of $\Phi_1^2$ defined by
        $\{ \langle \phi , \psi \rangle \mid \phi \xle{e} \psi \}$
        and $\iota : G_e \to \Phi_1^2$. We define $s_e,t_e : G_e \to \Phi_1$
        by $s_e \triangleq \pi_1 \circ \iota$
        and $t_e \triangleq \pi_2 \circ \iota$.
        It defines two predicates $s_e,t_e : \UFam(\ef_1)(G_e)$
        and $s_e \leq t_e$.
        By monotonicity of~$\tau$, $\tau (s_e) \leq \tau (t_e)$
        in $\UFam(\ef_2)(G_e)$. It means that there exists $e' \in E_2$
        such that for all $(\phi,\psi) \in G_e,
        F_{\tau}(\phi) \xle{e'} F_{\tau}(\psi)$.
        \item \underline{Top preservation:}
        Because $\tau$ is cartesian, in $\UFam(\ef_2)(1)$,
        we have $\lambda \_. \top_2 \leq F_{\tau} \circ (\lambda \_ . \top_1)$.
        It means that there exists $e \in E_2$
        such that $\top_2 \xle{e} F_{\tau}(\top_1)$.
        \item \underline{Implication distributivity:}
        We consider $\chi : \Phi_1^2 \to \Phi_1$ defined
        by $\chi(\phi,\psi) = (\phi \arrowE \psi)$.
        In~$\UFam(\ef_1)(\Phi_1^2)$, we have
        $\chi \wedge \pi_1 \leq \pi_2$ and therefore, in~$\UFam(\ef_2)(\Phi_1^2)$,
        we obtain $\tau (\chi \wedge \pi_1) \leq \tau(\pi_2)$.
        Because $\tau$ is cartesian, it gives,
        $\tau (\chi) \wedge \tau(\pi_1) \leq \tau(\pi_2)$
        and then
        $\tau (\chi) \leq \tau(\pi_1) \arrowE \tau(\pi_2)$.
        In particular, there exists $r \in E_2$ such that for all
        $(\phi,\psi) \in \Phi_1^2$, $F_{\tau}(\phi \arrowE \psi)  \xle{r}
        F_{\tau}(\phi) \arrowE F_{\tau}(\psi)$.
    \end{enumerate}
    We now assume that $\tau$ is regular.
    Consider the suboject $P$ of $\Phi \times \Omega^{\Phi}$
    defined by $P \triangleq \{ (\phi , \vec{\phi}) \mid \phi \in \vec{\phi}\}$ and $\iota : P \to \Phi \times \Omega^{\Phi}$.
    Write $\pi_1^{\iota}$ for $\pi_1 \circ \iota $ and
    $\pi_2^{\iota}$ for $\pi_2 \circ \iota $.
    Because $\tau$ is regular,
    $\tau (\coprod_{\pi_2^{\iota}}^1 (\pi_1^{\iota} )) \leq \coprod_{\pi_2^{\iota}}^2 (\tau (\pi_1^{\iota} ))$
    in $\UFam(\ef_2)(\Omega^{\Phi})$.
    It means that
    $F_{\tau} \circ \coprod_{\pi_2^{\iota}}^1 (\pi_1^{\iota}) \leq
    \coprod_{\pi_2^{\iota}}^2 (F_{\tau} \circ \pi_1^{\iota})$.
    This gives the uniform evidence $e_{\exists}$ needed to show
    that $F_{\tau}$ is regular.
\end{proof}

\proofappendix[prop:UFamisom]{ufamcorrespondence}
\begin{proof}
The map $\UFam(\_)$ sends a lax simulation $F$ to the cartesian transformation $\chi \mapsto F \circ \chi$
between the effectful triposes. Moreover, it respects the order (because both
of the orders are computed pointwise).
The map $F_{\_}$ sends a cartesian
morphism $\tau$ to the lax simulation
$\phi \mapsto \tau(\Phi_1)(\id)(\phi)$. Moreover, if $\tau_1 \leq \tau_2,
\tau_1(\Phi_1)(\id) \leq \tau_2(\Phi_1)(\id)$ and therefore
$F_{\tau_1} \leq F_{\tau_2}$. Finally, $\UFam(F_{\tau}) \simeq \tau$ and $F_{\UFam(F)} = F$.
\end{proof}

\proofappendix[thm:dense-ef-simulation-logical-transport]{geomtransport}
\begin{proof}
We show that the restriction of~$\UFam$ to geometric morphisms
is well-defined and
that the map $F_{\_} : \tripcart(\UFam(\ef_1),\UFam(\ef_2)) \to
\ef(\ef_1,\ef_2)$ sends a geometric transformation of triposes into a
geometric simulation of evidenced frames.
\begin{itemize}
    \item
    If $H$ is a right adjoint of $F$ then~$\UFam(H)$ is
    a right adjoint of $\UFam(F)$.
    Indeed the internal statements given by the assumption that $F$
    is geometric implies that $\id \leq \UFam(H) \circ \UFam(F)$
    and $\UFam(F) \circ \UFam(H) \leq \id$. Therefore, the functor $\UFam$
    sends a geometric EF simulation to a geometric tripos transformation.
    \item
    Take $\tau_1 : \UFam(\ef_1) \to \UFam(\ef_2)$
    and $\tau_2 : \UFam(\ef_2) \to \UFam(\ef_3)$.
    We apply lemma~\ref{lemma:triptrans}: $F_{\tau_2 \circ \tau_1} =
    (\tau_2 \circ \tau_1)_{\Phi_1}(\id)=
    \tau_2 (\tau_1(\Phi_1)(\id))
    \simeq (\tau_2(\Phi_2)(\id)) \circ (\tau_1(\Phi_1)(\id)) = F_{\tau_2}
    \circ F_{\tau_1}$.
    Moreover $F_{\_}$ sends the identity transformation to the identity
    simulation.
    It implies (using Theorem~\ref{prop:UFamisom}) that if
     $\tau_1 \dashv \tau_2$ then $F_{\tau_1} \dashv F_{\tau_2}$ (expressed
    with the unit and the counit of the adjunction). Therefore, if $\tau$
    is a geometric transformation of triposes, $F_{\tau}$ is a geometric
    $EF$ simulation. As a consequence,
    proposition~\ref{prop:UFamisom} implies that $\UFam$ is an
    isomorphism of preorders
    between $\efcd(\ef_1,\ef_2)$ and $\tripgeom(\UFam(\ef_1),\UFam(\ef_2))$.
\end{itemize}
\end{proof}

\proofappendix[prop:appx-ef-2]{cdgeometric}
\begin{proof}~
\begin{enumerate}
    \item
Let $\evCD  : 1 \to \Separator[2]$ be a witness of the computational density of $\Sim$.
    For $\phi_2 : \Omega^{\mca[2]}$,
    define $\Sim[\ast] (\phi_2) \triangleq
    \{\mathsf{a} \mid \Sim^*(\{\mathsf{a}\}) \xle{\evCD} \phi_2\} : \Omega^{\mca[1]}$.
    \begin{itemize}
        \item \underline{$\Sim[\ast]$ is monotone:}
        Let $\mathsf{b} : \mca[2]$ and apply computational
        density to the code $\evCD ; \mathsf{b}$ (where $\_;\_$
        is the composition in the corresponding EF) to
        get a code $\mathsf{a}$.
        Assume $\phi_2 \xle{\mathsf{b}} \psi_2$, we
        show that
        $\Sim[\ast](\phi_2) \xle{\mathsf{a}} \Sim[\ast](\psi_2)$. Therefore take $\mathsf{a'} : \Sim[\ast] (\phi_2)$, it means $\Sim^*(\{\mathsf{a'}\}) \xle{\evCD}  \phi_2$. But then, $\Sim^*(\{\mathsf{a'}\}) \xle{\evCD;
        \mathsf{b}}  \psi_2$. By computational density,
        we have $\after{\mathsf{c_a}}{\mathsf{a} \cdot
        \mathsf{a'}}{\Sim[\ast](\psi_2)(\mathsf{c_a})}$.
        \item \underline{$\Sim[\ast]$ is a right adjoint:}
        We need to find $\eta$ and $\epsilon$ such that
        for all $\phi_1 : \Omega^{\mca[1]}$ and
        $\phi_2 : \Omega^{\mca[2]}$:
        $$ \phi_1 \xle{\eta} \Sim[\ast] \Sim^{\ast} (\phi_1)
        \qquad \textrm{and} \qquad
        \Sim^{\ast} \Sim[\ast] (\phi_2)
        \xle{\epsilon} \phi_2.
        $$
        \begin{itemize}
            \item \underline{unit:}
            Take $\eta$ to be given by applying computational
            density to the code $\eid : \mca[2]$.
            Let~$ \mathsf{a} : \phi_1$ and observe that
            $\Sim^*(\{\mathsf{a}\}) \xle{\eid} \Sim^{\ast}(\phi_1)$
            (because $\Sim^{\ast}(\phi) = \underset{a \in \phi}{\bigcup}\Sim(\mathsf{a})$).
            Therefore, by computational density,
            $\after{\mathsf{c_a}}{\eta \cdot \mathsf{a}}{\Sim[\ast]\Sim^{\ast}(\phi)(\mathsf{c_a})}$.
            \item \underline{counit:} Define $\epsilon \define
            \evCD$ and unfold the definition of~$\Sim[\ast]$.
        \end{itemize}
    \end{itemize}
    \item
Let $F \dashv H$ witnessed by the unit $\eta \in \mca[1]$ and the
    counit $\epsilon \in \mca[2]$.
    Because $F$ is geometric, it is also regular and
    we know from Theorem~\ref{thm:ef-simulation-induces-mca-simulation} that $\Sim[F]$ is an applicative simulation. We show
    that it is computationally dense.
    First, consider an evidence $\mathsf{e} : \mca[2]$
    such that for all $\phi_1 : \Omega^{\mca[1]},
    \underset{\mathsf{c_a} \in \phi_1}{\bigcup}\Sim[F](c_a)
    \xle{\mathsf{e}}  F(\phi_1)$.
    The fact that  $\Sim[F]$ is computationally dense is witnessed
    by the code $\mathsf{e};\epsilon$, i.e. we define $\evCD \define \mathsf{e}; \epsilon$.
    Let $\mathsf{b} : \mca[2]$ and consider
    the code $\mathsf{a_b} : \mca[1]$ given by the monotonicity
    of $H$. We show that $\mathsf{a} \define \eta ; \mathsf{a_b}$ satisfies the required property
    of computational density.
    Let $\mathsf{a'}: \mca[1], \phi_2 : \Omega^{\mca[2]}$ and
    assume $\Sim[F]^*(\{\mathsf{a'}\}) \xle{b} \phi_2$.
    We therefore have $HF(\{\mathsf{a'}\}) \xle{\mathsf{a_b}} H(\phi_2)$.
    It follows that
    $\{\mathsf{a'}\} \xle{\eta ; \mathsf{a_b}} H(\phi_2)$.
    Finally, the assumption on the code~$\mathsf{e}$
    and the property of the counit~$\epsilon$ gives that if
    $c_a \in H(\phi_2)$, $\Sim[F](c_a) \xle{ e ; \epsilon}
    \phi_2$. All in all, $\after{c_a}{\mathsf{a} \cdot
    \mathsf{a'}}{\Sim[F](c_a) \xle{\evCD} \phi_2}$.
    \end{enumerate}
\end{proof}

\subsection{Proofs from Section~\ref{sec:transport-modalities}}

\proofappendix[prop:appx-dermod-1]{prop:nucleustoef}
\begin{proof}[Proof of Proposition~\ref{prop:appx-dermod-1}]
Let $\eR$ be the return evidence, $\eBind$ the bind evidence.
We first construct an evidence $e_s$ such that
 $ \vdashE{\E} \forall \phi_1, \phi_2 \! : \! \Phi. \;  j(\phi_1) \wedge j(\phi_2)
    \xle{e_s} j(\phi_1 \wedge \phi_2)$.
    With a multiple use of~$\eBind$, this statement reduces to
    $\phi_1 \wedge \phi_2
    \xle{e_s'} j(\phi_1 \wedge \phi_2)$, which is then obtained by~$\eR$.

The evidences witnessing that~$\efj$ is an EF are given below:
$$
\begin{array}{rcl}
        e_{\id,j} & \define & \eR \\
        e ;_j e' & \define & e ; \epair{\etrue ; \lambda (\esnd ; e'); \eBind}{\eid} ; \eeval \\
        e_{\top,j} & \define & \etrue ; \eR \\
        e_{\texttt{fst},j} & \define & \efst ; \eR \\
        e_{\texttt{snd},j} & \define & \esnd ; \eR \\
        \epair{e}{e'}_j & \define & \epair{e}{e'} ; e_s\\
         \lambda_j e & \define & \lambda e ;\eR \\
        \eevalj & \define & \eeval \\
\end{array}
$$
In the definition of $e ;_j e'$, the sequence of evidences
$\etrue ; \lambda (\esnd ; e'); \eBind$ satisfies :
$$
j(\phi_2) \xle{\etrue} \top \xle{\lambda \esnd ; e'} \phi_2 \supset j (\phi_3)
\xle{\eBind} j (\phi_2) \supset j(\phi_3).
$$
\end{proof}

\proofappendix[prop:appx-dermod-2]{prop:nucleusadj}
\begin{proof}[Proof of Proposition~\ref{prop:appx-dermod-2}]
To show that $id : \ef \to \efj$ is an EF simulation:
Monotonicity is obtained using composition and return, top preservation  is obtained using return.
For implication preservation,  we need evidence for
$(\phi \arrowE \psi) \xle{} j(\phi \arrowE j(\psi))$ in $\ef$.
Since $(\phi \arrowE \psi) \wedge \phi \xle{\ecomp{\eeval}{\eR}} j(\psi)$,
abstraction gives
$(\phi \arrowE \psi) \xle{\elambda{(\ecomp{\eeval}{\eR})}} \phi \arrowE j(\psi)$,
and composing with $\eR$ closes the conclusion under~$j$.
Similarly, $j : \efj \to \ef$ is an EF simulation.
Since $j(\phi) \xlej{\eid} \phi$ and $\phi \xle{\eR} j(\phi)$ (bind), we get $id \dashv j$.
\end{proof}

\proofappendix[prop:appx-dermod-2]{prop:adjnucleus}
\begin{proof}[Proof of Proposition~\ref{prop:appx-dermod-4}]
We give explicitly the return and bind evidence.
\begin{enumerate}
    \item $\eR$ is given by the evidence $e_\eta$ (witnessing
    the unit of the adjunction $F \dashv H$).
    \item We find
            $\eBind$ such that $
        \forall \phi_1 \phi_2,
        \phi_1 \arrowE HF(\phi_2) \xle{\eBind} HF(\phi_1) \arrowE HF(\phi_2)$.
        It is enough to show
        $
        \phi_1 \arrowE HF(\phi_2) \xle{e} H(F(\phi_1) \arrowE F(\phi_2))$
        for an evidence $e$ (because $H$ is a lax morphism).
    We will rather show that there exists $e$ such that
    $
        F(\phi_1 \arrowE HF(\phi_2)) \xle{e} F(\phi_1) \arrowE F(\phi_2)$.
        But, because $F$ is a lax morphism,
        $F(\phi_1 \arrowE HF(\phi_2)) \xle{r_F} F(\phi_1) \arrowE FHF(\phi_2)$.
    It remains to show $F(\phi_1) \arrowE FHF(\phi_2) \to  F(\phi_1) \arrowE F(\phi_2)$ which follows from $\eeval$ and the counit of the
    adjunction.
\end{enumerate}
\end{proof}

\proofappendix[prop:factor]{propfactor}
\begin{proof}[Proof of Proposition~\ref{prop:factor}]
We show that $F : \ef_{HF} \to \ef_2$ is a lax morphism. We write~$\eta$ and~$\epsilon$
for the unit and the counit of the adjunction~$ F \dashv H$.
\begin{itemize}
    \item \underline{Monotonicity:} Assume $\phi_1 \xle{e}_1 HF(\phi_2)$.
    By monotonicity of $F$, there exists $e'$ such that
    $F (\phi_1) \xle{e'}_2 FHF(\phi_2)$. We get monotonicity by composing
    by the counit~$\epsilon$ of the adjunction.
    \item \underline{Top preservation:} $F : \ef_1 \to \ef_2$ is lax and therefore preserves $\top$.
    \item \underline{Implication distributivity:} We need to show
    that $F(\phi_1 \supset HF (\phi_2)) \xle{r}_2 F(\phi_1) \supset F (\phi_2)$.
    It is obtained using that $F$ distributes over the implication
    and the counit~$\epsilon$ of the adjunction.
\end{itemize}
A similar proof shows that $H : \ef_2 \to \ef_{HF}$ is a lax simulation.
Moreover, $FH (\phi) \xle{\epsilon}_2 \phi$ as $F \dashv H$
and $\phi \xle{\eta ; \eta} HFHF(\phi)$, i.e. $\phi \xle{\eta ;\eta}_{HF} HF(\phi)$.
\end{proof}

\proofappendix[thm:modal-ef-sound]{modalefsound}
\begin{proof}
All reasoning is internal to $\E$, using the axioms of
\Cref{def:modality-nucleus}. By functional completeness of $S$ every closed term
built from application and the code pairing $\cpair{\cdot}{\cdot}$ is realized by a
code in $S$.
We use the projections $\mathsf{e}_{\pi_1},\mathsf{e}_{\pi_2}$ with
$\mathsf{e}_{\pi_i}\cdot\cpair{c_1}{c_2}=c_i$ and  realize the evidence combinators of
\Cref{fig:ef} by
\[
\begin{array}{lll}
\eid \define \llbracket\lambda x.\,x\rrbracket, &
\ecomp{e}{e'} \define \llbracket\lambda x.\,e'(e\,x)\rrbracket, &
\etrue \define \eid,\\[2pt]
\efst \define \mathsf{e}_{\pi_1}, &
\esnd \define \mathsf{e}_{\pi_2}, &
\epair{e_1}{e_2} \define \llbracket\lambda x.\,\cpair{e_1\,x}{e_2\,x}\rrbracket,\\[2pt]
\elambda{e} \define \llbracket\lambda a\,b.\,e\,\cpair{a}{b}\rrbracket, &
\eeval \define \llbracket\lambda p.\,(\efst\,p)(\esnd\,p)\rrbracket, &
\end{array}
\]
so that $\efst\cdot\cpair{c_1}{c_2}=c_1$, $\esnd\cdot\cpair{c_1}{c_2}=c_2$,
$\epair{e_1}{e_2}\cdot c=\cpair{e_1\cdot c}{e_2\cdot c}$,
$\ecomp{e}{e'}\cdot c=e'\cdot(e\cdot c)$,
$(\elambda{e}\cdot a)\cdot b=e\cdot\cpair{a}{b}$, and $\eeval\cdot\cpair{f}{c}=f\cdot c$.

\medskip\noindent\textsc{(refl)}\quad
$\eid\cdot c$ terminates with reduct $c$, so $\phi(\sigma,c)$ gives
$\eid\cdot c\Downarrow_\sigma\phi$, hence $\Box(\eid\cdot c\Downarrow_\sigma\phi)$
by extensivity.

\medskip\noindent\textsc{(trans)}\quad
Assume $\phi_1\xle{e}_\Box\phi_2$ and $\phi_2\xle{e'}_\Box\phi_3$, and let
$\phi_1(\sigma,c)$. We first show
$(e\cdot c\Downarrow_\sigma\phi_2)\sqsubseteq
\Box(\ecomp{e}{e'}\cdot c\Downarrow_\sigma\phi_3)$: assuming
$e\cdot c\Downarrow_\sigma\phi_2$, some reduct $(\sigma',t)$ satisfies
$\phi_2(\sigma',t)$, so $\phi_2\xle{e'}_\Box\phi_3$ gives
$\Box(e'\cdot t\Downarrow_{\sigma'}\phi_3)$.
Since
$\ecomp{e}{e'}\cdot c=e'\cdot(e\cdot c)$ reduces through $t$, the raw
$e'\cdot t\Downarrow_{\sigma'}\phi_3$ implies
$\ecomp{e}{e'}\cdot c\Downarrow_\sigma\phi_3$, and monotonicity gives the inclusion.
Now $\phi_1\xle{e}_\Box\phi_2$ yields $\Box(e\cdot c\Downarrow_\sigma\phi_2)$.
Applying $\Box$ to the inclusion and using idempotence,
$\Box(\ecomp{e}{e'}\cdot c\Downarrow_\sigma\phi_3)$.

\medskip\noindent\textsc{(top)}\quad
As for \textsc{(refl)}: $\etrue\cdot c=\eid\cdot c$ terminates and its reduct satisfies
$\top$, so $\Box(\etrue\cdot c\Downarrow_\sigma\top)$.

\medskip\noindent\textsc{(fst)}, \textsc{(snd)}\quad
If $(\phi_1\wedge\phi_2)(\sigma,p)$ then $p=\cpair{c_1}{c_2}$ with
$\phi_1(\sigma,c_1)$, $\phi_2(\sigma,c_2)$. As $\efst\cdot p$ terminates with
reduct $c_1$, $\Box(\efst\cdot p\Downarrow_\sigma\phi_1)$ by extensivity, i.e.\
$\phi_1\wedge\phi_2\xle{\efst}_\Box\phi_1$.  Symmetrically
$\phi_1\wedge\phi_2\xle{\esnd}_\Box\phi_2$.

\medskip\noindent\textsc{(pair)}\quad
Assume $\phi\xle{e_1}_\Box\phi_1$, $\phi\xle{e_2}_\Box\phi_2$, and
$\phi(\sigma,c)$. By finite-meet preservation,
$\Box(e_1\cdot c\Downarrow_\sigma\phi_1)\sqcap\Box(e_2\cdot c\Downarrow_\sigma\phi_2)
=\Box\big((e_1\cdot c\Downarrow_\sigma\phi_1)\sqcap(e_2\cdot c\Downarrow_\sigma\phi_2)\big)$.
Choosing accepting reducts $t_1,t_2$, the computation
$\epair{e_1}{e_2}\cdot c=\cpair{e_1\cdot c}{e_2\cdot c}$ has reduct $\cpair{t_1}{t_2}$
with $(\phi_1\wedge\phi_2)(\cpair{t_1}{t_2})$, so the conjoined judgement implies
$\epair{e_1}{e_2}\cdot c\Downarrow_\sigma(\phi_1\wedge\phi_2)$. Then, by monotonicity we get $\phi\xle{\epair{e_1}{e_2}}_\Box\phi_1\wedge\phi_2$.

\medskip\noindent\textsc{(curry)}\quad
Assume $\phi_1\wedge\phi_2\xle{e}_\Box\phi$ for all $\phi\in_{\Phi_0}\vec\phi$
and $\phi_1(\sigma,c_1)$.
As $\elambda{e}\cdot c_1$ terminates with reduct
$f\define\elambda{e}\cdot c_1$, by extensivity it suffices to show
$(\phi_2\imp\vec\phi)(\sigma,f)$. Take $\sigma'\ge\sigma$, $c_2$, and
$\phi\in_{\Phi_0}\vec\phi$ with $\phi_2(\sigma',c_2)$. Future stability gives
$\phi_1(\sigma',c_1)$, hence $(\phi_1\wedge\phi_2)(\sigma',\cpair{c_1}{c_2})$.
The premise yields $\Box(e\cdot\cpair{c_1}{c_2}\Downarrow_{\sigma'}\phi)$, and since
$f\cdot c_2$ behaves as $e\cdot\cpair{c_1}{c_2}$,
$\Box(f\cdot c_2\Downarrow_{\sigma'}\phi)$.

\medskip\noindent\textsc{(eval)}\quad
Let $\phi\in_{\Phi_0}\vec\phi$ and
$((\phi_1\imp\vec\phi)\wedge\phi_1)(\sigma,p)$, so $p=\cpair{f}{c}$ with
$(\phi_1\imp\vec\phi)(\sigma,f)$ and $\phi_1(\sigma,c)$. Instantiating $\imp$
at $\sigma'=\sigma$, $c'=c$, and $\phi$ gives $\Box(f\cdot c\Downarrow_\sigma\phi)$,
and since $\eeval\cdot p$ behaves as $f\cdot c$, $\Box(\eeval\cdot p\Downarrow_\sigma\phi)$.
\end{proof}

\end{document}